\documentclass[%
 reprint,
superscriptaddress,
 amsmath,amssymb,
 aps,
]{revtex4-2}

\def\doi#1{\relax}
\def\url#1{\relax}
\def\eprint#1{\relax}

\usepackage{graphicx}
\usepackage{dcolumn}
\usepackage{bm}
\usepackage{subcaption}
\usepackage{color}
\usepackage{float}

\usepackage[colorinlistoftodos,prependcaption]{todonotes}
\begin{document}

\preprint{APS/123-QED}

\title{
Lattice Thermal Conductivity from First Principles and Active Learning with Gaussian Process Regression
}


\author{Rasmus Tranås}

\affiliation{
 Department of Mechanical Engineering and Technology Management, \\ Norwegian University of Life Sciences, NO-1432 Ås, Norway
}%

\author{Ole Martin Løvvik}
\affiliation{Centre for Materials Science and Nanotechnology, Department of Physics, University of Oslo, NO-0316 Oslo, Norway
}%
\affiliation{ SINTEF Sustainable Energy Technology, NO-0314 Oslo, Norway
}%

\author{Kristian Berland}
\affiliation{  Department of Mechanical Engineering and Technology Management, \\ Norwegian University of Life Sciences, NO-1432 Ås, Norway
}%

\date{\today}

\begin{abstract}
The lattice thermal conductivity ($\kappa_{\ell}$) is a key materials property in power electronics, thermal barriers, and thermoelectric devices.
Identifying a wide pool of compounds with low $\kappa_{\ell}$ is 
particularly important in the development of materials with high thermoelectric efficiency.
The present study contributed to this with a reliable machine learning (ML) model based on a training set consisting of 268 cubic compounds. For those, $\kappa_{\ell}$ was calculated 
from first principles using the temperature-dependent effective potential (TDEP) method based on  forces and phonons calculated by density functional theory (DFT).
238 of these were preselected and used to train an initial ML model employing Gaussian process regression (GPR). The model was then improved with active learning (AL) by selecting the 30 compounds with the highest GPR uncertainty as new members of an expanded training set.
This was used to predict $\kappa_{\ell}$ of the 1574 cubic compounds in the \textsc{Materials Project} (MP) database with a validation R2-score of 0.81 and Spearman correlation of 0.93.
Out of these, 27 compounds were predicted to have very low values of $\kappa_{\ell}$ ($\leq 1.3$ at 300~K), which was verified by DFT calculations. 
Some of these have not previously been reported in the literature, suggesting further investigations of their electronic thermoelectric properties.
\end{abstract}

\maketitle


\section*{\label{sec:introduction}Introduction}

The thermal conductivity  
is an important materials parameter;
high thermal conductivity is, e.g.,\ needed to divert heat away from transistor components,
while low thermal conductivity is needed for thermal barrier materials in jet engines 
and for efficient thermoelectric devices. 
The latter can convert heat into electricity and vice versa
and are used both for cooling and harvesting of waste heat.\cite{SnyderComplexThermoelectric, RoomTempTEEnergyConsum, WasteHeatRegeneration}.

The importance of low lattice thermal conductivity $\kappa_\ell$ is evident from
the thermoelectric figure of merit,
\begin{equation}
    ZT = \dfrac{\sigma S^{2} T}{\kappa_{\ell} + \kappa_{e}}\,.
\end{equation}
\noindent 
A high $ZT$ requires simultaneously a high conductivity $\sigma$ and Seebeck coefficient $S$ as well as a low electronic, $\kappa_e$, and lattice thermal $\kappa_\ell$ conductivity. 
The electronic properties are interlinked and can be optimized by tuning the charge carrier concentration~\cite{SnyderComplexThermoelectric}.
Unfortunately, many materials with attractive electronic properties, 
such as the half-Heuslers suffer from large $\kappa_\ell$~\cite{gaultoisImprovementFromReducingKappa}. 
While the introduction of grain boundaries~\cite{RoweGBSiGe, SchradeGrainBoundary}, point defects~\cite{DefectEngineeringReview, DefectEngineeringTE2}, an alloying~\cite{IsovalentCdinZnSb, IsovalentSubstitutionsinChalcopyrite, berlandThermoelectricTransportTrends2019, eliassenLatticeThermalConductivity2017, TransAttaining}, 
can impede thermal transport, they also reduce the electronic mobility and hence $\sigma$~\cite{NbFeSbAlloyTewoReducingSigma, NbCoSnGBsIncreaseTransport, BiTePointDefect}.
Hence, it is not surprising that compounds with the highest thermoelectric performance typically
possess intrinsically low lattice thermal conductivity, i.e., $\kappa_\ell<1$~W/Km at room temperature. Examples include PbTe~\cite{PbTeExperiment, PbTeComputational}, $\rm{Bi}_{2}\rm{Te}_3$~\cite{BismuthTellurideFirstExp, BismuthTellurideReview}, and SnSe~\cite{SnSePolyCrystalline, SnSeComputational}. 
Obtaining a wide pool of 
compounds with low intrinsic $\kappa_{\ell}$ thus provides a good starting point for finding promising thermoelectric materials. 
Such a wide pool also makes it more likely to find compounds that also consist of non-toxic and abundant elements,\cite{EnvoronmentallyFriendlyTEMaterials, TEToxcicity} 
suited for a broader range of applications. 

Experimentally measuring $\kappa_{\ell}$ is time-consuming and requires expert training and costly equipment. 
The measured materials usually contain 
grain boundaries and other defects, including vacancies, pores, precipitations, and dislocations. 
All this influences the experimental results, making it harder to assess which materials have the most promise. 
Also, typically the total $\kappa = \kappa_e + \kappa_\ell$ is measured, and $\kappa_\ell$ is extracted using the Wiedemann-Franz law; $\kappa_e = \sigma LT$, where $L$ is the Lorentz number. Since $L$ can vary significantly between different materials, this adds to the uncertainty of experimentally reported values of $\kappa_\ell$.
So far, 
experimental $\kappa_{\ell}$ values of only $\sim 100$-$200$ compounds have been reported~\cite{ChenGPRLTC, gaultoisImprovementFromReducingKappa, MLinMaterialInformaticsMethodsandApplications}. Efficient and accurate methods to predict $\kappa_{\ell}$ can thus greatly advance the prediction of novel thermoelectrics. 

It has in recent years become possible to calculate $\kappa_\ell$ from first principles, for instance, by solving the Boltzmann transport equations with input from density functional theory (DFT).
This type of calculation bears a substantial computational cost because 
 calculating three-phonon scattering rates requires many DFT calculations of different atomic configurations in large supercells. 
Most software packages employ the relaxation-time approximation to predict $\kappa_\ell$, including \textsc{Phono3py}\cite{phono3py}, the temperature-dependent effective potential (\textsc{TDEP}) method~\cite{hellmanLatticeDynamicsAnharmonic2011, hellmanTemperaturedependentEffectiveThirdorder2013}, \textsc{hiPhive}~\cite{CompressiveSensing,hiphive}, and \textsc{shengBTE}~\cite{LI20141747}. 
Despite the high cost, screening studies based on DFT have
identified several new compounds with low $\kappa_{\ell}$~\cite{TeScreeningReview, LayereTEScreening, ScreeningTEBinary, LTCScreeningRattling, HighThroughputRattler} 
and the list of computed $\kappa_{\ell}$ values is 
growing~\cite{CarreteFindingLowLTCHHs, fengCharacterizationRattlingRelation2020}. 
Based on such input and experimental measurements, 
machine learning (ML) studies are also increasingly being used to predict $\kappa_{\ell}$~\cite{CarreteFindingLowLTCHHs, SekoBayesianLowLTC, LowLTCPerovskiteMLScreeening, JunejaHTGPR, WangMLScreeningLowLTCAflow, JaafrehAccelerated, WolvertonLTCML, MLPLTC, OleMartinScreeningSilicides, JunejaMLLTC, MLLTCMiniReview}. 
The advent of structured material databases, such as \textsc{Materials Project}~\cite{jainCommentaryMaterialsProject2013}, \textsc{OQMD}~\cite{OQMD}, and \textsc{NOMAD}~\cite{NOMAD} has furthered allowed high-throughput screening based on DFT to be conveniently extended with ML, as in this paper. 

Training sets 
used for making ML models for $\kappa_{\ell}$ have, so far, been quite modest in size, only covering $\sim40$-$200$ compounds.\cite{ChenGPRLTC, CarreteFindingLowLTCHHs, JaafrehAccelerated, MLinMaterialInformaticsMethodsandApplications}
Despite the small size, these models have been used to predict $\kappa_{\ell}$ 
of thousands or even tens of thousands of compounds.
Also, training sets that are taken from the literature or that are based on 
a limited set of materials types 
can be highly imbalanced.
It is thus likely that many predictions are poorly represented by the training set and thus suffering from the limited extrapolation power of ML~\cite{StrategyMLSmallDatasets,TranasHHActiveSampling}. 

We have in the present study addressed several of the previous shortcomings.
To confront the issue of a small training set, we first started with a large DFT-based training set of 
238 compounds. Second, to achieve a better balanced and diverse training set, we adopted an active learning (AL) procedure using Gaussian process regression (GPR). This method provides a prediction uncertainty, $\sigma^*_{GPR}$, and this uncertainty was used to expand the training set by an additional 30 compounds with high uncertainty. 
Third, the selection of features was limited to basic atomic and crystal-packing properties
using Voronoi cells, which facilitates using the model without any additional DFT calculations or experimental activities.
The ML model was finally used on 1574 cubic compounds to predict several hitherto unknown systems with very low $\kappa_{\ell}$, out of which the 29 lowest were verified with DFT using TDEP.
Details of our approach can be found in the Methods section.

\begin{table*}[t!]
\caption{Calculated $\kappa_\ell^{\mathrm{DFT}}$ at 300~K for 268 compounds used in training, the actively sampled shown in bold.
\label{tbl:kappa_tdep}}
\begin{ruledtabular}
\begin{tabular}{ll@{\hspace{0.8cm}}ll@{\hspace{0.8cm}} ll@{\hspace{0.8cm}} ll@{\hspace{0.8cm}} ll@{\hspace{0.8cm}} ll@{\hspace{0.8cm}} ll@{\hspace{0.8cm}} }
  \multicolumn{3}{l}{Compound\qquad  $\kappa_\ell$~[W/Km] } \\
  \hline
$\rm{Ag}_{2}\rm{S}$ & 0.19 & $\rm{Ca}_{2}\rm{Hg}\rm{Pb}$ & 2.06 & $\rm{Rb}\rm{Br}$ & 5.58 & $\rm{Zr}\rm{As}\rm{Ir}$ & 11.7 & $\rm{Ta}\rm{Rh}\rm{Sn}$ & 17.6 & $\rm{Ta}\rm{Os}\rm{Sb}$ & 24.7\\
$\rm{Cs}\rm{Rb}_{3}$ & 0.20 & $\rm{\bf{Ba}}\rm{\bf{Pd_{2}}}$ & 2.09 & $\rm{Ca}_{2}\rm{Ge}$ & 5.72 & $\rm{Zr}\rm{Rh}\rm{Bi}$ & 11.7 & $\rm{Nb}\rm{Os}\rm{Sb}$ & 17.6 & $\rm{Nb}\rm{Os}\rm{As}$ & 25.2\\
$\rm{K}\rm{Rb}_{3}$ & 0.24 & $\rm{Cs}\rm{F}$ & 2.20 & $\rm{\bf{Br}}\rm{\bf{K}}$ & 5.98 & $\rm{Sc}\rm{Ni}\rm{P}$ & 11.8 & $\rm{Zr}\rm{Ir}\rm{Sb}$ & 17.6 & $\rm{V}\rm{Co}\rm{Ge}$ & 25.3\\
$\rm{K}_{3}\rm{Na}$ & 0.29 & $\rm{Rb}\rm{I}$ & 2.22 & $\rm{Li}\rm{Zn}\rm{Sb}$ & 6.01 & $\rm{Nb}\rm{Rh}\rm{Pb}$ & 11.8 & $\rm{Zr}\rm{Rh}\rm{Sb}$ & 17.8 & $\rm{Ta}\rm{Co}\rm{Ge}$ & 25.5\\
$\rm{Cu}\rm{Cl}$ & 0.38 & $\rm{Rb}_{2}\rm{Te}$ & 2.26 & $\rm{\bf{Cu_{3}}}\rm{\bf{Li}}\rm{\bf{O_{3}}}$ & 6.50 & $\rm{V}\rm{Rh}\rm{Sn}$ & 11.9 & $\rm{Hf}\rm{Ni}\rm{Ge}$ & 17.8 & $\rm{Ta}\rm{Rh}\rm{Ge}$ & 25.7\\
$\rm{Na}_{2}\rm{Tl}\rm{Bi}$ & 0.39 & $\rm{Ba}\rm{Zn}$ & 2.29 & $\rm{Ca}_{3}\rm{Sb}\rm{N}$ & 7.02 & $\rm{Sc}\rm{Te}\rm{Rh}$ & 12.4 & $\rm{Hf}\rm{Rh}\rm{Sb}$ & 17.8 & $\rm{Ga}\rm{Ni}\rm{Nb}$ & 25.8\\
$\rm{\bf{I_{4}}}\rm{\bf{Sn}}$ & 0.41 & $\rm{\bf{Mo}}\rm{\bf{Zn_{6}}}$ & 2.44 & $\rm{Te}\rm{Pb}$ & 7.07 & $\rm{Sc}\rm{Sb}\rm{Pd}$ & 12.5 & $\rm{Hf}\rm{Ni}\rm{Sn}$ & 17.9 & $\rm{Nb}\rm{Ir}\rm{Ge}$ & 26.1\\
$\rm{Ba}_{2}\rm{Bi}\rm{Au}$ & 0.41 & $\rm{Cs}\rm{I}$ & 2.45 & $\rm{Hf}\rm{B}\rm{Rh}_{3}$ & 7.12 & $\rm{Li}_{2}\rm{Se}$ & 13.0 & $\rm{Nb}\rm{Ir}\rm{Sn}$ & 18.1 & $\rm{Ta}\rm{Ru}\rm{As}$ & 26.1\\
$\rm{Cs}\rm{Au}$ & 0.45 & $\rm{Al}_{3}\rm{Ga}$ & 2.59 & $\rm{Sn}\rm{Te}$ & 7.20 & $\rm{Al}\rm{Ge}\rm{Li}$ & 13.0 & $\rm{Nb}\rm{Fe}\rm{Bi}$ & 18.3 & $\rm{Al}\rm{Au}\rm{Hf}$ & 26.3\\
$\rm{Ba}_{2}\rm{Hg}\rm{Pb}$ & 0.49 & $\rm{In}\rm{Ag}_{3}$ & 2.63 & $\rm{Ge}\rm{Te}$ & 7.57 & $\rm{Li}\rm{Cd}\rm{As}$ & 13.0 & $\rm{Zr}\rm{Ni}\rm{Ge}$ & 18.3 & $\rm{Ge}\rm{Fe}\rm{W}$ & 26.5\\
$\rm{Cu}_{2}\rm{S}$ & 0.50 & $\rm{Cs}\rm{Br}$ & 2.66 & $\rm{\bf{Au}}\rm{\bf{Zr_{3}}}$ & 8.52 & $\rm{Ti}\rm{Rh}\rm{Bi}$ & 13.1 & $\rm{Hf}\rm{Pt}\rm{Ge}$ & 18.4 & $\rm{V}\rm{Os}\rm{As}$ & 26.8\\
$\rm{Cs}\rm{K}_{2}\rm{Bi}$ & 0.53 & $\rm{Cu}\rm{Cl}$ & 2.69 & $\rm{Bi}\rm{Pd}\rm{Sc}$ & 8.61 & $\rm{\bf{Cd}}\rm{\bf{S_{4}}}\rm{\bf{Zn_{3}}}$ & 13.1 & $\rm{Zr}\rm{Co}\rm{Bi}$ & 18.4 & $\rm{Ta}\rm{Fe}\rm{Sb}$ & 27.1\\
$\rm{Ba}_{2}\rm{Sb}\rm{Au}$ & 0.56 & $\rm{Rb}_{2}\rm{O}$ & 2.80 & $\rm{Ti}\rm{Pt}\rm{Pb}$ & 8.71 & $\rm{Zr}\rm{Pt}\rm{Ge}$ & 13.5 & $\rm{Zr}\rm{Sb}\rm{Rh}$ & 18.6 & $\rm{V}\rm{Rh}\rm{Ge}$ & 27.1\\
$\rm{Tl}\rm{Br}$ & 0.56 & $\rm{Rb}\rm{I}$ & 2.82 & $\rm{Ta}\rm{Ir}\rm{Pb}$ & 8.72 & $\rm{V}\rm{Ru}\rm{Sb}$ & 13.7 & $\rm{Ti}\rm{Te}\rm{Ru}$ & 18.6 & $\rm{Ta}\rm{Os}\rm{As}$ & 27.2\\
$\rm{\bf{In}}\rm{\bf{Tl_{3}}}$ & 0.59 & $\rm{Cs}\rm{I}$ & 2.93 & $\rm{Hf}\rm{Pt}\rm{Pb}$ & 8.75 & $\rm{Ti}\rm{Pd}\rm{Sn}$ & 13.8 & $\rm{Hf}\rm{Co}\rm{As}$ & 18.8 & $\rm{V}\rm{Ir}\rm{Ge}$ & 27.5\\
$\rm{Ag}\rm{I}$ & 0.61 & $\rm{Ca}\rm{Mg}_{3}$ & 3.10 & $\rm{Li}\rm{Mg}\rm{Sb}$ & 9.09 & $\rm{Nb}\rm{Ru}\rm{Bi}$ & 13.9 & $\rm{Zn}\rm{Se}$ & 18.9 & $\rm{Ga}\rm{Pt}\rm{Ta}$ & 27.9\\
$\rm{\bf{Bi_{2}}}\rm{\bf{Cs}}$ & 0.79 & $\rm{Rb}\rm{F}$ & 3.16 & $\rm{Bi}\rm{Ni}\rm{Y}$ & 9.12 & $\rm{Ti}\rm{Pt}\rm{Sn}$ & 14.1 & $\rm{Hf}\rm{Rh}\rm{As}$ & 18.9 & $\rm{Ti}\rm{Rh}\rm{As}$ & 28.4\\
$\rm{Sr}_{2}\rm{Sb}\rm{Au}$ & 0.82 & $\rm{Ba}\rm{Bi}\rm{K}$ & 3.17 & $\rm{Ti}\rm{Pd}\rm{Pb}$ & 9.22 & $\rm{Zr}\rm{Pd}\rm{Sn}$ & 14.2 & $\rm{Zr}\rm{Sn}\rm{Ru}_{2}$ & 18.9 & $\rm{Si}\rm{Co}\rm{Ta}$ & 28.8\\
$\rm{Mg}\rm{Hg}_{3}$ & 0.83 & $\rm{Li}_{2}\rm{Tl}\rm{Bi}$ & 3.18 & $\rm{Sc}_{3}\rm{B}\rm{Pb}$ & 9.28 & $\rm{V}\rm{Os}\rm{Sb}$ & 14.3 & $\rm{Ta}\rm{Fe}\rm{Bi}$ & 19.1 & $\rm{Ta}\rm{Ir}\rm{Ge}$ & 29.0\\
$\rm{Tl}\rm{Cl}$ & 0.84 & $\rm{Rb}\rm{Br}$ & 3.19 & $\rm{Cu}\rm{Br}$ & 9.31 & $\rm{Hf}\rm{Ir}\rm{As}$ & 14.4 & $\rm{Ti}\rm{Rh}\rm{Sb}$ & 19.1 & $\rm{Nb}\rm{Co}\rm{Ge}$ & 29.2\\
$\rm{Ag}\rm{Cl}$ & 0.93 & $\rm{Ag}\rm{I}$ & 3.20 & $\rm{Hf}\rm{Pd}\rm{Pb}$ & 9.33 & $\rm{Zr}\rm{Ni}\rm{Pb}$ & 14.4 & $\rm{Nb}\rm{Ru}\rm{Sb}$ & 19.2 & $\rm{\bf{Re}}\rm{\bf{W_{3}}}$ & 29.3\\
$\rm{Ca}_{2}\rm{Sb}\rm{Au}$ & 0.94 & $\rm{\bf{Li}}\rm{\bf{Pt_{2}}}$ & 3.23 & $\rm{Ta}\rm{Os}\rm{Bi}$ & 9.33 & $\rm{Ac}\rm{Al}\rm{O}_{3}$ & 14.5 & $\rm{Nb}\rm{Co}\rm{Sn}$ & 19.5 & $\rm{Ti}\rm{Co}\rm{As}$ & 29.5\\
$\rm{Ag}\rm{Br}$ & 0.96 & $\rm{Li}_{2}\rm{Ca}\rm{Si}$ & 3.41 & $\rm{Sr}\rm{F}_{2}$ & 9.54 & $\rm{Hf}\rm{Pd}\rm{Sn}$ & 14.6 & $\rm{La}\rm{P}$ & 19.8 & $\rm{Nb}\rm{Ru}\rm{As}$ & 29.6\\
$\rm{\bf{Hg_{4}}}\rm{\bf{Ni}}$ & 1.12 & $\rm{Rb}_{2}\rm{S}$ & 3.46 & $\rm{Ta}\rm{Rh}\rm{Pb}$ & 9.61 & $\rm{Hf}\rm{Si}\rm{Ru}_{2}$ & 14.7 & $\rm{Ti}\rm{Ir}\rm{Sb}$ & 19.9 & $\rm{Nb}\rm{Rh}\rm{Ge}$ & 29.9\\
$\rm{Ag}\rm{Br}$ & 1.13 & $\rm{Cu}_{3}\rm{As}$ & 3.48 & $\rm{Nb}\rm{Ir}\rm{Pb}$ & 9.74 & $\rm{V}\rm{Sb}\rm{Ru}$ & 14.8 & $\rm{Hf}\rm{Co}\rm{Bi}$ & 19.9 & $\rm{Ti}\rm{Co}\rm{Bi}$ & 30.5\\
$\rm{\bf{Mg}}\rm{\bf{Zn_{2}}}$ & 1.18 & $\rm{Li}_{2}\rm{In}\rm{Bi}$ & 3.59 & $\rm{Li}_{2}\rm{Te}$ & 9.90 & $\rm{Zr}\rm{Pt}\rm{Sn}$ & 14.8 & $\rm{Hf}\rm{Co}\rm{Sb}$ & 19.9 & $\rm{V}\rm{Fe}\rm{As}$ & 30.6\\
$\rm{\bf{Al_{2}}}\rm{\bf{Ba}}$ & 1.22 & $\rm{Sr}_{3}\rm{Bi}\rm{N}$ & 3.67 & $\rm{Zr}\rm{Pd}\rm{Pb}$ & 10.1 & $\rm{As}\rm{Ni}\rm{Sc}$ & 14.8 & $\rm{Hf}\rm{Sn}\rm{Ru}_{2}$ & 20.3 & $\rm{V}\rm{Ga}\rm{Fe}_{2}$ & 33.7\\
$\rm{K}\rm{Na}_{2}\rm{Bi}$ & 1.26 & $\rm{La}\rm{Bi}\rm{Pd}$ & 3.74 & $\rm{La}\rm{O}\rm{F}$ & 10.1 & $\rm{Zr}\rm{Si}\rm{Ru}_{2}$ & 14.9 & $\rm{V}\rm{Fe}\rm{Sb}$ & 20.4 & $\rm{La}\rm{As}$ & 34.5\\
$\rm{\bf{Cs_{2}}}\rm{\bf{Se}}$ & 1.35 & $\rm{\bf{Li_{3}}}\rm{\bf{S_{4}}}\rm{\bf{V}}$ & 3.81 & $\rm{Mg}\rm{Te}$ & 10.2 & $\rm{Al}\rm{Si}\rm{Li}$ & 15.3 & $\rm{Ta}\rm{Ru}\rm{Sb}$ & 21.2 & $\rm{Nb}\rm{Co}\rm{Si}$ & 34.8\\
$\rm{Yb}\rm{Pd}$ & 1.36 & $\rm{\bf{Ca_{3}}}\rm{\bf{As}}\rm{\bf{Br_{3}}}$ & 3.82 & $\rm{K}\rm{F}$ & 10.3 & $\rm{Nb}\rm{Rh}\rm{Sn}$ & 15.3 & $\rm{Hf}\rm{Ir}\rm{Sb}$ & 21.3 & $\rm{Nb}\rm{Fe}\rm{As}$ & 35.4\\
$\rm{La}\rm{Pt}\rm{Sb}$ & 1.37 & $\rm{Y}\rm{Ni}\rm{P}$ & 3.84 & $\rm{Zr}\rm{Pt}\rm{Pb}$ & 10.4 & $\rm{Ta}\rm{Co}\rm{Pb}$ & 15.6 & $\rm{Ta}\rm{Co}\rm{Sn}$ & 21.6 & $\rm{Ga}\rm{As}$ & 38.3\\
$\rm{Rb}\rm{Cl}$ & 1.38 & $\rm{Ag}\rm{Cl}$ & 3.88 & $\rm{Hf}\rm{Ir}\rm{Bi}$ & 10.4 & $\rm{Hf}\rm{Pt}\rm{Sn}$ & 15.7 & $\rm{\bf{B_{6}}}\rm{\bf{Ca}}$ & 22.0 & $\rm{V}\rm{Si}\rm{Rh}$ & 42.9\\
$\rm{\bf{Ba_{3}}}\rm{\bf{Sr}}$ & 1.50 & $\rm{Li}\rm{Br}$ & 4.21 & $\rm{\bf{Cr}}\rm{\bf{Ga_{4}}}$ & 10.4 & $\rm{Hf}\rm{Pd}\rm{Ge}$ & 15.7 & $\rm{Zr}\rm{Co}\rm{Sb}$ & 22.2 & $\rm{Li}\rm{B}\rm{Si}$ & 44.0\\
$\rm{K}\rm{Na}_{2}\rm{Sb}$ & 1.53 & $\rm{Cd}\rm{P}\rm{Na}$ & 4.51 & $\rm{Ti}\rm{Ir}\rm{Bi}$ & 10.6 & $\rm{Zr}\rm{Ge}\rm{Ru}_{2}$ & 15.9 & $\rm{Ti}\rm{Ni}\rm{Ge}$ & 22.2 & $\rm{Al}\rm{V}\rm{Fe}_{2}$ & 45.9\\
$\rm{\bf{Hg_{4}}}\rm{\bf{Pt}}$ & 1.54 & $\rm{Rb}\rm{F}$ & 4.60 & $\rm{\bf{Ni}}\rm{\bf{Zn}}$ & 10.7 & $\rm{Hf}\rm{Ni}\rm{Pb}$ & 16.0 & $\rm{Ti}\rm{Ni}\rm{Pb}$ & 22.2 & $\rm{\bf{Ge}}\rm{\bf{Si_{7}}}$ & 59.5\\
$\rm{Cd}_{3}\rm{Pd}$ & 1.71 & $\rm{Hg}\rm{Se}$ & 4.64 & $\rm{\bf{Be_{2}}}\rm{\bf{Cu}}$ & 11.0 & $\rm{Ti}\rm{Sn}\rm{Ru}_{2}$ & 16.1 & $\rm{Zr}\rm{Rh}\rm{As}$ & 22.2 & $\rm{Al}\rm{Bi}$ & 61.1\\
$\rm{La}\rm{Rh}\rm{Te}$ & 1.72 & $\rm{Ca}_{2}\rm{Cd}\rm{Pb}$ & 4.83 & $\rm{Ta}\rm{Ru}\rm{Bi}$ & 11.1 & $\rm{Zr}\rm{Ni}\rm{Sn}$ & 16.2 & $\rm{Zr}\rm{Co}\rm{As}$ & 22.3 & $\rm{Ru}\rm{O}_{2}$ & 78.9\\
$\rm{\bf{Ca}}\rm{\bf{Li_{2}}}$ & 1.77 & $\rm{Ba}\rm{O}$ & 4.88 & $\rm{Hf}\rm{Rh}\rm{Bi}$ & 11.1 & $\rm{Te}\rm{Ru}\rm{Zr}$ & 16.4 & $\rm{Te}\rm{Fe}\rm{Ti}$ & 22.5 & $\rm{Be}\rm{Te}$ & 89.1\\
$\rm{Tl}\rm{Pd}\rm{F}_{3}$ & 1.77 & $\rm{Li}\rm{Ca}\rm{As}$ & 4.99 & $\rm{Zr}\rm{Ir}\rm{Bi}$ & 11.1 & $\rm{Zr}\rm{Pd}\rm{Ge}$ & 16.4 & $\rm{Ti}\rm{Co}\rm{Sb}$ & 22.7 & $\rm{Li}\rm{Be}\rm{Sb}$ & 92.5\\
$\rm{\bf{Au}}\rm{\bf{Be}}$ & 1.78 & $\rm{\bf{Au_{2}}}\rm{\bf{S}}$ & 5.15 & $\rm{V}\rm{Ir}\rm{Sn}$ & 11.2 & $\rm{Nb}\rm{Co}\rm{Pb}$ & 16.4 & $\rm{Ti}\rm{Pt}\rm{Ge}$ & 22.7 & $\rm{Si}\rm{C}$ & 410\\
$\rm{Tl}\rm{Zn}\rm{F}_{3}$ & 1.86 & $\rm{Cd}\rm{F}_{2}$ & 5.17 & $\rm{Bi}\rm{Ni}\rm{Sc}$ & 11.4 & $\rm{Hf}\rm{Ge}\rm{Ru}_{2}$ & 16.5 & $\rm{Ti}\rm{Pd}\rm{Ge}$ & 23.1 & $\rm{B}\rm{P}$ & 443\\
$\rm{Pt}_{3}\rm{Pb}$ & 2.02 & $\rm{Ce}\rm{Pt}_{3}$ & 5.20 & $\rm{Zr}\rm{Ir}\rm{As}$ & 11.4 & $\rm{\bf{F_{3}}}\rm{\bf{Mg_{3}}}\rm{\bf{N}}$ & 16.5 & $\rm{V}\rm{Ru}\rm{As}$ & 23.2 & $\rm{B}\rm{N}$ & 659\\
$\rm{Ca}\rm{Cl}_{2}$ & 2.04 & $\rm{\bf{Cu_{3}}}\rm{\bf{Ta}}\rm{\bf{Te_{4}}}$ & 5.23 & $\rm{Sc}\rm{Sb}\rm{Pt}$ & 11.5 & $\rm{Ti}\rm{Ni}\rm{Sn}$ & 16.6 & $\rm{Ti}\rm{Ir}\rm{As}$ & 23.3 & $\rm{C}$ & 2040\\
$\rm{Sr}_{3}\rm{Ca}$ & 2.04 & $\rm{\bf{Cu_{2}}}\rm{\bf{O}}$ & 5.49 & $\rm{Nb}\rm{Os}\rm{Bi}$ & 11.6 & $\rm{V}\rm{Co}\rm{Sn}$ & 17.3 & $\rm{Ta}\rm{Fe}\rm{As}$ & 24.1 &  & \\
$\rm{Na}\rm{Zn}\rm{As}$ & 2.06 & $\rm{La}\rm{Y}_{3}$ & 5.55 & $\rm{Ta}\rm{Tl}\rm{O}_{3}$ & 11.7 & $\rm{Ta}\rm{Ir}\rm{Sn}$ & 17.5 & $\rm{Nb}\rm{Fe}\rm{Sb}$ & 24.1 &  & \\
\end{tabular}
\end{ruledtabular}
\end{table*}

\section*{\label{sec:results} Results and discussion}

\subsection*{DFT based values of $\kappa_\ell$}

The $\kappa_\ell$ of the 268 compounds computed with DFT are listed in Table~\ref{tbl:kappa_tdep}. 
Only the compound names are listed, further information can be retrieved from the MP-codes given in the supporting information (SI). 
Out of these, 122 are half-Heuslers~\cite{TranasHHActiveSampling, TransAttaining}
and ten are compounds for which low values of $\kappa_\ell$ have been 
reported earlier: (CsI~\cite{CsI}, CuCl~\cite{CuClF43m}, TlCl~\cite{TransferLearningExploring}, TlBr~\cite{TlBrLowLTC}, $\rm{Cu}_2\rm{S}$~\cite{Cu2S}, $\rm{Ag}_2\rm{S}$~\cite{Ag2SLowLTC}, $\rm{Ba}_{2}BiAu$~\cite{Ba2BiAuLowLTC, Ba2BiAuLowLTCComp}, $\rm{Ba}_{2}SbAu$~\cite{Ba2SbAu}, $\rm{Sr}_{2}SbAu$~\cite{Ba2SbAu}, and $\rm{Ba}_{2}HgPb$~\cite{Ba2BiAuLowLTC}).
The remaining 106 were randomly selected from the total set of cubic compounds.
These 238 compounds were used in the initial ML models. 
Finally, an additional 30 were selected in an AL procedure as part of the construction of the final ML model (highlighted in boldface) 

The training set holds several compounds which could hold potential as thermoelectric materials,
featuring both a very low $\kappa_\ell$ and a quite low band gap.
This motivates further 
computational and experimental assessment. 

\subsection*{Machine learning model}

\begin{figure*}[t!]
\includegraphics[width=0.99\linewidth]{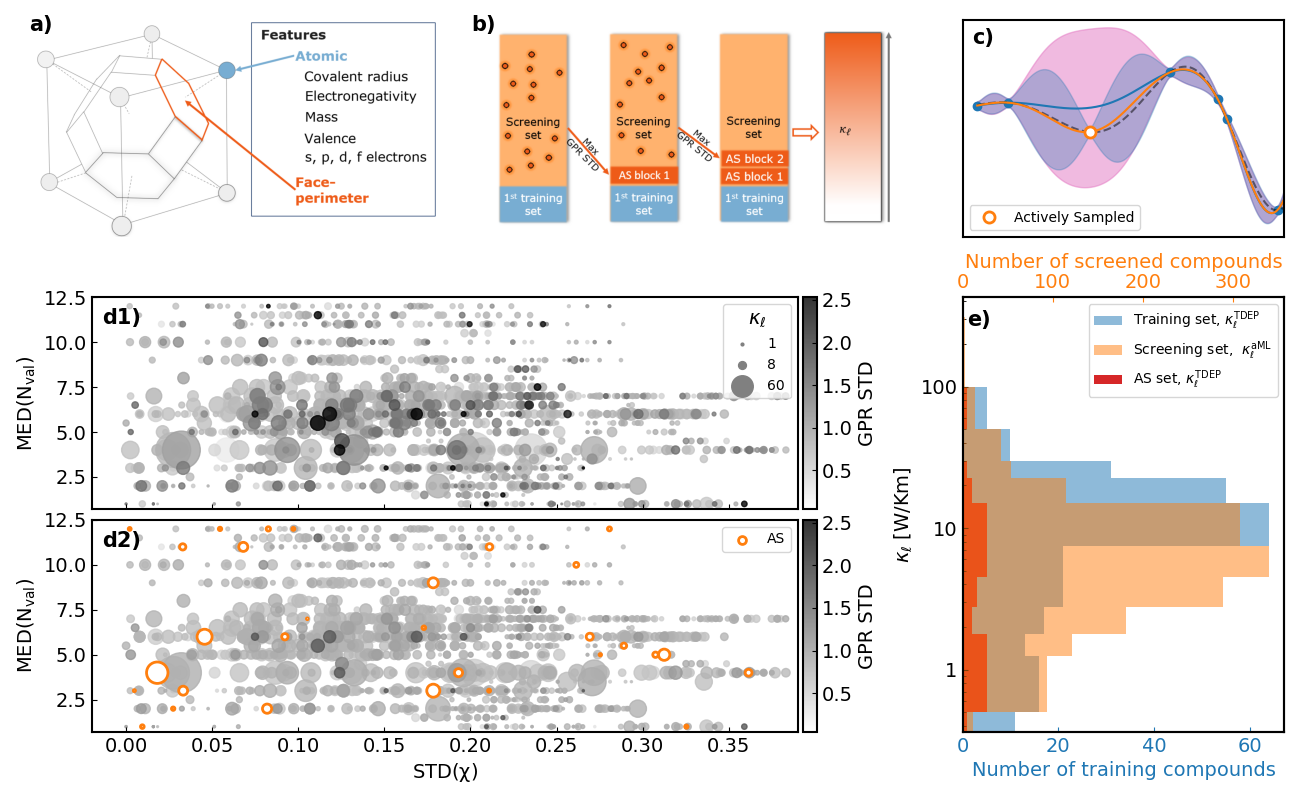}
\caption{Various concepts and preliminary results of the ML model:
{\bf a} Features from atomic properties and the Voronoi cell of atoms.
{\bf b} Workflow of the active learning (AL) and ML procedure.
{\bf c} Illustration of AL with GPR for 1d case: The shaded regions indicate the GPR-uncertainty 
before and after AL.
{\bf d1/d2} The GPR uncertainty as given by the shading and the corresponding predicted $\kappa_\ell$ given by the size of the disk for initial/final data set, as spanned by two most important features (see the main text and Table~\ref{tbl:features2}).
Orange circles indicate the actively selected compounds.
{\bf e} Distribution of $\kappa_{\ell}^{\rm{DFT}}$ for training compounds, compounds selected in the AL (bottom horizontal axis), and the $\kappa_{\ell}^{\rm{aML}}$ for screened compounds (top horizontal axis).}
\label{fig:composite}
\end{figure*}

Fig.~\ref{fig:composite} details the AL procedure used to train the final ML model. 
The ML model is based on features derived from atomic properties and the Voronoi cell (panel {\bf{a}}), 
using the values of $\sigma^*_{\rm GPR}$ to expand the data set, as conceptually illustrated by a one-dimensional toy model in 
 panel {{\bf c}}. 
In AL, there is a tradeoff between accuracy and simplicity.
For instance, simply expanding the training set
based on the $\sigma^*_{\rm GPR}$ values of the initial model,
would result in 
clusters of compounds with very similar features to be selected. 
On the other hand, selecting one compound at a time and retraining the model in each step),
enforces serial $\kappa_\ell$ computation of the selected compounds, which is impractical and time-consuming. 
One strategy to circumvent the serialization issue is to use $\kappa_{\ell}^{\rm ML}$ as a surrogate for $\kappa_{ \ell}$, but this strategy introduces a gradual degradation of the AL scheme as the fraction of surrogates increases. 
We, therefore, adopted the hybrid approach 
outlined in ({\bf b}): 
The surrogate approach was used to identify 15 compounds,
for which $\kappa_\ell^{\rm DFT}$ was computed.
This was then used to retrain the model and select another 
15 compounds, followed by subsequent retraining and feature selection. 

Panels {\bf d1} and {\bf d2} 
span out the compounds of the screening pool in feature space 
before and after the AL procedure, 
 respectively. 
The two most important features of the final model are plotted: the median of the atomic valence numbers $N_{\rm val}$, and the STD of the electronegativity $\chi$ of the atoms in the unit cell. 
The shade of gray indicates $\sigma^*_{\rm GPR}$; it becomes lighter in the final model compared to the first, since the overall model uncertainty is lower. 
The size of the scatter disks corresponds to the predicted $\kappa_\ell$ of the ML models.
The fact that the distribution of $\kappa_\ell$ values shows no strong directional dependence or clustering, 
highlights the non-linear nature of lattice thermal conductivity.
Furthermore, the fact that selected compounds (orange circles) are widely spread out in the feature space, shows that the iterative procedure enhances the model diversity.

Panel {\bf e} compares the
the distribution of computed $\kappa_{\ell}$ values in the total training set
to that of the predicted in the screening set. 
Adequate model diversity should also ensure that all $\kappa_{\ell}$-ranges are 
adequately sampled.
The initial training set had comparably many compounds with $\kappa_\ell$ values in the 10-30~W/Km range and compounds with very low values. 
However, very few compounds in the training set had low-intermediate values.
After the AL procedure, 
the overall distribution of $\kappa_{\ell}$ values
in the training set became more similar to that of the screening set. 
In addition, relatively more compounds with low $\kappa_{\ell}$ were selected compared to the distribution of $\kappa_{\ell}^{\rm{aML}}$. This may be explained by the great span in features giving rise to low $\kappa_{\ell}$.  

A remarkable finding from the AL procedure was that roughly half of the compounds with high model uncertainty were found to be dynamically unstable --- i.e., having
imaginary phonon frequencies ---, a fraction significantly higher than for the randomly selected systems from the initial training set.  
One possible explanation could be that the
compounds differ not substantially from those of the training set, but also from "typical" compounds of the MP database, which presumably are dynamically stable. 
Thus, the AL procedure can also serve to identify compounds that are likely to be dynamically unstable---or close to dynamically unstable. 
The dynamically unstable compounds were removed from the screening set and are listed in the SI.

\subsection*{\label{sec:ml_compound_and_feature} Compound and feature selection}

\begin{figure*}[t!]
\includegraphics[width=.90\linewidth]{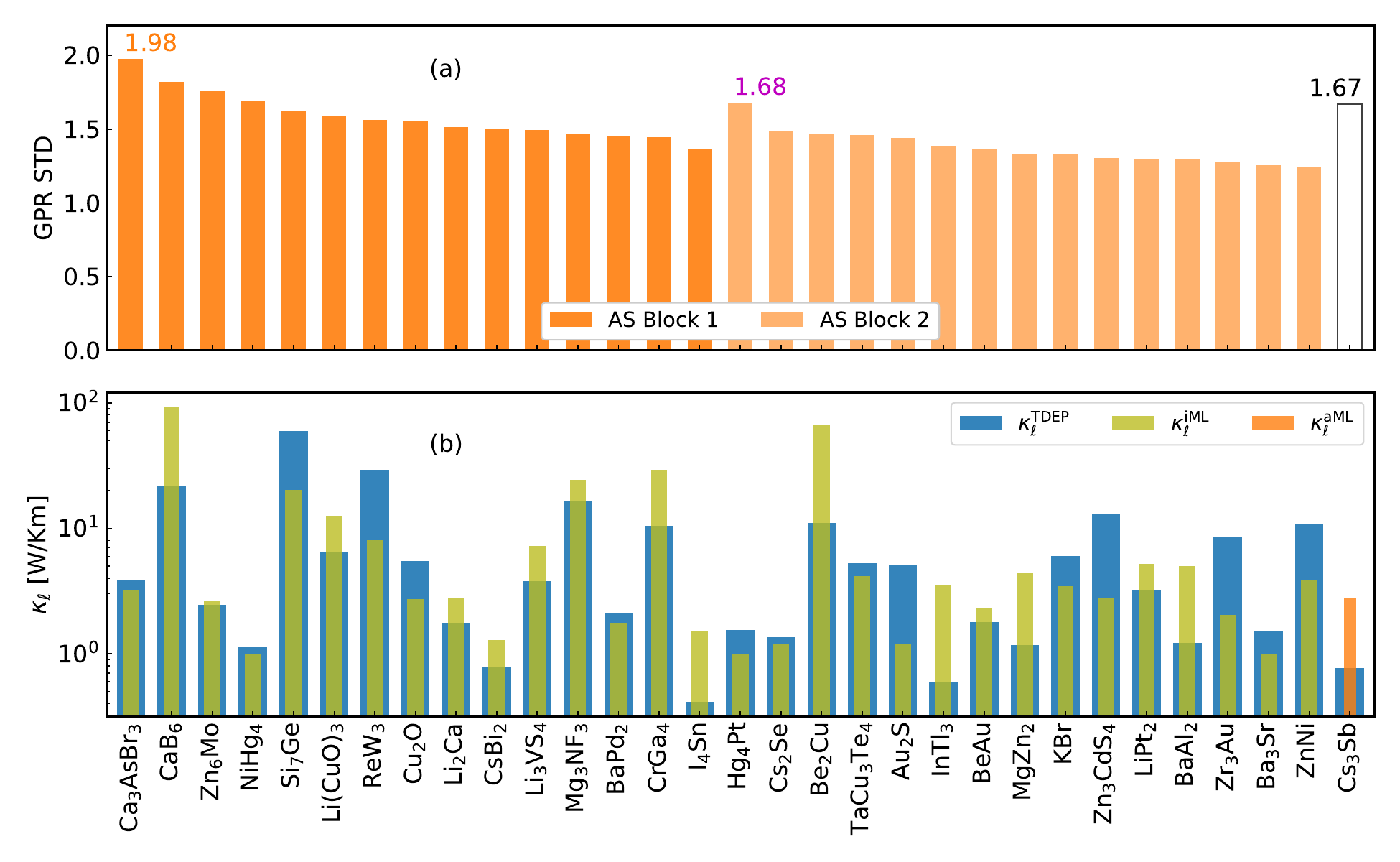}
\caption{\label{fig:active_sampling}\bf{a} Bars show the model uncertainty ($\sigma^*_{\rm GPR}$) after each actively sampled compound is added to the training set. Orange bars correspond to the first active learning (AL) block of 15 compounds, and purple the second. The white bar shows the 
 $\sigma^*_{\rm GPR}$ once the model has been restrained. {\bf b} Corresponding, $\kappa_{\ell}^{\rm{DFT}}$ , $\kappa_{\ell}^{\rm{iML}}$ for the actively sampled compounds, and the value of $\kappa_{\ell}^{\rm{aML}}$. }
\end{figure*}

The compounds selected by the AL procedure are shown in Fig.~\ref{fig:active_sampling},
ordered from left-to-right according to their inclusion in the ML model. 
In panel {\bf a}, Block 1 indicates the $\sigma^*_{\rm GPR}$ based on the initial training set 
with successive retraining based on the surrogate ML predictions. 
Block 2 indicates the same after the inclusion of the first block of 15 $\kappa_{\ell}^{\rm{DFT}}$.
Finally, the open bar shows $\sigma^*_{\rm GPR}$ of the compound with the highest model uncertainty of the final ML model. 
The corresponding $\kappa_{\ell}^{\rm{iML}}$, $\kappa_{\ell}^{\rm{aML}}$ and $\kappa_{\ell}^{\rm{DFT}}$ are shown in panel {\bf b}.
The initial model only showed a  log-log correlation with a r2-score of 0.33 for these 30 compounds. 
The weak-to-moderate correlation is not surprising, given that these compounds were among those exhibiting the largest $\sigma^*_{\rm GPR}$ of the screening set (1574 compounds). 

\begin{figure*}[t!]
\includegraphics[width=.99\linewidth]{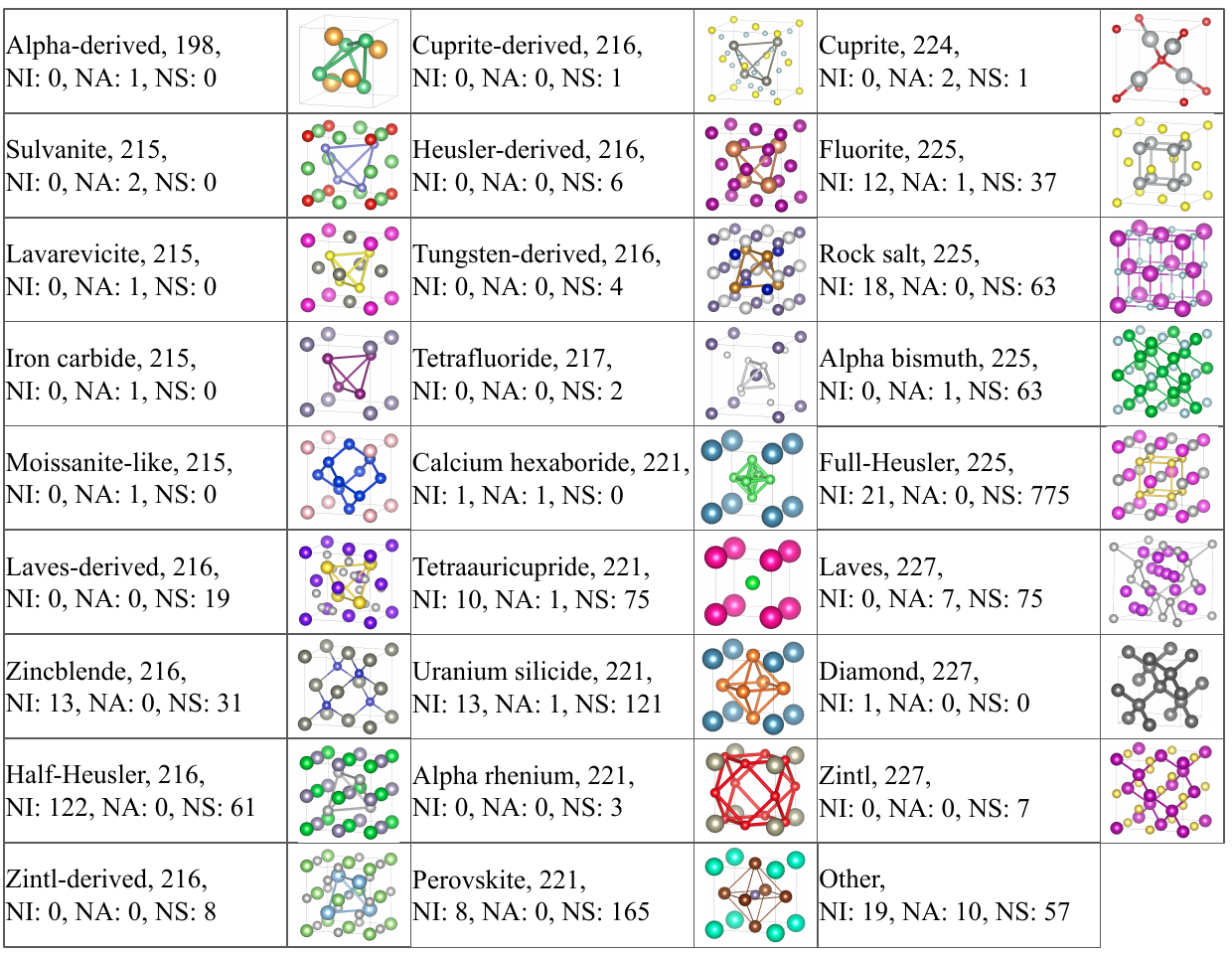}
\caption{\label{fig:ML_compounds}Overview of compound sets with structure, space group, and the number of initial (NI), actively sampled (NA), and screened (NS) compounds listed. The "Other" block (bottom right) displays compounds with non-existing structure labels from Robocrys~\cite{robocrys}.}
\end{figure*}

Fig.\ref{fig:ML_compounds} presents how the compounds in the training, screening, and AL sets are distributed on different crystal structures and shows how the AL procedure ensures a better training diversity 
For instance, none of the 61 half-Heuslers or 775 full-Heuslers in the screening set 
which were well sampled in the initial training set, were selected, while seven of the cubic Laves structures were selected.

A full list of the features explored, selected, and their feature importance is given in Table~\ref{tbl:features2}, 
for both the initial and final models. 
The most important feature, the $\rm{STD}(\chi)$, reflects the ionicity of the bonds, which tends to reduce $\kappa_{\ell}$. 
The median valence, $\rm{Med}(N_{\rm{val}})$, is the second most important. It distinguishes 
compounds such as the half-Heusler TiNiSn with ${\rm Med}(N_{\rm{val}})=4$
from compounds such as the zincblende AgBr and CuCl with ${\rm Med}(N_{\rm{val}})=9$.
Another selected feature is the median mass, ${\rm Med}(m)$. 
One might expect this to correlate strongly with $\kappa_{\ell}^{\rm{DFT}}$~\cite{ChenGPRLTC, JunejaHTGPR}, since heavy atoms reduce the phonon group velocity and hence $\kappa_{\ell}$~\cite{HighMassLowVelocity}. However, its importance was significantly lower than what one might expect: 
high mass, while beneficial, is not essential to obtain low intrinsic $\kappa_{\ell}$.

The final model achieved a validation R2-score of 0.81 and a Spearman correlation of 0.93,
compared to 0.83 and 0.91 in the initial model. 
The fact that the R2-metric was slightly lower in the final model is counterintuitive --- one could expect it to increase with more data --- can be 
understood from the exposure to a more diverse training set that differs from that of the initial training set, from which the validation set is drawn from,
and the ability to rank low-to-high Spearman correlation has slightly improved. 
Nevertheless, the relatively high validation scores of the AL procedure are impressive, when taking into account the low cost and simplicity of the features used in the model. 
This is further illustrated by the validation study in the Methods section (Fig.~\ref{fig:R2_active_sampling_case}), which further quantifies the efficacy of the AL scheme.


\subsection*{\label{sec:screening_low_lattice_thermal_conductivity_compounds} The lattice thermal conductivity from machine learning}

\begin{figure}
\includegraphics[width=1.00\linewidth]{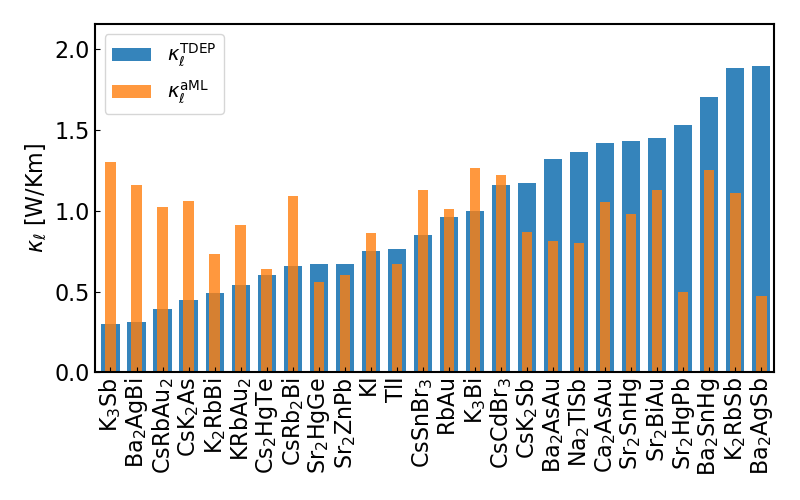}
\caption{\label{fig:low_LTC}  Low $\kappa_\ell$ materials predicted and verified with DFT.} 
\end{figure}

The key goal of the ML screening was to identify compounds with low $\kappa_\ell$. 
DFT calculations were used to verify the prediction of the compounds with the lowest $\kappa_\ell$ values.
Thus, for the 32 compounds with $\kappa_{\ell}^{\rm{aML}} \leq 1.30$~W/Km
 and with a non-zero bandgap at the GGA level, $E_{g}^{\rm{MP}}>0$, $\kappa_{\ell}^{\rm{DFT}}$ was calculated. In this assessment, four compounds (RbCaBr$_3$, RbGeI$_3$, NaTl$_2$Bi, Na$_2$TlSb)
were found to be dynamically unstable and excluded. 
Fig.\ \ref{fig:low_LTC} compares $\kappa_{\ell}^{\rm{aML}}$ and $\kappa_{\ell}^{\rm{DFT}}$ 
for the remaining 28 compounds. 
While the quantitative agreement did vary among the compounds,
it is reassuring that all these 28 compounds were found to have 
a very low $\kappa_\ell^{\rm{DFT}}$, i.e., between 0.30 and 2.05 W/Km.

There is an overall good agreement between our results and the available data in the literature.
 Wang et al.~\cite{Ba2AgSbLowLTC}
 computed a value of $\kappa_{\ell}=$ 1.4~W/Km
 for the full-Heusler $\mathrm{Ba}_2\mathrm{AgSb}$,
while He et al.~\cite{Ba2BiAuLowLTC} 
predicted 1 W/Km for the full-Heusler Sr$_2$BiAu.
Many of the compounds with low predicted $\kappa_{\ell}$, in particular the Cs-containing compounds, coincide with those predicted in the ML study of Jaafreh et al.~\cite{JaafrehAccelerated}. 
The compound $\rm{Cs}_{2}\rm{Se}$, with $\kappa_{\ell}^{\rm{DFT}}=1.35$~W/Km of Table~\ref{tbl:kappa_tdep} was also previously identified in the screening study of Juneja et al.~\cite{JunejaHTGPR}. 

For ten of the compounds --- 
Cs$_{2}$HgTe,
 K$_{2}$RbSb, 
  TlI, 
  Sr$_{2}$ZnPb,  
  CsRb$_{2}$Bi, 
  CsRbAu$_{2}$, 
  KRbAu$_{2}$,   
  Sr$_{2}$SnHg,
  Sr$_{2}$HgGe, 
  K$_{2}$RbBi
we did not find earlier reported $\kappa_{\ell}$ values or
reports on their potential as thermoelectric materials. 
All but the first three have bandgaps below 0.44~eV, and among these, the five last listed are at or 0.01 eV per atom or less above the convex hull. 

Interestingly, the compounds with low predicted $\kappa_{\ell}$ show a significant span in their characteristics:
the bandgaps, $E_{g}^{\rm{MP}}$, range from 0.04~eV ($\rm{Ba}_{2}\rm{SnHg}$) to 3.24~eV ($\rm{KI}$) and the average atomic mass ranges from 59.8~u ($\rm{K}_{3}\rm{Sb}$) to 165.6~u (TlI); the mass ratio between the lightest and heaviest elements ranges from 0.11 ($\mathrm{Na}_{2}\mathrm{Tl}\rm{Bi}$) to 0.79 ($\mathrm{Ba}_2\mathrm{AgSb}$);
the number of atoms ranges from 2 to 5 in the primitive cell; and they span three space groups: 221 (perovskite, tetraauricupride), 225 (full-Heusler), and 227 (Laves). 
The crystal stacking thus varies; as an example,
in the tetraauricupride-structured KI, the light K element has eight I atoms as nearest neighbors, while in Laves structure $\rm{K}\rm{Bi}_2$, the K atoms have twelve Bi atoms as nearest neighbors. 
In line with the low feature importance of the mass, light atomic masses did not necessarily result in high lattice thermal conductivity, i.e., both heavy compounds, such as $\rm{Ba}_{2}\rm{Ag}\rm{Sb}$ and $\rm{Ba}_{2}\rm{Ag}\rm{Bi}$, and lighter ones such as $\rm{K}_{3}\rm{Sb}$ and $\rm{K}\rm{I}$ can give low $\kappa_{\ell}$.  

The SI provides predicted $\kappa_{\ell}^{\rm{aML}}$ for all the compounds screened.
The list of identified compounds includes 89 compounds 
with $E_{g}^{\rm{MP}}=0$~eV.
(Semi-)metals generally have low thermoelectric efficiency, 
as bipolar transport reduces $S$. 
However, some of the compounds might be incorrectly labeled metallic since standard 
DFT calculations underestimate bandgaps,  
and new compounds can be uncovered by using high-level theory, as illustrated in Ref.~\cite{DiscardedGemsBerland}.
For instance, the compound $\rm{Ba}_2\rm{HgPb}$, which has a low $\kappa_{\ell}$ in the training set (Table~\ref{tbl:kappa_tdep}), obtained a non-zero bandgap at the hybrid functional level.

Many of the identified compounds with low $\kappa_\ell$ contain toxic elements. 
Future studies should explore whether replacing such elements, 
e.g., by interchanging Tl, As, and Hg for the less toxic In, Sb, and Cd,
would retain structural stability and low lattice thermal conductivity.
Further, the thermoelectric properties of several of the identified compounds should be 
assessed  
by computing electronic transport properties. 

Limited training data is a key challenge for making reliable machine learning (ML) models 
for predicting materials properties like $\kappa_{\ell}$. 
In this paper, we overcame this issue using AL based on the GPR method.
All the 28 compounds with $\kappa_\ell  \leq 1.3$~W/Km were verified as compounds with low  $\kappa_\ell$ with DFT. 
In the AL procedure, 
about half of the compounds samples were dynamically unstable,
which was linked to the fact that the procedure picks up usual compounds. 
Whether some of the corresponding dynamically stable non-cubic phases 
could have low  $\kappa_{\ell}$ of 
and whether some of them might become cubic at higher temperatures,
merits further investigation. 
The model we developed here can be used to screen cubic compounds 
without the need for any DFT input and relies only on the crystal structure
and elemental information. 
Using features requiring DFT calculations or experimental input, such as the bulk modulus~\cite{ChenGPRLTC, TranasHHActiveSampling} or the phonon dispersion,
would, however, would likely make the model more accurate\cite{AnharmMeasureForMaterials, CarreteFindingLowLTCHHs,TranasHHActiveSampling}.
In this work, we chose to make a more general model that also can be applied to the screening of hypothetical prototype compounds~\cite{CarreteHHStabilityScreening, JaafrehAccelerated, JaafrehZTDeepLearning}, that can be used 
in tandem with ML-based crystal structure prediction. 
Our study highlights that many kinds of materials can have 
low $\kappa_{\ell}$, reflected in the great span of feature values of such compounds,
and this brings much promise for the discovery of novel thermoelectric materials. 


\section*{\label{sec:methods}Methods}

\subsection*{\label{sec:lattice_thermal_conductivity}The lattice thermal conductivity from DFT and TDEP calculations}

The lattice thermal conductivity, $\kappa_{\ell}$, was calculated from TDEP~\cite{hellmanLatticeDynamicsAnharmonic2011, hellmanTemperaturedependentEffectiveThirdorder2013} which solves the
phonon Boltzmann transport equation within the relaxation-time approximation.
In this approximation, $\kappa_{\ell}$ can be expressed as follows,
\begin{equation}\label{eq:ltc}
    \kappa_{\ell}^{\rm DFT} = \dfrac{1}{V}\sum_{\mathbf{q}s}c_{\mathbf{q}s}v_{\mathbf{q}s}^{2}\tau_{\mathbf{q}s}.
\end{equation}
Here, $V$ is the volume, $\mathbf{q}$, the phonon wave vector, $s$, the phonon mode, $c_{\mathbf{q}s}$, the phonon heat capacity, $v_{\mathbf{q}s}$, the phonon group velocity, and $\tau_{\mathbf{q}s}$, the phonon relaxation time. 
TDEP was also used to generate configurations of supercell structures
based on a canonical ensemble of the Debye model at $T=300$~K~\cite{shulumbaIntrinsicLocalized,AndersonSimplifiedMethodDebye}.
To obtain the finite-temperature second- and third-order force constants, we computed the forces with DFT using 50 such configurations of supercells with $4\times 4 \times 4$ repetitions of the primitive cell for compounds with two atoms in the primitive cell, and $3\times 3 \times 3$ for the rest. 
Radial cutoffs for the second-and third-order force constants were set, respectively, to be $\sim 2$~\% larger and $\sim2$~\% smaller than half the width of the supercells,
to ensure a consistent number of force constants for similar compounds. 
The $\mathbf{q}$-point grid was sampled on a $35 \times 35 \times 35$ grid for compounds with less than 5 atoms in the primitive cell and $30 \times 30 \times 30$ for the rest.

The DFT calculations were done with \textsc{VASP}~\cite{kresseInitioMolecularDynamics1993, kresseEfficientIterativeSchemes1996, kresseEfficiencyAbinitioTotal1996}. 
For systems with only four or fewer atoms in the primitive cell,
we used the Perdew–Burke–Ernzerhof generalized gradient approximation (GGA) for solids (PBEsol)~\cite{perdewRestoringDensityGradientExpansion2008, csonkaAssessingPerformanceRecent2009};
but for systems with more atoms, we switched to 
 the consistent-exchange van der Waals density functional (vdW-DF-cx) functional,\cite{cx1, cx2}
to account for the effect of dispersion forces, which can play a role in more open systems even if the dominant binding force is ionic or covalent. Moreover, vdW-DF-cx 
accurately predict properties of cubic perovskites~\cite{VDWAccuracyPerovskite, CXLatticeParameter, VDWAccuracy}. 
This inconsistent selection of functionals introduces a small source of noise in the training data; however, the initial dataset of Half Heuslers was based on PBEsol. Both vdW-DF-cx and PBEsol provide accurate lattice parameters for the systems studied here.
Our testing for the smaller systems revealed only small differences in the predicted $\kappa_\ell$ values between vdW-DF-cx and PBEsol, 
which is much smaller than the difference between $\kappa_\ell$ of PBE and PBEsol.
%
\cite{PBESolAccuracy}. 

The plane-wave energy cutoff was set to 500~eV, and
electronic self-consistency was reached with energy differences less than $10^{-6}$~eV.
The $\mathbf{k}$-point sampling of the Brillouin zone
used an $11 \times 11 \times 11$  ($9 \times 9 \times 9$) $\Gamma$-centered mesh
for relaxation of the primitive unit cells. 
Atomic positions were relaxed until all forces fell below 1~meV/Å.

\subsection*{\label{sec:ml_models} Machine learning model}

\begin{figure}[t!]
\includegraphics[width=0.95\linewidth]{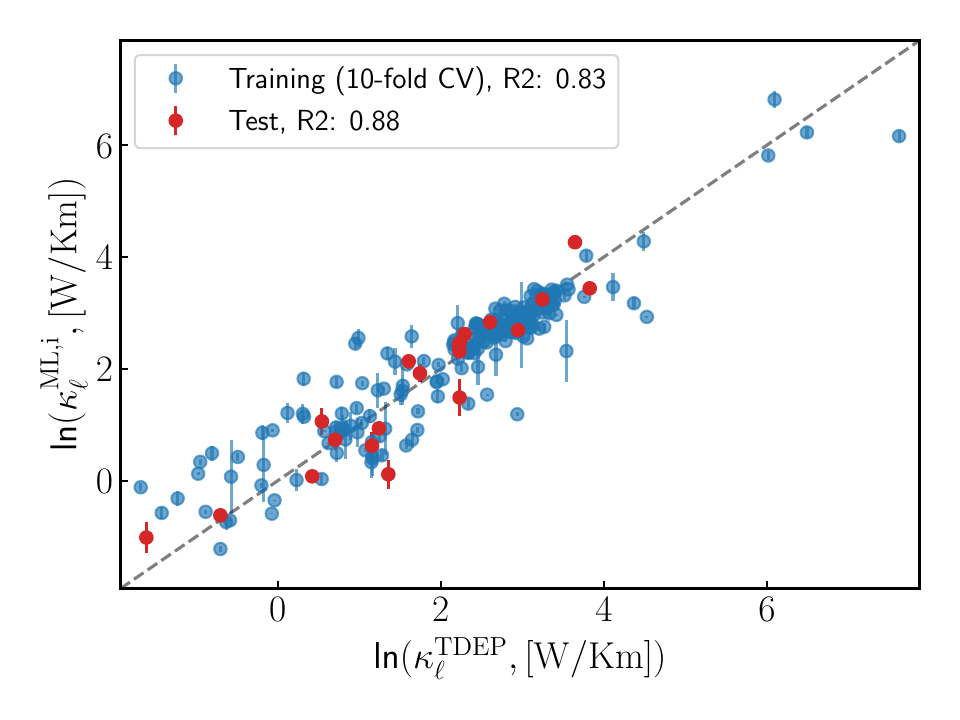}
\caption{\label{fig:ml_predictions_initial} Parity plot for the validation of the initial ML model. 
}

\includegraphics[width=0.95\linewidth]{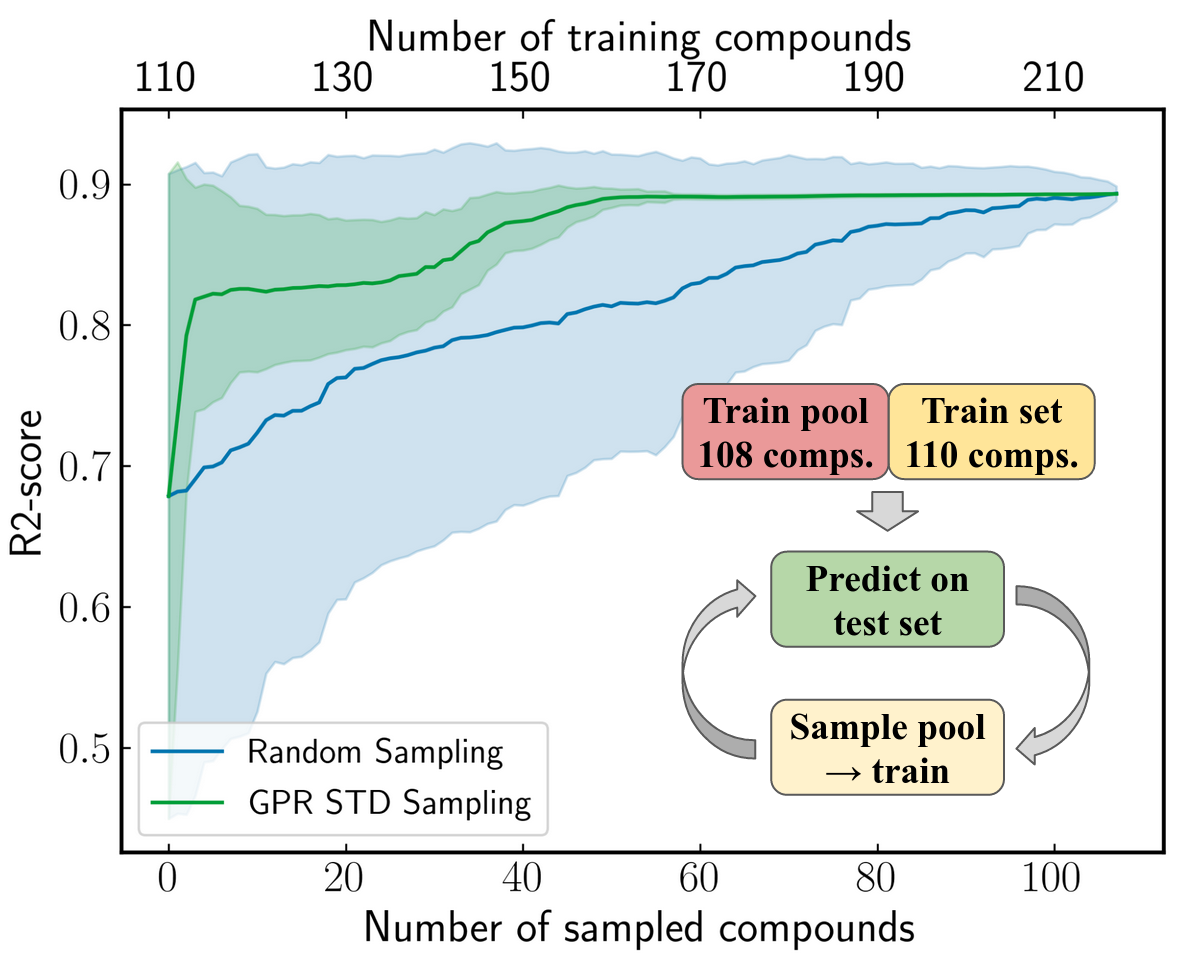}
\caption{\label{fig:R2_active_sampling_case}
Validation of the AL procedure and ML with GPR based on a test study. 
In each step, the compounds with largest $\sigma^*_{\rm GPR}$ in the training pool is moved to the training set. 
The shaded regions indicate the STD of the predictions obtained from 100 random splits of the initial training set and training pool.}
\end{figure}

Gaussian process regression (GPR)\cite{GPRMaterialsDiscovery,GPRAerogels,GPRForMaterialsAndMolecules, GPRCompundRepresentation, GPRSurrogateModel,SekoBayesianLowLTC, ChenGPRLTC,
PerspectiveMLThermalTransport} 
is a Bayesian approach based on the construction of a distribution function $p$. 
In the case of $\kappa_\ell$, a predicted value $\kappa^{*}_{\ell}$ for a given feature ($\mathbf{x}_{*}$) is given by 
$p(\kappa^{*}_{\ell}|\{\kappa_{\ell}\}, \{X\}, \mathbf{x}_{*})$, where the given ("prior") training set is  $(\{X\}$, $\{\kappa_{\ell}\})$. 
The GPR prediction is then given by the mean of the distribution, 
while the standard deviation (STD) provides a model uncertainty.
By identifying compounds with high ML model uncertainty,
$\kappa_{\ell}$ of these compounds can be computed from DFT and given to the model as additional training. 
The concept is exemplified in Fig. \ref{fig:composite}{\bf a} for a one-dimensional regression. 

We used the GPR model implemented in
\textsc{Scikit-learn}~\cite{pedregosaScikitlearnMachineLearning2011} with the Matern kernel, given as
\begin{equation}
    k(x,x') = \dfrac{1}{\Gamma(\nu)2^{\nu-1}}\left(r d(x,x')\right)^{\nu}K_{\nu}\left(r d(x,x')\right)\,,
\end{equation}
where $\Gamma(\nu)$ is the gamma function, $d(x,x')$, the Euclidean distance, $K_{\nu}$, a modified Bessel function of the second kind, and $r = \sqrt{2\nu}/\ell$, where $\nu$ is a hyperparameter controlling the smoothness. The hyperparameter $\nu$ is often set to $3/2$ or $5/2$, both of which were investigated in the hyperparameter tuning. 
The length-scale parameter, $\ell$, another hyperparameter, sets how fast $k(x, x')$ drops off for increasing dissimilarity between compounds. 
Forward-sequential feature floating selection (SFFS)~\cite{SFFS} based on  \textsc{Mlxtend}~\cite{raschkaMLxtendProvidingMachine2018} 
was used for feature selection.
Model hyperparameters were tuned using a 10-fold cross-validation. 
Feature importance was assessed with feature permutation and was defined as the reduction in R2-score when 
a given feature is shuffled, i.e., a compound is assigned the feature value of another compound. 
The model performance was evaluated using a 10-times repeated 10-fold cross-validation with random shuffling of training data.

\subsection*{\label{sec:ml_initial_model} Validation of the initial model}

A comparison of the predictions of the initial ML model  
$\kappa_{\ell}^{\rm{ML, i}}$ with the corresponding DFT results $\kappa_{\ell}^{\rm{DFT}}$
is shown in Fig.~\ref{fig:ml_predictions_initial}.
The R2-score of the training set cross-validation was 0.83 and 0.88 for the test set,
with corresponding Spearman correlations of 0.91 and 0.95. 
The fact that the initial ML model shows good model perform, makes it a good starting point for further active learning.  

\subsection*{\label{sec:dataset} Selection of materials}

In the screening for low $\kappa_{\ell}$ materials, we limited the scope to the 
$1633$ cubic compounds listed in the \textsc{Materials Project} database\cite{jainCommentaryMaterialsProject2013}
with 2-8 atoms in the primitive cell, 
a vanishing magnetic moment, and a cohesion energy less than 0.1~eV/atom above the convex hull. 
Compounds containing H, He, atoms with atomic numbers between 58 (Ce) and 71 (Lu), and those with numbers above 83 (Bi) were also excluded.

\subsection*{\label{sec:features} Selection of features used in the ML model}
In the ML model, we used solely features taken from 
atomic properties and the crystal structure~\cite{CarreteHHStabilityScreening, JaafrehAccelerated, JaafrehZTDeepLearning}. This is summarized in Fig.~\ref{fig:composite}a) and details are shown in Table~\ref{tbl:features2}. 
The features were based on statistical properties of the atoms in the unit cell, i.e., the average $\rm{Avg}()$, median $\rm{Med}()$, and standard deviation $\rm{STD}()$ (not to be confused with the GRP predicted STD, $\sigma^*_{\rm GPR}$).
 The dipole polarizabilities of the atoms~\cite{dipolepolarizability}, $\mu_{\rm{dip}}$, also  used in the study of Ju et al.~\cite{TransferLearningExploring}, 
 were based on the experimental or calculated values recommended by Schwerdtfenger et al.~\cite{dipolepolarizability}. 
An effective representation of crystal structures was done using features derived from Voronoi diagrams~\cite{VoronoiOriginal, VoronoiAsMlFeatures, JaafrehAccelerated} as shown for the body-centered cubic lattice in Fig.~\ref{fig:composite}{\bf a}.
The face-perimeter circumference and area, and the volume of the Voronoi polyhedron were used as inputs in the feature generation.
The Voronoi-based surfaces and Voronoi weighted Steinhardt's parameter
were obtained using \textsc{Pyscal}~\cite{pyscal}.

\begin{table*}
\caption{Features used for training the GPR model. Values in the average (Avg), median (Med), and standard deviation (STD) columns display the feature importance of the initial model, with importance after active sampling shown in parentheses. Features not selected, indicated by a dash.  
}
\centering
\label{tbl:features2}
\begin{tabular}{p{6.65cm}|p{1.50cm}|p{3.45cm}p{3.45cm}p{3.45cm}}
\toprule
                                                      Description &                                                              Symbol &                                                  Avg &   Med &                                                                              STD \\
\toprule
                                                                                                                   Covalent radius &                  $r$ & 0.098 (0.232) & 0.098  &       (0.045) \\
                                                                                                                 Electronegativity &               $\chi$ &             - &             - & 0.328 (0.126) \\
                                                                                                             Dipole polarizability &     $\mu_{\rm{dip}}$ & 0.097 (0.060) &         0.045 &             - \\
                                                                                                                       atomic mass &                  $m$ &             - &         0.034 &             - \\
                                                                                                       Num. valence electrons &       $N_{\rm{val}}$ & 0.092 (0.062) & 0.187 (0.146) &             - \\
                                                                                                             Num s. electrons &         $N_{\rm{s}}$ &             - &       (0.037) &             - \\
                                                                                                             Num. p electrons &         $N_{\rm{p}}$ &             - &         0.066 &             - \\
                                                                                                             Num. d electrons &         $N_{\rm{d}}$ &             - &             - &             - \\
                                                                                                             Num. f electrons &         $N_{\rm{f}}$ &             - &             - &             - \\
                                                                                    Vol. Voronoi polyhedrons &       $V_{\rm{Vor}}$ &        (0.00) &             (0.140) &             - \\
Vol. cov. spheres over $V_{\rm Vor}$ $\Pi = V_{\rm{cov.}}/V_{\rm{Vor}}$. &                $\Pi$ & 0.069 (0.127) & 0.033 (0.027) &       (0.045) \\
                                                       Sum of face perimeters of Voronoi polyhedrons, $l=\sum_{n}L_{\rm{per}, n}$. &                  $l$ &       (0.044) & 0.051 (0.054) &         0.058 \\
  $\Lambda=\sum_{n}L_{\rm{per}, n}(r+r_{n})$. &            $\Lambda$ & 0.032 (0.182) &    (0.082)         & 0.146 (0.092) \\
                              Effective coordination number, $\rm{ECN}=(\sum_{n}L_{\rm{per}, n})^{2}/\sum_{n}L_{\rm{per}, n}^{2}$. &                  ECN & 0.071 (0.093) &         0.060 & 0.054 \\
                                                                                                               Coordination number &       $N_{\rm{cor}}$ &   -            &               &               \\
                                                                                                               \toprule
                                                                                                      & & Value         \\
                                                                                                      \toprule
                                                                                              Numb.  elements in primitive cell &  $N_{\rm{ele}}$ &  -  0.048             &         -      &           -    \\
                                                                                       1st Voronoi weighted Steinhardt's parameter &          $S_1$  &           0.030    &    -           &   -            \\
                                                                                       2nd-4th Voronoi weighted Steinhardt's parameters &                $S_2 - S_4$ &   -            &       -        &  -             \\
\hline 

\end{tabular}
\end{table*}

\subsection*{\label{sec:ml_active_sampling} Assessment of GPR active learning}

Fig.~\ref{fig:R2_active_sampling_case} 
details a case study illustrating the potency of the GPR-based active learning,
compared to a random enlargement of a training set. 
 The comparison was based on 238 compounds with $\kappa_\ell$ values obtained from DFT, and did not include the additional compounds identified in the final AL procedure. 
In this case study, we randomly divided the set of compounds into an initial training set of 110 compounds, a training pool of 108 compounds, and a test set of 20 compounds.
After training the initial model, 
we iteratively moved the compound with the largest   $\sigma^*_{\rm GPR}$ 
from the training pool to the training set, retraining the model in each iteration. 
For comparison, we also made a reference scheme, where we enlarged the training set by randomly moving compounds from the training pool to the training set.
Comparing the blue and green curve demonstrates the superior performance with AL. 
Interestingly, the ML model generated with the AL scheme stopped improving after 50 compounds had been added to the training set, and adding more data did not change the model performance. 
This is likely simply because the training set envelopes all the remaining data point in the test set.

\bibliography{references}

\begin{thebibliography}{96}%
\makeatletter
\providecommand \@ifxundefined [1]{%
 \@ifx{#1\undefined}
}%
\providecommand \@ifnum [1]{%
 \ifnum #1\expandafter \@firstoftwo
 \else \expandafter \@secondoftwo
 \fi
}%
\providecommand \@ifx [1]{%
 \ifx #1\expandafter \@firstoftwo
 \else \expandafter \@secondoftwo
 \fi
}%
\providecommand \natexlab [1]{#1}%
\providecommand \enquote  [1]{``#1''}%
\providecommand \bibnamefont  [1]{#1}%
\providecommand \bibfnamefont [1]{#1}%
\providecommand \citenamefont [1]{#1}%
\providecommand \href@noop [0]{\@secondoftwo}%
\providecommand \href [0]{\begingroup \@sanitize@url \@href}%
\providecommand \@href[1]{\@@startlink{#1}\@@href}%
\providecommand \@@href[1]{\endgroup#1\@@endlink}%
\providecommand \@sanitize@url [0]{\catcode `\\12\catcode `\$12\catcode
  `\&12\catcode `\#12\catcode `\^12\catcode `\_12\catcode `\%12\relax}%
\providecommand \@@startlink[1]{}%
\providecommand \@@endlink[0]{}%
\providecommand \url  [0]{\begingroup\@sanitize@url \@url }%
\providecommand \@url [1]{\endgroup\@href {#1}{\urlprefix }}%
\providecommand \urlprefix  [0]{URL }%
\providecommand \Eprint [0]{\href }%
\providecommand \doibase [0]{https://doi.org/}%
\providecommand \selectlanguage [0]{\@gobble}%
\providecommand \bibinfo  [0]{\@secondoftwo}%
\providecommand \bibfield  [0]{\@secondoftwo}%
\providecommand \translation [1]{[#1]}%
\providecommand \BibitemOpen [0]{}%
\providecommand \bibitemStop [0]{}%
\providecommand \bibitemNoStop [0]{.\EOS\space}%
\providecommand \EOS [0]{\spacefactor3000\relax}%
\providecommand \BibitemShut  [1]{\csname bibitem#1\endcsname}%
\let\auto@bib@innerbib\@empty
\bibitem [{\citenamefont {Snyder}\ and\ \citenamefont
  {Toberer}(2008)}]{SnyderComplexThermoelectric}%
  \BibitemOpen
  \bibfield  {author} {\bibinfo {author} {\bibfnamefont {G.}~\bibnamefont
  {Snyder}}\ and\ \bibinfo {author} {\bibfnamefont {E.}~\bibnamefont
  {Toberer}},\ }\bibfield  {title} {\bibinfo {title} {Complex thermoelectric
  materials},\ }\href {https://doi.org/10.1038/nmat2090} {\bibfield  {journal}
  {\bibinfo  {journal} {Nat. Mater.}\ }\textbf {\bibinfo {volume} {7}},\
  \bibinfo {pages} {105} (\bibinfo {year} {2008})}\BibitemShut {NoStop}%
\bibitem [{\citenamefont {Soleimani}\ \emph {et~al.}(2020)\citenamefont
  {Soleimani}, \citenamefont {Zoras}, \citenamefont {Ceranic}, \citenamefont
  {Shahzad},\ and\ \citenamefont {Cui}}]{RoomTempTEEnergyConsum}%
  \BibitemOpen
  \bibfield  {author} {\bibinfo {author} {\bibfnamefont {Z.}~\bibnamefont
  {Soleimani}}, \bibinfo {author} {\bibfnamefont {S.}~\bibnamefont {Zoras}},
  \bibinfo {author} {\bibfnamefont {B.}~\bibnamefont {Ceranic}}, \bibinfo
  {author} {\bibfnamefont {S.}~\bibnamefont {Shahzad}},\ and\ \bibinfo {author}
  {\bibfnamefont {Y.}~\bibnamefont {Cui}},\ }\bibfield  {title} {\bibinfo
  {title} {A review on recent developments of thermoelectric materials for
  room-temperature applications},\ }\href
  {https://doi.org/https://doi.org/10.1016/j.seta.2019.100604} {\bibfield
  {journal} {\bibinfo  {journal} {Sustain. Energy Technol. Assess.}\ }\textbf
  {\bibinfo {volume} {37}},\ \bibinfo {pages} {100604} (\bibinfo {year}
  {2020})}\BibitemShut {NoStop}%
\bibitem [{\citenamefont {Farhat}\ \emph {et~al.}(2022)\citenamefont {Farhat},
  \citenamefont {Faraj}, \citenamefont {Hachem}, \citenamefont {Castelain},\
  and\ \citenamefont {Khaled}}]{WasteHeatRegeneration}%
  \BibitemOpen
  \bibfield  {author} {\bibinfo {author} {\bibfnamefont {O.}~\bibnamefont
  {Farhat}}, \bibinfo {author} {\bibfnamefont {J.}~\bibnamefont {Faraj}},
  \bibinfo {author} {\bibfnamefont {F.}~\bibnamefont {Hachem}}, \bibinfo
  {author} {\bibfnamefont {C.}~\bibnamefont {Castelain}},\ and\ \bibinfo
  {author} {\bibfnamefont {M.}~\bibnamefont {Khaled}},\ }\bibfield  {title}
  {\bibinfo {title} {A recent review on waste heat recovery methodologies and
  applications: Comprehensive review, critical analysis and potential
  recommendations},\ }\href
  {https://doi.org/https://doi.org/10.1016/j.clet.2021.100387} {\bibfield
  {journal} {\bibinfo  {journal} {Cleaner Eng. Technol.}\ }\textbf {\bibinfo
  {volume} {6}},\ \bibinfo {pages} {100387} (\bibinfo {year}
  {2022})}\BibitemShut {NoStop}%
\bibitem [{\citenamefont {Gaultois}\ and\ \citenamefont
  {Sparks}(2014)}]{gaultoisImprovementFromReducingKappa}%
  \BibitemOpen
  \bibfield  {author} {\bibinfo {author} {\bibfnamefont {M.~W.}\ \bibnamefont
  {Gaultois}}\ and\ \bibinfo {author} {\bibfnamefont {T.~D.}\ \bibnamefont
  {Sparks}},\ }\bibfield  {title} {\bibinfo {title} {How much improvement in
  thermoelectric performance can come from reducing thermal conductivity?},\
  }\href {https://doi.org/10.1063/1.4869232} {\bibfield  {journal} {\bibinfo
  {journal} {Appl. Phys. Lett.}\ }\textbf {\bibinfo {volume} {104}},\ \bibinfo
  {pages} {113906} (\bibinfo {year} {2014})}\BibitemShut {NoStop}%
\bibitem [{\citenamefont {Rowe}\ and\ \citenamefont
  {Shukla}(1981)}]{RoweGBSiGe}%
  \BibitemOpen
  \bibfield  {author} {\bibinfo {author} {\bibfnamefont {D.~M.}\ \bibnamefont
  {Rowe}}\ and\ \bibinfo {author} {\bibfnamefont {V.~S.}\ \bibnamefont
  {Shukla}},\ }\bibfield  {title} {\bibinfo {title} {The effect of phonon-grain
  boundary scattering on the lattice thermal conductivity and thermoelectric
  conversion efficiency of heavily doped fine-grained, hot-pressed silicon
  germanium alloy},\ }\href {https://doi.org/10.1063/1.328733} {\bibfield
  {journal} {\bibinfo  {journal} {J. Appl. Phys.}\ }\textbf {\bibinfo {volume}
  {52}},\ \bibinfo {pages} {7421} (\bibinfo {year} {1981})}\BibitemShut
  {NoStop}%
\bibitem [{\citenamefont {Schrade}\ \emph {et~al.}(2017)\citenamefont
  {Schrade}, \citenamefont {Berland}, \citenamefont {Eliassen}, \citenamefont
  {Guzik}, \citenamefont {Echevarria-Bonet}, \citenamefont {Sørby},
  \citenamefont {Jenus}, \citenamefont {Hauback}, \citenamefont {Tofan},
  \citenamefont {Gunnæs}, \citenamefont {Persson}, \citenamefont {Løvvik},\
  and\ \citenamefont {Finstad}}]{SchradeGrainBoundary}%
  \BibitemOpen
  \bibfield  {author} {\bibinfo {author} {\bibfnamefont {M.}~\bibnamefont
  {Schrade}}, \bibinfo {author} {\bibfnamefont {K.}~\bibnamefont {Berland}},
  \bibinfo {author} {\bibfnamefont {S.}~\bibnamefont {Eliassen}}, \bibinfo
  {author} {\bibfnamefont {M.}~\bibnamefont {Guzik}}, \bibinfo {author}
  {\bibfnamefont {C.}~\bibnamefont {Echevarria-Bonet}}, \bibinfo {author}
  {\bibfnamefont {M.}~\bibnamefont {Sørby}}, \bibinfo {author} {\bibfnamefont
  {P.}~\bibnamefont {Jenus}}, \bibinfo {author} {\bibfnamefont
  {B.}~\bibnamefont {Hauback}}, \bibinfo {author} {\bibfnamefont
  {R.}~\bibnamefont {Tofan}}, \bibinfo {author} {\bibfnamefont
  {A.}~\bibnamefont {Gunnæs}}, \bibinfo {author} {\bibfnamefont
  {C.}~\bibnamefont {Persson}}, \bibinfo {author} {\bibfnamefont
  {O.}~\bibnamefont {Løvvik}},\ and\ \bibinfo {author} {\bibfnamefont
  {T.}~\bibnamefont {Finstad}},\ }\bibfield  {title} {\bibinfo {title} {The
  role of grain boundary scattering in reducing the thermal conductivity of
  polycrystalline {XNiSn} ({X} = {Hf}, {Zr}, {Ti}) half-{H}eusler alloys},\
  }\href {https://doi.org/10.1038/s41598-017-14013-8} {\bibfield  {journal}
  {\bibinfo  {journal} {Sci. Rep.}\ }\textbf {\bibinfo {volume} {7}} (\bibinfo
  {year} {2017})}\BibitemShut {NoStop}%
\bibitem [{\citenamefont {Zheng}\ \emph {et~al.}(2021)\citenamefont {Zheng},
  \citenamefont {Slade}, \citenamefont {Hu}, \citenamefont {Tan}, \citenamefont
  {Luo}, \citenamefont {Luo}, \citenamefont {Xu}, \citenamefont {Yan},\ and\
  \citenamefont {Kanatzidis}}]{DefectEngineeringReview}%
  \BibitemOpen
  \bibfield  {author} {\bibinfo {author} {\bibfnamefont {Y.}~\bibnamefont
  {Zheng}}, \bibinfo {author} {\bibfnamefont {T.~J.}\ \bibnamefont {Slade}},
  \bibinfo {author} {\bibfnamefont {L.}~\bibnamefont {Hu}}, \bibinfo {author}
  {\bibfnamefont {X.~Y.}\ \bibnamefont {Tan}}, \bibinfo {author} {\bibfnamefont
  {Y.}~\bibnamefont {Luo}}, \bibinfo {author} {\bibfnamefont {Z.-Z.}\
  \bibnamefont {Luo}}, \bibinfo {author} {\bibfnamefont {J.}~\bibnamefont
  {Xu}}, \bibinfo {author} {\bibfnamefont {Q.}~\bibnamefont {Yan}},\ and\
  \bibinfo {author} {\bibfnamefont {M.~G.}\ \bibnamefont {Kanatzidis}},\
  }\bibfield  {title} {\bibinfo {title} {Defect engineering in thermoelectric
  materials: what have we learned?},\ }\href
  {https://doi.org/10.1039/D1CS00347J} {\bibfield  {journal} {\bibinfo
  {journal} {Chem. Soc. Rev.}\ }\textbf {\bibinfo {volume} {50}},\ \bibinfo
  {pages} {9022} (\bibinfo {year} {2021})}\BibitemShut {NoStop}%
\bibitem [{\citenamefont {Zhao}\ \emph {et~al.}(2020)\citenamefont {Zhao},
  \citenamefont {Li}, \citenamefont {Fan}, \citenamefont {Xiao},\ and\
  \citenamefont {Xie}}]{DefectEngineeringTE2}%
  \BibitemOpen
  \bibfield  {author} {\bibinfo {author} {\bibfnamefont {C.}~\bibnamefont
  {Zhao}}, \bibinfo {author} {\bibfnamefont {Z.}~\bibnamefont {Li}}, \bibinfo
  {author} {\bibfnamefont {T.}~\bibnamefont {Fan}}, \bibinfo {author}
  {\bibfnamefont {C.}~\bibnamefont {Xiao}},\ and\ \bibinfo {author}
  {\bibfnamefont {Y.}~\bibnamefont {Xie}},\ }\bibfield  {title} {\bibinfo
  {title} {Defects engineering with multiple dimensions in thermoelectric
  materials},\ }\href {https://doi.org/10.34133/2020/9652749} {\bibfield
  {journal} {\bibinfo  {journal} {Research}\ }\textbf {\bibinfo {volume}
  {2020}},\ \bibinfo {pages} {1} (\bibinfo {year} {2020})}\BibitemShut
  {NoStop}%
\bibitem [{\citenamefont {Zhou}\ \emph {et~al.}(2016)\citenamefont {Zhou},
  \citenamefont {Huang}, \citenamefont {Wang}, \citenamefont {Liu},
  \citenamefont {Cai},\ and\ \citenamefont {Sui}}]{IsovalentCdinZnSb}%
  \BibitemOpen
  \bibfield  {author} {\bibinfo {author} {\bibfnamefont {J.}~\bibnamefont
  {Zhou}}, \bibinfo {author} {\bibfnamefont {L.}~\bibnamefont {Huang}},
  \bibinfo {author} {\bibfnamefont {Z.}~\bibnamefont {Wang}}, \bibinfo {author}
  {\bibfnamefont {Z.}~\bibnamefont {Liu}}, \bibinfo {author} {\bibfnamefont
  {W.}~\bibnamefont {Cai}},\ and\ \bibinfo {author} {\bibfnamefont
  {J.}~\bibnamefont {Sui}},\ }\bibfield  {title} {\bibinfo {title} {Effect of
  {Cd} isoelectronic substitution on thermoelectric properties of
  $\mathrm{Zn}_{0.995}\mathrm{Na}_{0.005}\mathrm{Sb}$},\ }\href
  {https://doi.org/https://doi.org/10.1016/j.jmat.2016.08.003} {\bibfield
  {journal} {\bibinfo  {journal} {J. Materiomics}\ }\textbf {\bibinfo {volume}
  {2}},\ \bibinfo {pages} {324} (\bibinfo {year} {2016})}\BibitemShut {NoStop}%
\bibitem [{\citenamefont {Zhang}\ \emph {et~al.}(2019)\citenamefont {Zhang},
  \citenamefont {Huang}, \citenamefont {Zhu}, \citenamefont {Zhou},
  \citenamefont {Jabar}, \citenamefont {Li}, \citenamefont {Zhu}, \citenamefont
  {Wang}, \citenamefont {Song}, \citenamefont {Xin}, \citenamefont {Li},\ and\
  \citenamefont {Qin}}]{IsovalentSubstitutionsinChalcopyrite}%
  \BibitemOpen
  \bibfield  {author} {\bibinfo {author} {\bibfnamefont {J.}~\bibnamefont
  {Zhang}}, \bibinfo {author} {\bibfnamefont {L.}~\bibnamefont {Huang}},
  \bibinfo {author} {\bibfnamefont {C.}~\bibnamefont {Zhu}}, \bibinfo {author}
  {\bibfnamefont {C.}~\bibnamefont {Zhou}}, \bibinfo {author} {\bibfnamefont
  {B.}~\bibnamefont {Jabar}}, \bibinfo {author} {\bibfnamefont
  {J.}~\bibnamefont {Li}}, \bibinfo {author} {\bibfnamefont {X.}~\bibnamefont
  {Zhu}}, \bibinfo {author} {\bibfnamefont {L.}~\bibnamefont {Wang}}, \bibinfo
  {author} {\bibfnamefont {C.}~\bibnamefont {Song}}, \bibinfo {author}
  {\bibfnamefont {S.}~\bibnamefont {Xin}}, \bibinfo {author} {\bibfnamefont
  {D.}~\bibnamefont {Li}},\ and\ \bibinfo {author} {\bibfnamefont {X.-y.}\
  \bibnamefont {Qin}},\ }\bibfield  {title} {\bibinfo {title} {Design of domain
  structure and realization of ultralow thermal conductivity for record‐high
  thermoelectric performance in chalcopyrite},\ }\href
  {https://doi.org/10.1002/adma.201905210} {\bibfield  {journal} {\bibinfo
  {journal} {Adv. Mater.}\ }\textbf {\bibinfo {volume} {31}} (\bibinfo {year}
  {2019})}\BibitemShut {NoStop}%
\bibitem [{\citenamefont {Berland}\ \emph {et~al.}(2019)\citenamefont
  {Berland}, \citenamefont {Shulumba}, \citenamefont {Hellman}, \citenamefont
  {Persson},\ and\ \citenamefont
  {Løvvik}}]{berlandThermoelectricTransportTrends2019}%
  \BibitemOpen
  \bibfield  {author} {\bibinfo {author} {\bibfnamefont {K.}~\bibnamefont
  {Berland}}, \bibinfo {author} {\bibfnamefont {N.}~\bibnamefont {Shulumba}},
  \bibinfo {author} {\bibfnamefont {O.}~\bibnamefont {Hellman}}, \bibinfo
  {author} {\bibfnamefont {C.}~\bibnamefont {Persson}},\ and\ \bibinfo {author}
  {\bibfnamefont {O.~M.}\ \bibnamefont {Løvvik}},\ }\bibfield  {title}
  {\bibinfo {title} {Thermoelectric transport trends in group 4 half-{H}eusler
  alloys},\ }\href {https://doi.org/https://doi.org/10.1063/1.5117288}
  {\bibfield  {journal} {\bibinfo  {journal} {J. Appl. Phys.}\ }\textbf
  {\bibinfo {volume} {126}},\ \bibinfo {pages} {145102} (\bibinfo {year}
  {2019})}\BibitemShut {NoStop}%
\bibitem [{\citenamefont {Eliassen}\ \emph {et~al.}(2017)\citenamefont
  {Eliassen}, \citenamefont {Katre}, \citenamefont {Madsen}, \citenamefont
  {Persson}, \citenamefont {L\o{}vvik},\ and\ \citenamefont
  {Berland}}]{eliassenLatticeThermalConductivity2017}%
  \BibitemOpen
  \bibfield  {author} {\bibinfo {author} {\bibfnamefont {S.~N.~H.}\
  \bibnamefont {Eliassen}}, \bibinfo {author} {\bibfnamefont {A.}~\bibnamefont
  {Katre}}, \bibinfo {author} {\bibfnamefont {G.~K.~H.}\ \bibnamefont
  {Madsen}}, \bibinfo {author} {\bibfnamefont {C.}~\bibnamefont {Persson}},
  \bibinfo {author} {\bibfnamefont {O.~M.}\ \bibnamefont {L\o{}vvik}},\ and\
  \bibinfo {author} {\bibfnamefont {K.}~\bibnamefont {Berland}},\ }\bibfield
  {title} {\bibinfo {title} {Lattice thermal conductivity of
  $\mathrm{Ti}_{x}\mathrm{Zr}_{y}\mathrm{Hf}_{1\ensuremath{-}x\ensuremath{-}y}\mathrm{NiSn}$
  half-{H}eusler alloys calculated from first principles: Key role of nature of
  phonon modes},\ }\href {https://doi.org/10.1103/PhysRevB.95.045202}
  {\bibfield  {journal} {\bibinfo  {journal} {Phys. Rev. B}\ }\textbf {\bibinfo
  {volume} {95}},\ \bibinfo {pages} {045202} (\bibinfo {year}
  {2017})}\BibitemShut {NoStop}%
\bibitem [{\citenamefont {Tran{\aa}s}\ \emph {et~al.}(2022)\citenamefont
  {Tran{\aa}s}, \citenamefont {L{\o}vvik},\ and\ \citenamefont
  {Berland}}]{TransAttaining}%
  \BibitemOpen
  \bibfield  {author} {\bibinfo {author} {\bibfnamefont {R.}~\bibnamefont
  {Tran{\aa}s}}, \bibinfo {author} {\bibfnamefont {O.~M.}\ \bibnamefont
  {L{\o}vvik}},\ and\ \bibinfo {author} {\bibfnamefont {K.}~\bibnamefont
  {Berland}},\ }\bibfield  {title} {\bibinfo {title} {Attaining low lattice
  thermal conductivity in half-heusler sublattice solid solutions: Which
  substitution site is most effective?},\ }\href
  {https://doi.org/10.3390/electronicmat3010001} {\bibfield  {journal}
  {\bibinfo  {journal} {Electron. Mater.}\ }\textbf {\bibinfo {volume} {3}},\
  \bibinfo {pages} {1} (\bibinfo {year} {2022})}\BibitemShut {NoStop}%
\bibitem [{\citenamefont {Yu}\ \emph {et~al.}(2017)\citenamefont {Yu},
  \citenamefont {Fu}, \citenamefont {Liu}, \citenamefont {Xia}, \citenamefont
  {Aydemir}, \citenamefont {Chasapis}, \citenamefont {Snyder}, \citenamefont
  {Zhao},\ and\ \citenamefont {Zhu}}]{NbFeSbAlloyTewoReducingSigma}%
  \BibitemOpen
  \bibfield  {author} {\bibinfo {author} {\bibfnamefont {J.}~\bibnamefont
  {Yu}}, \bibinfo {author} {\bibfnamefont {C.}~\bibnamefont {Fu}}, \bibinfo
  {author} {\bibfnamefont {Y.}~\bibnamefont {Liu}}, \bibinfo {author}
  {\bibfnamefont {K.}~\bibnamefont {Xia}}, \bibinfo {author} {\bibfnamefont
  {U.}~\bibnamefont {Aydemir}}, \bibinfo {author} {\bibfnamefont
  {T.}~\bibnamefont {Chasapis}}, \bibinfo {author} {\bibfnamefont
  {G.}~\bibnamefont {Snyder}}, \bibinfo {author} {\bibfnamefont
  {X.}~\bibnamefont {Zhao}},\ and\ \bibinfo {author} {\bibfnamefont
  {T.}~\bibnamefont {Zhu}},\ }\bibfield  {title} {\bibinfo {title} {Unique role
  of refractory {Ta} alloying in enhancing the figure of merit of {NbFeSb}
  thermoelectric materials},\ }\href {https://doi.org/10.1002/aenm.201701313}
  {\bibfield  {journal} {\bibinfo  {journal} {Adv. Energy Mater.}\ }\textbf
  {\bibinfo {volume} {8}},\ \bibinfo {pages} {1701313} (\bibinfo {year}
  {2017})}\BibitemShut {NoStop}%
\bibitem [{\citenamefont {Luo}\ \emph {et~al.}(2021)\citenamefont {Luo},
  \citenamefont {Serrano-Sánchez}, \citenamefont {Bishara}, \citenamefont
  {Zhang}, \citenamefont {{Bueno Villoro}}, \citenamefont {Kuo}, \citenamefont
  {Felser}, \citenamefont {Scheu}, \citenamefont {Snyder}, \citenamefont
  {Best}, \citenamefont {Dehm}, \citenamefont {Yu}, \citenamefont {Raabe},
  \citenamefont {Fu},\ and\ \citenamefont
  {Gault}}]{NbCoSnGBsIncreaseTransport}%
  \BibitemOpen
  \bibfield  {author} {\bibinfo {author} {\bibfnamefont {T.}~\bibnamefont
  {Luo}}, \bibinfo {author} {\bibfnamefont {F.}~\bibnamefont
  {Serrano-Sánchez}}, \bibinfo {author} {\bibfnamefont {H.}~\bibnamefont
  {Bishara}}, \bibinfo {author} {\bibfnamefont {S.}~\bibnamefont {Zhang}},
  \bibinfo {author} {\bibfnamefont {R.}~\bibnamefont {{Bueno Villoro}}},
  \bibinfo {author} {\bibfnamefont {J.~J.}\ \bibnamefont {Kuo}}, \bibinfo
  {author} {\bibfnamefont {C.}~\bibnamefont {Felser}}, \bibinfo {author}
  {\bibfnamefont {C.}~\bibnamefont {Scheu}}, \bibinfo {author} {\bibfnamefont
  {G.~J.}\ \bibnamefont {Snyder}}, \bibinfo {author} {\bibfnamefont {J.~P.}\
  \bibnamefont {Best}}, \bibinfo {author} {\bibfnamefont {G.}~\bibnamefont
  {Dehm}}, \bibinfo {author} {\bibfnamefont {Y.}~\bibnamefont {Yu}}, \bibinfo
  {author} {\bibfnamefont {D.}~\bibnamefont {Raabe}}, \bibinfo {author}
  {\bibfnamefont {C.}~\bibnamefont {Fu}},\ and\ \bibinfo {author}
  {\bibfnamefont {B.}~\bibnamefont {Gault}},\ }\bibfield  {title} {\bibinfo
  {title} {Dopant-segregation to grain boundaries controls electrical
  conductivity of n-type {NbCo(Pt)Sn} half-{H}eusler alloy mediating
  thermoelectric performance},\ }\href
  {https://doi.org/https://doi.org/10.1016/j.actamat.2021.117147} {\bibfield
  {journal} {\bibinfo  {journal} {Acta Mater.}\ }\textbf {\bibinfo {volume}
  {217}},\ \bibinfo {pages} {117147} (\bibinfo {year} {2021})}\BibitemShut
  {NoStop}%
\bibitem [{\citenamefont {Hu}\ \emph {et~al.}(2014)\citenamefont {Hu},
  \citenamefont {Zhu}, \citenamefont {Liu},\ and\ \citenamefont
  {Zhao}}]{BiTePointDefect}%
  \BibitemOpen
  \bibfield  {author} {\bibinfo {author} {\bibfnamefont {L.}~\bibnamefont
  {Hu}}, \bibinfo {author} {\bibfnamefont {T.}~\bibnamefont {Zhu}}, \bibinfo
  {author} {\bibfnamefont {X.}~\bibnamefont {Liu}},\ and\ \bibinfo {author}
  {\bibfnamefont {X.}~\bibnamefont {Zhao}},\ }\bibfield  {title} {\bibinfo
  {title} {Point defect engineering of high-performance bismuth-telluride-based
  thermoelectric materials},\ }\href
  {https://doi.org/https://doi.org/10.1002/adfm.201400474} {\bibfield
  {journal} {\bibinfo  {journal} {Adv. Funct. Mater.}\ }\textbf {\bibinfo
  {volume} {24}},\ \bibinfo {pages} {5211} (\bibinfo {year}
  {2014})}\BibitemShut {NoStop}%
\bibitem [{\citenamefont {Su}(2018)}]{PbTeExperiment}%
  \BibitemOpen
  \bibfield  {author} {\bibinfo {author} {\bibfnamefont {C.-H.}\ \bibnamefont
  {Su}},\ }\bibfield  {title} {\bibinfo {title} {Experimental determination of
  lattice thermal conductivity and {L}orenz number as functions of temperature
  for n-type {PbTe}},\ }\href
  {https://doi.org/https://doi.org/10.1016/j.mtphys.2018.05.005} {\bibfield
  {journal} {\bibinfo  {journal} {Mater. Today Phys.}\ }\textbf {\bibinfo
  {volume} {5}},\ \bibinfo {pages} {58} (\bibinfo {year} {2018})}\BibitemShut
  {NoStop}%
\bibitem [{\citenamefont {Lu}\ \emph {et~al.}(2018)\citenamefont {Lu},
  \citenamefont {Sun},\ and\ \citenamefont {Zhang}}]{PbTeComputational}%
  \BibitemOpen
  \bibfield  {author} {\bibinfo {author} {\bibfnamefont {Y.}~\bibnamefont
  {Lu}}, \bibinfo {author} {\bibfnamefont {T.}~\bibnamefont {Sun}},\ and\
  \bibinfo {author} {\bibfnamefont {D.-B.}\ \bibnamefont {Zhang}},\ }\bibfield
  {title} {\bibinfo {title} {Lattice anharmonicity, phonon dispersion, and
  thermal conductivity of {PbTe} studied by the phonon quasiparticle
  approach},\ }\href {https://doi.org/10.1103/PhysRevB.97.174304} {\bibfield
  {journal} {\bibinfo  {journal} {Phys. Rev. B}\ }\textbf {\bibinfo {volume}
  {97}} (\bibinfo {year} {2018})}\BibitemShut {NoStop}%
\bibitem [{\citenamefont {Satterthwaite}\ and\ \citenamefont
  {Ure}(1957)}]{BismuthTellurideFirstExp}%
  \BibitemOpen
  \bibfield  {author} {\bibinfo {author} {\bibfnamefont {C.~B.}\ \bibnamefont
  {Satterthwaite}}\ and\ \bibinfo {author} {\bibfnamefont {R.~W.}\ \bibnamefont
  {Ure}},\ }\bibfield  {title} {\bibinfo {title} {Electrical and thermal
  properties of $\rm{Bi}_{2}\rm{Te}_{3}$},\ }\href
  {https://doi.org/10.1103/PhysRev.108.1164} {\bibfield  {journal} {\bibinfo
  {journal} {Phys. Rev.}\ }\textbf {\bibinfo {volume} {108}},\ \bibinfo {pages}
  {1164} (\bibinfo {year} {1957})}\BibitemShut {NoStop}%
\bibitem [{\citenamefont {Witting}\ \emph {et~al.}(2019)\citenamefont
  {Witting}, \citenamefont {Chasapis}, \citenamefont {Ricci}, \citenamefont
  {Peters}, \citenamefont {Heinz}, \citenamefont {Hautier},\ and\ \citenamefont
  {Snyder}}]{BismuthTellurideReview}%
  \BibitemOpen
  \bibfield  {author} {\bibinfo {author} {\bibfnamefont {I.~T.}\ \bibnamefont
  {Witting}}, \bibinfo {author} {\bibfnamefont {T.~C.}\ \bibnamefont
  {Chasapis}}, \bibinfo {author} {\bibfnamefont {F.}~\bibnamefont {Ricci}},
  \bibinfo {author} {\bibfnamefont {M.}~\bibnamefont {Peters}}, \bibinfo
  {author} {\bibfnamefont {N.~A.}\ \bibnamefont {Heinz}}, \bibinfo {author}
  {\bibfnamefont {G.}~\bibnamefont {Hautier}},\ and\ \bibinfo {author}
  {\bibfnamefont {G.~J.}\ \bibnamefont {Snyder}},\ }\bibfield  {title}
  {\bibinfo {title} {The thermoelectric properties of bismuth telluride},\
  }\href {https://doi.org/https://doi.org/10.1002/aelm.201800904} {\bibfield
  {journal} {\bibinfo  {journal} {Adv. Electron. Mater.}\ }\textbf {\bibinfo
  {volume} {5}},\ \bibinfo {pages} {1800904} (\bibinfo {year}
  {2019})}\BibitemShut {NoStop}%
\bibitem [{\citenamefont {Zhou}\ \emph {et~al.}(2021)\citenamefont {Zhou},
  \citenamefont {Lee}, \citenamefont {Yu}, \citenamefont {Byun}, \citenamefont
  {Luo}, \citenamefont {Lee}, \citenamefont {Ge}, \citenamefont {Lee},
  \citenamefont {Chen}, \citenamefont {Lee}, \citenamefont {Cojocaru-Mirédin},
  \citenamefont {Chang}, \citenamefont {Im}, \citenamefont {Cho}, \citenamefont
  {Wuttig}, \citenamefont {Dravid}, \citenamefont {Kanatzidis},\ and\
  \citenamefont {Chung}}]{SnSePolyCrystalline}%
  \BibitemOpen
  \bibfield  {author} {\bibinfo {author} {\bibfnamefont {C.}~\bibnamefont
  {Zhou}}, \bibinfo {author} {\bibfnamefont {Y.~K.}\ \bibnamefont {Lee}},
  \bibinfo {author} {\bibfnamefont {Y.}~\bibnamefont {Yu}}, \bibinfo {author}
  {\bibfnamefont {S.}~\bibnamefont {Byun}}, \bibinfo {author} {\bibfnamefont
  {Z.-Z.}\ \bibnamefont {Luo}}, \bibinfo {author} {\bibfnamefont
  {H.}~\bibnamefont {Lee}}, \bibinfo {author} {\bibfnamefont {B.}~\bibnamefont
  {Ge}}, \bibinfo {author} {\bibfnamefont {Y.-L.}\ \bibnamefont {Lee}},
  \bibinfo {author} {\bibfnamefont {X.}~\bibnamefont {Chen}}, \bibinfo {author}
  {\bibfnamefont {J.~Y.}\ \bibnamefont {Lee}}, \bibinfo {author} {\bibfnamefont
  {O.}~\bibnamefont {Cojocaru-Mirédin}}, \bibinfo {author} {\bibfnamefont
  {H.}~\bibnamefont {Chang}}, \bibinfo {author} {\bibfnamefont
  {J.}~\bibnamefont {Im}}, \bibinfo {author} {\bibfnamefont {S.-P.}\
  \bibnamefont {Cho}}, \bibinfo {author} {\bibfnamefont {M.}~\bibnamefont
  {Wuttig}}, \bibinfo {author} {\bibfnamefont {V.~P.}\ \bibnamefont {Dravid}},
  \bibinfo {author} {\bibfnamefont {M.~G.}\ \bibnamefont {Kanatzidis}},\ and\
  \bibinfo {author} {\bibfnamefont {I.}~\bibnamefont {Chung}},\ }\bibfield
  {title} {\bibinfo {title} {Polycrystalline {SnSe} with a thermoelectric
  figure of merit greater than the single crystal},\ }\bibfield  {journal}
  {\bibinfo  {journal} {Nat. Mater.}\ }\textbf {\bibinfo {volume} {20}},\ \href
  {https://doi.org/10.1038/s41563-021-01064-6} {10.1038/s41563-021-01064-6}
  (\bibinfo {year} {2021})\BibitemShut {NoStop}%
\bibitem [{\citenamefont {Xiao}\ \emph {et~al.}(2016)\citenamefont {Xiao},
  \citenamefont {Chang}, \citenamefont {Pei}, \citenamefont {Wu}, \citenamefont
  {Peng}, \citenamefont {Zhou}, \citenamefont {Gong}, \citenamefont {He},
  \citenamefont {Zhang}, \citenamefont {Zeng},\ and\ \citenamefont
  {Zhao}}]{SnSeComputational}%
  \BibitemOpen
  \bibfield  {author} {\bibinfo {author} {\bibfnamefont {Y.}~\bibnamefont
  {Xiao}}, \bibinfo {author} {\bibfnamefont {C.}~\bibnamefont {Chang}},
  \bibinfo {author} {\bibfnamefont {Y.}~\bibnamefont {Pei}}, \bibinfo {author}
  {\bibfnamefont {D.}~\bibnamefont {Wu}}, \bibinfo {author} {\bibfnamefont
  {K.}~\bibnamefont {Peng}}, \bibinfo {author} {\bibfnamefont {X.}~\bibnamefont
  {Zhou}}, \bibinfo {author} {\bibfnamefont {S.}~\bibnamefont {Gong}}, \bibinfo
  {author} {\bibfnamefont {J.}~\bibnamefont {He}}, \bibinfo {author}
  {\bibfnamefont {Y.}~\bibnamefont {Zhang}}, \bibinfo {author} {\bibfnamefont
  {Z.}~\bibnamefont {Zeng}},\ and\ \bibinfo {author} {\bibfnamefont {L.-D.}\
  \bibnamefont {Zhao}},\ }\bibfield  {title} {\bibinfo {title} {Origin of low
  thermal conductivity in {SnSe}},\ }\href
  {https://doi.org/10.1103/PhysRevB.94.125203} {\bibfield  {journal} {\bibinfo
  {journal} {Phys. Rev. B}\ }\textbf {\bibinfo {volume} {94}},\ \bibinfo
  {pages} {125203} (\bibinfo {year} {2016})}\BibitemShut {NoStop}%
\bibitem [{\citenamefont {Caballero-Calero}\ \emph {et~al.}(2021)\citenamefont
  {Caballero-Calero}, \citenamefont {Ares},\ and\ \citenamefont
  {Martín-González}}]{EnvoronmentallyFriendlyTEMaterials}%
  \BibitemOpen
  \bibfield  {author} {\bibinfo {author} {\bibfnamefont {O.}~\bibnamefont
  {Caballero-Calero}}, \bibinfo {author} {\bibfnamefont {J.~R.}\ \bibnamefont
  {Ares}},\ and\ \bibinfo {author} {\bibfnamefont {M.}~\bibnamefont
  {Martín-González}},\ }\bibfield  {title} {\bibinfo {title} {Environmentally
  friendly thermoelectric materials: High performance from inorganic components
  with low toxicity and abundance in the earth},\ }\href
  {https://doi.org/https://doi.org/10.1002/adsu.202100095} {\bibfield
  {journal} {\bibinfo  {journal} {Adv. Sustain. Syst.}\ }\textbf {\bibinfo
  {volume} {5}},\ \bibinfo {pages} {2100095} (\bibinfo {year}
  {2021})}\BibitemShut {NoStop}%
\bibitem [{\citenamefont {Ren}\ \emph {et~al.}(2018)\citenamefont {Ren},
  \citenamefont {Liu}, \citenamefont {He}, \citenamefont {Lv}, \citenamefont
  {Gao},\ and\ \citenamefont {Xu}}]{TEToxcicity}%
  \BibitemOpen
  \bibfield  {author} {\bibinfo {author} {\bibfnamefont {P.}~\bibnamefont
  {Ren}}, \bibinfo {author} {\bibfnamefont {Y.}~\bibnamefont {Liu}}, \bibinfo
  {author} {\bibfnamefont {J.}~\bibnamefont {He}}, \bibinfo {author}
  {\bibfnamefont {T.}~\bibnamefont {Lv}}, \bibinfo {author} {\bibfnamefont
  {J.}~\bibnamefont {Gao}},\ and\ \bibinfo {author} {\bibfnamefont
  {G.}~\bibnamefont {Xu}},\ }\bibfield  {title} {\bibinfo {title} {Recent
  advances in inorganic material thermoelectrics},\ }\href
  {https://doi.org/10.1039/C8QI00366A} {\bibfield  {journal} {\bibinfo
  {journal} {Inorg. Chem. Front.}\ }\textbf {\bibinfo {volume} {5}},\ \bibinfo
  {pages} {2380} (\bibinfo {year} {2018})}\BibitemShut {NoStop}%
\bibitem [{\citenamefont {Chen}\ \emph {et~al.}(2019)\citenamefont {Chen},
  \citenamefont {Tran}, \citenamefont {Batra}, \citenamefont {Kim},\ and\
  \citenamefont {Ramprasad}}]{ChenGPRLTC}%
  \BibitemOpen
  \bibfield  {author} {\bibinfo {author} {\bibfnamefont {L.}~\bibnamefont
  {Chen}}, \bibinfo {author} {\bibfnamefont {H.}~\bibnamefont {Tran}}, \bibinfo
  {author} {\bibfnamefont {R.}~\bibnamefont {Batra}}, \bibinfo {author}
  {\bibfnamefont {C.}~\bibnamefont {Kim}},\ and\ \bibinfo {author}
  {\bibfnamefont {R.}~\bibnamefont {Ramprasad}},\ }\bibfield  {title} {\bibinfo
  {title} {Machine learning models for the lattice thermal conductivity
  prediction of inorganic materials},\ }\href
  {https://doi.org/https://doi.org/10.1016/j.commatsci.2019.109155} {\bibfield
  {journal} {\bibinfo  {journal} {Comput. Mater. Sci.}\ }\textbf {\bibinfo
  {volume} {170}},\ \bibinfo {pages} {109155} (\bibinfo {year}
  {2019})}\BibitemShut {NoStop}%
\bibitem [{\citenamefont {Antunes}\ \emph {et~al.}(2022)\citenamefont
  {Antunes}, \citenamefont {Vikram}, \citenamefont {Plata}, \citenamefont
  {Powell}, \citenamefont {Butler},\ and\ \citenamefont
  {Grau-Crespo}}]{MLinMaterialInformaticsMethodsandApplications}%
  \BibitemOpen
  \bibfield  {author} {\bibinfo {author} {\bibfnamefont {L.~M.}\ \bibnamefont
  {Antunes}}, \bibinfo {author} {\bibnamefont {Vikram}}, \bibinfo {author}
  {\bibfnamefont {J.~J.}\ \bibnamefont {Plata}}, \bibinfo {author}
  {\bibfnamefont {A.~V.}\ \bibnamefont {Powell}}, \bibinfo {author}
  {\bibfnamefont {K.~T.}\ \bibnamefont {Butler}},\ and\ \bibinfo {author}
  {\bibfnamefont {R.}~\bibnamefont {Grau-Crespo}},\ }\bibinfo {title} {Machine
  learning approaches for accelerating the discovery of thermoelectric
  materials},\ in\ \href {https://doi.org/10.1021/bk-2022-1416.ch001} {\emph
  {\bibinfo {booktitle} {Machine Learning in Materials Informatics: Methods and
  Applications}}}\ (\bibinfo {year} {2022})\ Chap.~\bibinfo {chapter} {1}, pp.\
  \bibinfo {pages} {1--32}\BibitemShut {NoStop}%
\bibitem [{\citenamefont {Togo}\ \emph {et~al.}(2015)\citenamefont {Togo},
  \citenamefont {Chaput},\ and\ \citenamefont {Tanaka}}]{phono3py}%
  \BibitemOpen
  \bibfield  {author} {\bibinfo {author} {\bibfnamefont {A.}~\bibnamefont
  {Togo}}, \bibinfo {author} {\bibfnamefont {L.}~\bibnamefont {Chaput}},\ and\
  \bibinfo {author} {\bibfnamefont {I.}~\bibnamefont {Tanaka}},\ }\bibfield
  {title} {\bibinfo {title} {Distributions of phonon lifetimes in brillouin
  zones},\ }\href {https://doi.org/10.1103/PhysRevB.91.094306} {\bibfield
  {journal} {\bibinfo  {journal} {Phys. Rev. B}\ }\textbf {\bibinfo {volume}
  {91}},\ \bibinfo {pages} {094306} (\bibinfo {year} {2015})}\BibitemShut
  {NoStop}%
\bibitem [{\citenamefont {Hellman}\ \emph {et~al.}(2011)\citenamefont
  {Hellman}, \citenamefont {Abrikosov},\ and\ \citenamefont
  {Simak}}]{hellmanLatticeDynamicsAnharmonic2011}%
  \BibitemOpen
  \bibfield  {author} {\bibinfo {author} {\bibfnamefont {O.}~\bibnamefont
  {Hellman}}, \bibinfo {author} {\bibfnamefont {I.~A.}\ \bibnamefont
  {Abrikosov}},\ and\ \bibinfo {author} {\bibfnamefont {S.~I.}\ \bibnamefont
  {Simak}},\ }\bibfield  {title} {\bibinfo {title} {Lattice dynamics of
  anharmonic solids from first principles},\ }\href
  {https://doi.org/10.1103/PhysRevB.84.180301} {\bibfield  {journal} {\bibinfo
  {journal} {Phys. Rev. B}\ }\textbf {\bibinfo {volume} {84}},\ \bibinfo
  {pages} {180301} (\bibinfo {year} {2011})}\BibitemShut {NoStop}%
\bibitem [{\citenamefont {Hellman}\ and\ \citenamefont
  {Abrikosov}(2013)}]{hellmanTemperaturedependentEffectiveThirdorder2013}%
  \BibitemOpen
  \bibfield  {author} {\bibinfo {author} {\bibfnamefont {O.}~\bibnamefont
  {Hellman}}\ and\ \bibinfo {author} {\bibfnamefont {I.~A.}\ \bibnamefont
  {Abrikosov}},\ }\bibfield  {title} {\bibinfo {title} {Temperature-dependent
  effective third-order interatomic force constants from first principles},\
  }\href {https://doi.org/10.1103/PhysRevB.88.144301} {\bibfield  {journal}
  {\bibinfo  {journal} {Phys. Rev. B}\ }\textbf {\bibinfo {volume} {88}},\
  \bibinfo {pages} {144301} (\bibinfo {year} {2013})}\BibitemShut {NoStop}%
\bibitem [{\citenamefont {Zhou}\ \emph {et~al.}(2014)\citenamefont {Zhou},
  \citenamefont {Nielson}, \citenamefont {Xia},\ and\ \citenamefont
  {Ozoli\ifmmode \mbox{\c{n}}\else \c{n}\fi{}\ifmmode~\check{s}\else
  \v{s}\fi{}}}]{CompressiveSensing}%
  \BibitemOpen
  \bibfield  {author} {\bibinfo {author} {\bibfnamefont {F.}~\bibnamefont
  {Zhou}}, \bibinfo {author} {\bibfnamefont {W.}~\bibnamefont {Nielson}},
  \bibinfo {author} {\bibfnamefont {Y.}~\bibnamefont {Xia}},\ and\ \bibinfo
  {author} {\bibfnamefont {V.}~\bibnamefont {Ozoli\ifmmode \mbox{\c{n}}\else
  \c{n}\fi{}\ifmmode~\check{s}\else \v{s}\fi{}}},\ }\bibfield  {title}
  {\bibinfo {title} {Lattice anharmonicity and thermal conductivity from
  compressive sensing of first-principles calculations},\ }\href
  {https://doi.org/10.1103/PhysRevLett.113.185501} {\bibfield  {journal}
  {\bibinfo  {journal} {Phys. Rev. Lett.}\ }\textbf {\bibinfo {volume} {113}},\
  \bibinfo {pages} {185501} (\bibinfo {year} {2014})}\BibitemShut {NoStop}%
\bibitem [{\citenamefont {Eriksson}\ \emph {et~al.}(2019)\citenamefont
  {Eriksson}, \citenamefont {Fransson},\ and\ \citenamefont
  {Erhart}}]{hiphive}%
  \BibitemOpen
  \bibfield  {author} {\bibinfo {author} {\bibfnamefont {F.}~\bibnamefont
  {Eriksson}}, \bibinfo {author} {\bibfnamefont {E.}~\bibnamefont {Fransson}},\
  and\ \bibinfo {author} {\bibfnamefont {P.}~\bibnamefont {Erhart}},\
  }\bibfield  {title} {\bibinfo {title} {The hiphive package for the extraction
  of high-order force constants by machine learning},\ }\href
  {https://doi.org/10.1002/adts.201800184} {\bibfield  {journal} {\bibinfo
  {journal} {Adv. Theory Simul.}\ }\textbf {\bibinfo {volume} {2}},\ \bibinfo
  {pages} {1800184} (\bibinfo {year} {2019})}\BibitemShut {NoStop}%
\bibitem [{\citenamefont {Li}\ \emph {et~al.}(2014)\citenamefont {Li},
  \citenamefont {Carrete}, \citenamefont {{A. Katcho}},\ and\ \citenamefont
  {Mingo}}]{LI20141747}%
  \BibitemOpen
  \bibfield  {author} {\bibinfo {author} {\bibfnamefont {W.}~\bibnamefont
  {Li}}, \bibinfo {author} {\bibfnamefont {J.}~\bibnamefont {Carrete}},
  \bibinfo {author} {\bibfnamefont {N.}~\bibnamefont {{A. Katcho}}},\ and\
  \bibinfo {author} {\bibfnamefont {N.}~\bibnamefont {Mingo}},\ }\bibfield
  {title} {\bibinfo {title} {Shengbte: A solver of the boltzmann transport
  equation for phonons},\ }\href
  {https://doi.org/https://doi.org/10.1016/j.cpc.2014.02.015} {\bibfield
  {journal} {\bibinfo  {journal} {Computer Physics Communications}\ }\textbf
  {\bibinfo {volume} {185}},\ \bibinfo {pages} {1747} (\bibinfo {year}
  {2014})}\BibitemShut {NoStop}%
\bibitem [{\citenamefont {Gorai}\ \emph {et~al.}(2017)\citenamefont {Gorai},
  \citenamefont {Stevanovic},\ and\ \citenamefont
  {Toberer}}]{TeScreeningReview}%
  \BibitemOpen
  \bibfield  {author} {\bibinfo {author} {\bibfnamefont {P.}~\bibnamefont
  {Gorai}}, \bibinfo {author} {\bibfnamefont {V.}~\bibnamefont {Stevanovic}},\
  and\ \bibinfo {author} {\bibfnamefont {E.}~\bibnamefont {Toberer}},\
  }\bibfield  {title} {\bibinfo {title} {Computationally guided discovery of
  thermoelectric materials},\ }\href
  {https://doi.org/10.1038/natrevmats.2017.53} {\bibfield  {journal} {\bibinfo
  {journal} {Nat. Rew. Mater.}\ }\textbf {\bibinfo {volume} {2}},\ \bibinfo
  {pages} {17503} (\bibinfo {year} {2017})}\BibitemShut {NoStop}%
\bibitem [{\citenamefont {Gan}\ \emph {et~al.}(2021)\citenamefont {Gan},
  \citenamefont {Wang}, \citenamefont {Zhou},\ and\ \citenamefont
  {Sun}}]{LayereTEScreening}%
  \BibitemOpen
  \bibfield  {author} {\bibinfo {author} {\bibfnamefont {Y.}~\bibnamefont
  {Gan}}, \bibinfo {author} {\bibfnamefont {G.}~\bibnamefont {Wang}}, \bibinfo
  {author} {\bibfnamefont {J.}~\bibnamefont {Zhou}},\ and\ \bibinfo {author}
  {\bibfnamefont {Z.}~\bibnamefont {Sun}},\ }\bibfield  {title} {\bibinfo
  {title} {Prediction of thermoelectric performance for layered {IV-V-VI}
  semiconductors by high-throughput ab initio calculations and machine
  learning},\ }\bibfield  {journal} {\bibinfo  {journal} {npj Comput. Mater.}\
  }\textbf {\bibinfo {volume} {7}},\ \href
  {https://doi.org/10.1038/s41524-021-00645-y} {10.1038/s41524-021-00645-y}
  (\bibinfo {year} {2021})\BibitemShut {NoStop}%
\bibitem [{\citenamefont {Jia}\ \emph {et~al.}(2020)\citenamefont {Jia},
  \citenamefont {Feng}, \citenamefont {Guo}, \citenamefont {Zhang},\ and\
  \citenamefont {Zhang}}]{ScreeningTEBinary}%
  \BibitemOpen
  \bibfield  {author} {\bibinfo {author} {\bibfnamefont {T.}~\bibnamefont
  {Jia}}, \bibinfo {author} {\bibfnamefont {Z.}~\bibnamefont {Feng}}, \bibinfo
  {author} {\bibfnamefont {S.}~\bibnamefont {Guo}}, \bibinfo {author}
  {\bibfnamefont {X.}~\bibnamefont {Zhang}},\ and\ \bibinfo {author}
  {\bibfnamefont {Y.}~\bibnamefont {Zhang}},\ }\bibfield  {title} {\bibinfo
  {title} {Screening promising thermoelectric materials in binary chalcogenides
  through high-throughput computations},\ }\href
  {https://doi.org/10.1021/acsami.9b23297} {\bibfield  {journal} {\bibinfo
  {journal} {ACS Appl. Mater. Interfaces.}\ }\textbf {\bibinfo {volume} {12}},\
  \bibinfo {pages} {11852} (\bibinfo {year} {2020})},\ \bibinfo {note} {pMID:
  32069390}\BibitemShut {NoStop}%
\bibitem [{\citenamefont {Li}\ \emph {et~al.}(2022)\citenamefont {Li},
  \citenamefont {Hu},\ and\ \citenamefont {Yang}}]{LTCScreeningRattling}%
  \BibitemOpen
  \bibfield  {author} {\bibinfo {author} {\bibfnamefont {J.}~\bibnamefont
  {Li}}, \bibinfo {author} {\bibfnamefont {W.}~\bibnamefont {Hu}},\ and\
  \bibinfo {author} {\bibfnamefont {J.}~\bibnamefont {Yang}},\ }\bibfield
  {title} {\bibinfo {title} {High-throughput screening of rattling-induced
  ultralow lattice thermal conductivity in semiconductors},\ }\href
  {https://doi.org/10.1021/jacs.1c11887} {\bibfield  {journal} {\bibinfo
  {journal} {J. Am. Chem. Soc.}\ }\textbf {\bibinfo {volume} {144}},\ \bibinfo
  {pages} {4448} (\bibinfo {year} {2022})},\ \bibinfo {note} {pMID:
  35230828}\BibitemShut {NoStop}%
\bibitem [{\citenamefont {{Pal}}\ \emph {et~al.}(2021)\citenamefont {{Pal}},
  \citenamefont {{Xia}}, \citenamefont {{Shen}}, \citenamefont {{He}},
  \citenamefont {{Luo}}, \citenamefont {{Kanatzidis}},\ and\ \citenamefont
  {{Wolverton}}}]{HighThroughputRattler}%
  \BibitemOpen
  \bibfield  {author} {\bibinfo {author} {\bibfnamefont {K.}~\bibnamefont
  {{Pal}}}, \bibinfo {author} {\bibfnamefont {Y.}~\bibnamefont {{Xia}}},
  \bibinfo {author} {\bibfnamefont {J.}~\bibnamefont {{Shen}}}, \bibinfo
  {author} {\bibfnamefont {J.}~\bibnamefont {{He}}}, \bibinfo {author}
  {\bibfnamefont {Y.}~\bibnamefont {{Luo}}}, \bibinfo {author} {\bibfnamefont
  {M.~G.}\ \bibnamefont {{Kanatzidis}}},\ and\ \bibinfo {author} {\bibfnamefont
  {C.}~\bibnamefont {{Wolverton}}},\ }\bibfield  {title} {\bibinfo {title}
  {{Accelerated discovery of a large family of quaternary chalcogenides with
  very low lattice thermal conductivity}},\ }\href
  {https://doi.org/10.1038/s41524-021-00549-x} {\bibfield  {journal} {\bibinfo
  {journal} {npj Comput. Mater}\ }\textbf {\bibinfo {volume} {7}},\ \bibinfo
  {eid} {82} (\bibinfo {year} {2021})}\BibitemShut {NoStop}%
\bibitem [{\citenamefont {Carrete}\ \emph {et~al.}(2014)\citenamefont
  {Carrete}, \citenamefont {Li}, \citenamefont {Mingo}, \citenamefont {Wang},\
  and\ \citenamefont {Curtarolo}}]{CarreteFindingLowLTCHHs}%
  \BibitemOpen
  \bibfield  {author} {\bibinfo {author} {\bibfnamefont {J.}~\bibnamefont
  {Carrete}}, \bibinfo {author} {\bibfnamefont {W.}~\bibnamefont {Li}},
  \bibinfo {author} {\bibfnamefont {N.}~\bibnamefont {Mingo}}, \bibinfo
  {author} {\bibfnamefont {S.}~\bibnamefont {Wang}},\ and\ \bibinfo {author}
  {\bibfnamefont {S.}~\bibnamefont {Curtarolo}},\ }\bibfield  {title} {\bibinfo
  {title} {Finding unprecedentedly low-thermal-conductivity half-{H}eusler
  semiconductors via high-throughput materials modeling},\ }\href
  {https://doi.org/10.1103/PhysRevX.4.011019} {\bibfield  {journal} {\bibinfo
  {journal} {Phys. Rev. X}\ }\textbf {\bibinfo {volume} {4}} (\bibinfo {year}
  {2014})}\BibitemShut {NoStop}%
\bibitem [{\citenamefont {Feng}\ \emph {et~al.}(2020)\citenamefont {Feng},
  \citenamefont {Fu}, \citenamefont {Zhang},\ and\ \citenamefont
  {Singh}}]{fengCharacterizationRattlingRelation2020}%
  \BibitemOpen
  \bibfield  {author} {\bibinfo {author} {\bibfnamefont {Z.}~\bibnamefont
  {Feng}}, \bibinfo {author} {\bibfnamefont {Y.}~\bibnamefont {Fu}}, \bibinfo
  {author} {\bibfnamefont {Y.}~\bibnamefont {Zhang}},\ and\ \bibinfo {author}
  {\bibfnamefont {D.~J.}\ \bibnamefont {Singh}},\ }\bibfield  {title} {\bibinfo
  {title} {Characterization of rattling in relation to thermal conductivity:
  ordered half-{Heusler} semiconductors},\ }\href
  {https://doi.org/10.1103/PhysRevB.101.064301} {\bibfield  {journal} {\bibinfo
   {journal} {Phys. Rev. B}\ }\textbf {\bibinfo {volume} {101}},\ \bibinfo
  {pages} {064301} (\bibinfo {year} {2020})}\BibitemShut {NoStop}%
\bibitem [{\citenamefont {Seko}\ \emph {et~al.}(2015)\citenamefont {Seko},
  \citenamefont {Togo}, \citenamefont {Hayashi}, \citenamefont {Tsuda},
  \citenamefont {Chaput},\ and\ \citenamefont {Tanaka}}]{SekoBayesianLowLTC}%
  \BibitemOpen
  \bibfield  {author} {\bibinfo {author} {\bibfnamefont {A.}~\bibnamefont
  {Seko}}, \bibinfo {author} {\bibfnamefont {A.}~\bibnamefont {Togo}}, \bibinfo
  {author} {\bibfnamefont {H.}~\bibnamefont {Hayashi}}, \bibinfo {author}
  {\bibfnamefont {K.}~\bibnamefont {Tsuda}}, \bibinfo {author} {\bibfnamefont
  {L.}~\bibnamefont {Chaput}},\ and\ \bibinfo {author} {\bibfnamefont
  {I.}~\bibnamefont {Tanaka}},\ }\bibfield  {title} {\bibinfo {title}
  {Prediction of low-thermal-conductivity compounds with first-principles
  anharmonic lattice-dynamics calculations and bayesian optimization},\ }\href
  {https://doi.org/10.1103/PhysRevLett.115.205901} {\bibfield  {journal}
  {\bibinfo  {journal} {Phys. Rev. Lett.}\ }\textbf {\bibinfo {volume} {115}},\
  \bibinfo {pages} {205901} (\bibinfo {year} {2015})}\BibitemShut {NoStop}%
\bibitem [{\citenamefont {Wang}\ \emph
  {et~al.}(2022{\natexlab{a}})\citenamefont {Wang}, \citenamefont {Zhao},
  \citenamefont {Zeng}, \citenamefont {Wang}, \citenamefont {Chen},\ and\
  \citenamefont {Ni}}]{LowLTCPerovskiteMLScreeening}%
  \BibitemOpen
  \bibfield  {author} {\bibinfo {author} {\bibfnamefont {X.}~\bibnamefont
  {Wang}}, \bibinfo {author} {\bibfnamefont {Y.}~\bibnamefont {Zhao}}, \bibinfo
  {author} {\bibfnamefont {S.}~\bibnamefont {Zeng}}, \bibinfo {author}
  {\bibfnamefont {Z.}~\bibnamefont {Wang}}, \bibinfo {author} {\bibfnamefont
  {Y.}~\bibnamefont {Chen}},\ and\ \bibinfo {author} {\bibfnamefont
  {J.}~\bibnamefont {Ni}},\ }\bibfield  {title} {\bibinfo {title} {Cubic halide
  perovskites as potential low thermal conductivity materials: A combined
  approach of machine learning and first-principles calculations},\ }\href
  {https://doi.org/10.1103/PhysRevB.105.014310} {\bibfield  {journal} {\bibinfo
   {journal} {Phys. Rev. B}\ }\textbf {\bibinfo {volume} {105}},\ \bibinfo
  {pages} {014310} (\bibinfo {year} {2022}{\natexlab{a}})}\BibitemShut
  {NoStop}%
\bibitem [{\citenamefont {Juneja}\ \emph {et~al.}(2019)\citenamefont {Juneja},
  \citenamefont {Yumnam}, \citenamefont {Satsangi},\ and\ \citenamefont
  {Singh}}]{JunejaHTGPR}%
  \BibitemOpen
  \bibfield  {author} {\bibinfo {author} {\bibfnamefont {R.}~\bibnamefont
  {Juneja}}, \bibinfo {author} {\bibfnamefont {G.}~\bibnamefont {Yumnam}},
  \bibinfo {author} {\bibfnamefont {S.}~\bibnamefont {Satsangi}},\ and\
  \bibinfo {author} {\bibfnamefont {A.~K.}\ \bibnamefont {Singh}},\ }\bibfield
  {title} {\bibinfo {title} {Coupling the high-throughput property map to
  machine learning for predicting lattice thermal conductivity},\ }\href
  {https://doi.org/10.1021/acs.chemmater.9b01046} {\bibfield  {journal}
  {\bibinfo  {journal} {Chem. Mater.}\ }\textbf {\bibinfo {volume} {31}},\
  \bibinfo {pages} {5145} (\bibinfo {year} {2019})}\BibitemShut {NoStop}%
\bibitem [{\citenamefont {Wang}\ \emph {et~al.}(2020)\citenamefont {Wang},
  \citenamefont {Zeng}, \citenamefont {Wang},\ and\ \citenamefont
  {Ni}}]{WangMLScreeningLowLTCAflow}%
  \BibitemOpen
  \bibfield  {author} {\bibinfo {author} {\bibfnamefont {X.}~\bibnamefont
  {Wang}}, \bibinfo {author} {\bibfnamefont {S.}~\bibnamefont {Zeng}}, \bibinfo
  {author} {\bibfnamefont {Z.}~\bibnamefont {Wang}},\ and\ \bibinfo {author}
  {\bibfnamefont {J.}~\bibnamefont {Ni}},\ }\bibfield  {title} {\bibinfo
  {title} {Identification of crystalline materials with ultra-low thermal
  conductivity based on machine learning study},\ }\href
  {https://doi.org/10.1021/acs.jpcc.9b11610} {\bibfield  {journal} {\bibinfo
  {journal} {J. Phys. Chem. C}\ }\textbf {\bibinfo {volume} {124}},\ \bibinfo
  {pages} {8488} (\bibinfo {year} {2020})}\BibitemShut {NoStop}%
\bibitem [{\citenamefont {Jaafreh}\ \emph {et~al.}(2021)\citenamefont
  {Jaafreh}, \citenamefont {Kang},\ and\ \citenamefont
  {Hamad}}]{JaafrehAccelerated}%
  \BibitemOpen
  \bibfield  {author} {\bibinfo {author} {\bibfnamefont {R.}~\bibnamefont
  {Jaafreh}}, \bibinfo {author} {\bibfnamefont {Y.~S.}\ \bibnamefont {Kang}},\
  and\ \bibinfo {author} {\bibfnamefont {K.}~\bibnamefont {Hamad}},\ }\bibfield
   {title} {\bibinfo {title} {Lattice thermal conductivity: An accelerated
  discovery guided by machine learning},\ }\href
  {https://doi.org/10.1021/acsami.1c17378} {\bibfield  {journal} {\bibinfo
  {journal} {ACS Appl. Mater. Interfaces}\ }\textbf {\bibinfo {volume} {13}},\
  \bibinfo {pages} {57204} (\bibinfo {year} {2021})}\BibitemShut {NoStop}%
\bibitem [{\citenamefont {Pal}\ \emph {et~al.}(2022)\citenamefont {Pal},
  \citenamefont {Park}, \citenamefont {Xia}, \citenamefont {Shen},\ and\
  \citenamefont {Wolverton}}]{WolvertonLTCML}%
  \BibitemOpen
  \bibfield  {author} {\bibinfo {author} {\bibfnamefont {K.}~\bibnamefont
  {Pal}}, \bibinfo {author} {\bibfnamefont {C.}~\bibnamefont {Park}}, \bibinfo
  {author} {\bibfnamefont {Y.}~\bibnamefont {Xia}}, \bibinfo {author}
  {\bibfnamefont {J.}~\bibnamefont {Shen}},\ and\ \bibinfo {author}
  {\bibfnamefont {C.}~\bibnamefont {Wolverton}},\ }\bibfield  {title} {\bibinfo
  {title} {Scale-invariant machine-learning model accelerates the discovery of
  quaternary chalcogenides with ultralow lattice thermal conductivity},\
  }\href@noop {} {\bibfield  {journal} {\bibinfo  {journal} {npj Comput.
  Mater.}\ }\textbf {\bibinfo {volume} {8}} (\bibinfo {year}
  {2022})}\BibitemShut {NoStop}%
\bibitem [{\citenamefont {Choi}\ \emph {et~al.}(2022)\citenamefont {Choi},
  \citenamefont {Lee}, \citenamefont {Kim}, \citenamefont {Moon}, \citenamefont
  {Jeong},\ and\ \citenamefont {Han}}]{MLPLTC}%
  \BibitemOpen
  \bibfield  {author} {\bibinfo {author} {\bibfnamefont {J.~M.}\ \bibnamefont
  {Choi}}, \bibinfo {author} {\bibfnamefont {K.}~\bibnamefont {Lee}}, \bibinfo
  {author} {\bibfnamefont {S.}~\bibnamefont {Kim}}, \bibinfo {author}
  {\bibfnamefont {M.}~\bibnamefont {Moon}}, \bibinfo {author} {\bibfnamefont
  {W.}~\bibnamefont {Jeong}},\ and\ \bibinfo {author} {\bibfnamefont
  {S.}~\bibnamefont {Han}},\ }\bibfield  {title} {\bibinfo {title} {Accelerated
  computation of lattice thermal conductivity using neural network interatomic
  potentials},\ }\href
  {https://doi.org/https://doi.org/10.1016/j.commatsci.2022.111472} {\bibfield
  {journal} {\bibinfo  {journal} {Comput. Mater. Sci.}\ }\textbf {\bibinfo
  {volume} {211}},\ \bibinfo {pages} {111472} (\bibinfo {year}
  {2022})}\BibitemShut {NoStop}%
\bibitem [{\citenamefont {Løvvik}\ \emph {et~al.}(2020)\citenamefont
  {Løvvik}, \citenamefont {Flage-Larsen},\ and\ \citenamefont
  {Skomedal}}]{OleMartinScreeningSilicides}%
  \BibitemOpen
  \bibfield  {author} {\bibinfo {author} {\bibfnamefont {O.~M.}\ \bibnamefont
  {Løvvik}}, \bibinfo {author} {\bibfnamefont {E.}~\bibnamefont
  {Flage-Larsen}},\ and\ \bibinfo {author} {\bibfnamefont {G.}~\bibnamefont
  {Skomedal}},\ }\bibfield  {title} {\bibinfo {title} {Screening of
  thermoelectric silicides with atomistic transport calculations},\ }\href
  {https://doi.org/10.1063/5.0008198} {\bibfield  {journal} {\bibinfo
  {journal} {J. Appl. Phys.}\ }\textbf {\bibinfo {volume} {128}},\ \bibinfo
  {pages} {125105} (\bibinfo {year} {2020})}\BibitemShut {NoStop}%
\bibitem [{\citenamefont {Juneja}\ and\ \citenamefont
  {Singh}(2020)}]{JunejaMLLTC}%
  \BibitemOpen
  \bibfield  {author} {\bibinfo {author} {\bibfnamefont {R.}~\bibnamefont
  {Juneja}}\ and\ \bibinfo {author} {\bibfnamefont {A.~K.}\ \bibnamefont
  {Singh}},\ }\bibfield  {title} {\bibinfo {title} {Guided patchwork kriging to
  develop highly transferable thermal conductivity prediction models},\ }\href
  {https://doi.org/10.1088/2515-7639/ab78f2} {\bibfield  {journal} {\bibinfo
  {journal} {J. Phys. Mater.}\ }\textbf {\bibinfo {volume} {3}},\ \bibinfo
  {pages} {024006} (\bibinfo {year} {2020})}\BibitemShut {NoStop}%
\bibitem [{\citenamefont {Luo}\ \emph {et~al.}(2023)\citenamefont {Luo},
  \citenamefont {Li}, \citenamefont {Yuan}, \citenamefont {Liu},\ and\
  \citenamefont {Fang}}]{MLLTCMiniReview}%
  \BibitemOpen
  \bibfield  {author} {\bibinfo {author} {\bibfnamefont {Y.}~\bibnamefont
  {Luo}}, \bibinfo {author} {\bibfnamefont {M.}~\bibnamefont {Li}}, \bibinfo
  {author} {\bibfnamefont {H.}~\bibnamefont {Yuan}}, \bibinfo {author}
  {\bibfnamefont {H.}~\bibnamefont {Liu}},\ and\ \bibinfo {author}
  {\bibfnamefont {Y.}~\bibnamefont {Fang}},\ }\bibfield  {title} {\bibinfo
  {title} {Predicting lattice thermal conductivity via machine learning: a mini
  review},\ }\bibfield  {journal} {\bibinfo  {journal} {npj Comput. Mater.}\
  }\textbf {\bibinfo {volume} {9}},\ \href
  {https://doi.org/10.1038/s41524-023-00964-2} {10.1038/s41524-023-00964-2}
  (\bibinfo {year} {2023})\BibitemShut {NoStop}%
\bibitem [{\citenamefont {Jain}\ \emph {et~al.}(2013)\citenamefont {Jain},
  \citenamefont {Ong}, \citenamefont {Hautier}, \citenamefont {Chen},
  \citenamefont {Richards}, \citenamefont {Dacek}, \citenamefont {Cholia},
  \citenamefont {Gunter}, \citenamefont {Skinner}, \citenamefont {Ceder},\ and\
  \citenamefont {Persson}}]{jainCommentaryMaterialsProject2013}%
  \BibitemOpen
  \bibfield  {author} {\bibinfo {author} {\bibfnamefont {A.}~\bibnamefont
  {Jain}}, \bibinfo {author} {\bibfnamefont {S.~P.}\ \bibnamefont {Ong}},
  \bibinfo {author} {\bibfnamefont {G.}~\bibnamefont {Hautier}}, \bibinfo
  {author} {\bibfnamefont {W.}~\bibnamefont {Chen}}, \bibinfo {author}
  {\bibfnamefont {W.~D.}\ \bibnamefont {Richards}}, \bibinfo {author}
  {\bibfnamefont {S.}~\bibnamefont {Dacek}}, \bibinfo {author} {\bibfnamefont
  {S.}~\bibnamefont {Cholia}}, \bibinfo {author} {\bibfnamefont
  {D.}~\bibnamefont {Gunter}}, \bibinfo {author} {\bibfnamefont
  {D.}~\bibnamefont {Skinner}}, \bibinfo {author} {\bibfnamefont
  {G.}~\bibnamefont {Ceder}},\ and\ \bibinfo {author} {\bibfnamefont {K.~A.}\
  \bibnamefont {Persson}},\ }\bibfield  {title} {\bibinfo {title} {Commentary:
  {The} {Materials} {Project}: {A} materials genome approach to accelerating
  materials innovation},\ }\href {https://doi.org/10.1063/1.4812323} {\bibfield
   {journal} {\bibinfo  {journal} {APL Mater.}\ }\textbf {\bibinfo {volume}
  {1}},\ \bibinfo {pages} {011002} (\bibinfo {year} {2013})}\BibitemShut
  {NoStop}%
\bibitem [{\citenamefont {Saal}\ \emph {et~al.}(2013)\citenamefont {Saal},
  \citenamefont {Kirklin}, \citenamefont {Aykol}, \citenamefont {Meredig},\
  and\ \citenamefont {Wolverton}}]{OQMD}%
  \BibitemOpen
  \bibfield  {author} {\bibinfo {author} {\bibfnamefont {J.}~\bibnamefont
  {Saal}}, \bibinfo {author} {\bibfnamefont {S.}~\bibnamefont {Kirklin}},
  \bibinfo {author} {\bibfnamefont {M.}~\bibnamefont {Aykol}}, \bibinfo
  {author} {\bibfnamefont {B.}~\bibnamefont {Meredig}},\ and\ \bibinfo {author}
  {\bibfnamefont {C.}~\bibnamefont {Wolverton}},\ }\bibfield  {title} {\bibinfo
  {title} {Materials design and discovery with high-throughput density
  functional theory: The open quantum materials database ({OQMD})},\ }\href
  {https://doi.org/10.1007/s11837-013-0755-4} {\bibfield  {journal} {\bibinfo
  {journal} {JOM}\ }\textbf {\bibinfo {volume} {65}} (\bibinfo {year}
  {2013})}\BibitemShut {NoStop}%
\bibitem [{\citenamefont {Draxl}\ and\ \citenamefont
  {Scheffler}(2019)}]{NOMAD}%
  \BibitemOpen
  \bibfield  {author} {\bibinfo {author} {\bibfnamefont {C.}~\bibnamefont
  {Draxl}}\ and\ \bibinfo {author} {\bibfnamefont {M.}~\bibnamefont
  {Scheffler}},\ }\bibfield  {title} {\bibinfo {title} {The {NOMAD} laboratory:
  from data sharing to artificial intelligence},\ }\href
  {https://doi.org/10.1088/2515-7639/ab13bb} {\bibfield  {journal} {\bibinfo
  {journal} {JPhys Mater.}\ }\textbf {\bibinfo {volume} {2}},\ \bibinfo {pages}
  {036001} (\bibinfo {year} {2019})}\BibitemShut {NoStop}%
\bibitem [{\citenamefont {Zhang}\ and\ \citenamefont
  {Ling}(2018)}]{StrategyMLSmallDatasets}%
  \BibitemOpen
  \bibfield  {author} {\bibinfo {author} {\bibfnamefont {Y.}~\bibnamefont
  {Zhang}}\ and\ \bibinfo {author} {\bibfnamefont {C.}~\bibnamefont {Ling}},\
  }\bibfield  {title} {\bibinfo {title} {A strategy to apply machine learning
  to small datasets in materials science},\ }\href
  {https://doi.org/10.1038/s41524-018-0081-z} {\bibfield  {journal} {\bibinfo
  {journal} {npj Comput. Mater.}\ }\textbf {\bibinfo {volume} {4}} (\bibinfo
  {year} {2018})}\BibitemShut {NoStop}%
\bibitem [{\citenamefont {Tranås}\ \emph {et~al.}(2022)\citenamefont
  {Tranås}, \citenamefont {Løvvik}, \citenamefont {Tomic},\ and\
  \citenamefont {Berland}}]{TranasHHActiveSampling}%
  \BibitemOpen
  \bibfield  {author} {\bibinfo {author} {\bibfnamefont {R.}~\bibnamefont
  {Tranås}}, \bibinfo {author} {\bibfnamefont {O.~M.}\ \bibnamefont
  {Løvvik}}, \bibinfo {author} {\bibfnamefont {O.}~\bibnamefont {Tomic}},\
  and\ \bibinfo {author} {\bibfnamefont {K.}~\bibnamefont {Berland}},\
  }\bibfield  {title} {\bibinfo {title} {Lattice thermal conductivity of
  half-{H}euslers with density functional theory and machine learning:
  Enhancing predictivity by active sampling with principal component
  analysis},\ }\href
  {https://doi.org/https://doi.org/10.1016/j.commatsci.2021.110938} {\bibfield
  {journal} {\bibinfo  {journal} {Comput. Mater. Sci.}\ }\textbf {\bibinfo
  {volume} {202}},\ \bibinfo {pages} {110938} (\bibinfo {year}
  {2022})}\BibitemShut {NoStop}%
\bibitem [{\citenamefont {Wei}\ \emph {et~al.}(2019)\citenamefont {Wei},
  \citenamefont {Yu}, \citenamefont {Yang}, \citenamefont {Rao}, \citenamefont
  {Wang}, \citenamefont {Chi}, \citenamefont {Sun},\ and\ \citenamefont
  {Hong}}]{CsI}%
  \BibitemOpen
  \bibfield  {author} {\bibinfo {author} {\bibfnamefont {B.}~\bibnamefont
  {Wei}}, \bibinfo {author} {\bibfnamefont {X.}~\bibnamefont {Yu}}, \bibinfo
  {author} {\bibfnamefont {C.}~\bibnamefont {Yang}}, \bibinfo {author}
  {\bibfnamefont {X.}~\bibnamefont {Rao}}, \bibinfo {author} {\bibfnamefont
  {X.}~\bibnamefont {Wang}}, \bibinfo {author} {\bibfnamefont {S.}~\bibnamefont
  {Chi}}, \bibinfo {author} {\bibfnamefont {X.}~\bibnamefont {Sun}},\ and\
  \bibinfo {author} {\bibfnamefont {J.}~\bibnamefont {Hong}},\ }\bibfield
  {title} {\bibinfo {title} {Low-temperature anharmonicity and the thermal
  conductivity of {C}esium {I}odide},\ }\href
  {https://doi.org/10.1103/PhysRevB.99.184301} {\bibfield  {journal} {\bibinfo
  {journal} {Phys. Rev. B}\ }\textbf {\bibinfo {volume} {99}},\ \bibinfo
  {pages} {184301} (\bibinfo {year} {2019})}\BibitemShut {NoStop}%
\bibitem [{\citenamefont {Mukhopadhyay}\ \emph {et~al.}(2017)\citenamefont
  {Mukhopadhyay}, \citenamefont {Bansal}, \citenamefont {Delaire},
  \citenamefont {Perrodin}, \citenamefont {Bourret-Courchesne}, \citenamefont
  {Singh},\ and\ \citenamefont {Lindsay}}]{CuClF43m}%
  \BibitemOpen
  \bibfield  {author} {\bibinfo {author} {\bibfnamefont {S.}~\bibnamefont
  {Mukhopadhyay}}, \bibinfo {author} {\bibfnamefont {D.}~\bibnamefont
  {Bansal}}, \bibinfo {author} {\bibfnamefont {O.}~\bibnamefont {Delaire}},
  \bibinfo {author} {\bibfnamefont {D.}~\bibnamefont {Perrodin}}, \bibinfo
  {author} {\bibfnamefont {E.}~\bibnamefont {Bourret-Courchesne}}, \bibinfo
  {author} {\bibfnamefont {D.~J.}\ \bibnamefont {Singh}},\ and\ \bibinfo
  {author} {\bibfnamefont {L.}~\bibnamefont {Lindsay}},\ }\bibfield  {title}
  {\bibinfo {title} {The curious case of cuprous chloride: Giant thermal
  resistance and anharmonic quasiparticle spectra driven by dispersion
  nesting},\ }\href {https://doi.org/10.1103/PhysRevB.96.100301} {\bibfield
  {journal} {\bibinfo  {journal} {Phys. Rev. B}\ }\textbf {\bibinfo {volume}
  {96}},\ \bibinfo {pages} {100301} (\bibinfo {year} {2017})}\BibitemShut
  {NoStop}%
\bibitem [{\citenamefont {Ju}\ \emph {et~al.}(2021)\citenamefont {Ju},
  \citenamefont {Yoshida}, \citenamefont {Liu}, \citenamefont {Wu},
  \citenamefont {Hongo}, \citenamefont {Tadano},\ and\ \citenamefont
  {Shiomi}}]{TransferLearningExploring}%
  \BibitemOpen
  \bibfield  {author} {\bibinfo {author} {\bibfnamefont {S.}~\bibnamefont
  {Ju}}, \bibinfo {author} {\bibfnamefont {R.}~\bibnamefont {Yoshida}},
  \bibinfo {author} {\bibfnamefont {C.}~\bibnamefont {Liu}}, \bibinfo {author}
  {\bibfnamefont {S.}~\bibnamefont {Wu}}, \bibinfo {author} {\bibfnamefont
  {K.}~\bibnamefont {Hongo}}, \bibinfo {author} {\bibfnamefont
  {T.}~\bibnamefont {Tadano}},\ and\ \bibinfo {author} {\bibfnamefont
  {J.}~\bibnamefont {Shiomi}},\ }\bibfield  {title} {\bibinfo {title}
  {Exploring diamond-like lattice thermal conductivity crystals via
  feature-based transfer learning},\ }\href
  {https://doi.org/10.1103/PhysRevMaterials.5.053801} {\bibfield  {journal}
  {\bibinfo  {journal} {Phys. Rev. Mater.}\ }\textbf {\bibinfo {volume} {5}},\
  \bibinfo {pages} {053801} (\bibinfo {year} {2021})}\BibitemShut {NoStop}%
\bibitem [{\citenamefont {Pandey}\ \emph {et~al.}(2020)\citenamefont {Pandey},
  \citenamefont {Lindsay}, \citenamefont {Sales},\ and\ \citenamefont
  {Parker}}]{TlBrLowLTC}%
  \BibitemOpen
  \bibfield  {author} {\bibinfo {author} {\bibfnamefont {T.}~\bibnamefont
  {Pandey}}, \bibinfo {author} {\bibfnamefont {L.}~\bibnamefont {Lindsay}},
  \bibinfo {author} {\bibfnamefont {B.~C.}\ \bibnamefont {Sales}},\ and\
  \bibinfo {author} {\bibfnamefont {D.~S.}\ \bibnamefont {Parker}},\ }\bibfield
   {title} {\bibinfo {title} {Lattice instabilities and phonon thermal
  transport in {TlBr}},\ }\href
  {https://doi.org/10.1103/PhysRevMaterials.4.045403} {\bibfield  {journal}
  {\bibinfo  {journal} {Phys. Rev. Mater.}\ }\textbf {\bibinfo {volume} {4}},\
  \bibinfo {pages} {045403} (\bibinfo {year} {2020})}\BibitemShut {NoStop}%
\bibitem [{\citenamefont {Yao}\ \emph {et~al.}(2017)\citenamefont {Yao},
  \citenamefont {Zhang}, \citenamefont {Pei}, \citenamefont {Liu},\ and\
  \citenamefont {Li}}]{Cu2S}%
  \BibitemOpen
  \bibfield  {author} {\bibinfo {author} {\bibfnamefont {Y.}~\bibnamefont
  {Yao}}, \bibinfo {author} {\bibfnamefont {B.-P.}\ \bibnamefont {Zhang}},
  \bibinfo {author} {\bibfnamefont {J.}~\bibnamefont {Pei}}, \bibinfo {author}
  {\bibfnamefont {Y.-C.}\ \bibnamefont {Liu}},\ and\ \bibinfo {author}
  {\bibfnamefont {J.-F.}\ \bibnamefont {Li}},\ }\bibfield  {title} {\bibinfo
  {title} {Thermoelectric performance enhancement of
  {$\mathrm{Cu}_{2}\mathrm{S}$} by {S}e doping leading to a simultaneous power
  factor increase and thermal conductivity reduction},\ }\href
  {https://doi.org/10.1039/C7TC01937H} {\bibfield  {journal} {\bibinfo
  {journal} {J. Mater. Chem. C}\ }\textbf {\bibinfo {volume} {5}},\ \bibinfo
  {pages} {7845} (\bibinfo {year} {2017})}\BibitemShut {NoStop}%
\bibitem [{\citenamefont {Chen}\ \emph {et~al.}(2018)\citenamefont {Chen},
  \citenamefont {Yue}, \citenamefont {Ren}, \citenamefont {Zeng}, \citenamefont
  {Wei}, \citenamefont {Zhao}, \citenamefont {Yang}, \citenamefont {Qiu},
  \citenamefont {Chen},\ and\ \citenamefont {Shi}}]{Ag2SLowLTC}%
  \BibitemOpen
  \bibfield  {author} {\bibinfo {author} {\bibfnamefont {H.}~\bibnamefont
  {Chen}}, \bibinfo {author} {\bibfnamefont {Z.}~\bibnamefont {Yue}}, \bibinfo
  {author} {\bibfnamefont {D.}~\bibnamefont {Ren}}, \bibinfo {author}
  {\bibfnamefont {H.}~\bibnamefont {Zeng}}, \bibinfo {author} {\bibfnamefont
  {T.}~\bibnamefont {Wei}}, \bibinfo {author} {\bibfnamefont {K.}~\bibnamefont
  {Zhao}}, \bibinfo {author} {\bibfnamefont {R.}~\bibnamefont {Yang}}, \bibinfo
  {author} {\bibfnamefont {P.}~\bibnamefont {Qiu}}, \bibinfo {author}
  {\bibfnamefont {L.}~\bibnamefont {Chen}},\ and\ \bibinfo {author}
  {\bibfnamefont {X.}~\bibnamefont {Shi}},\ }\bibfield  {title} {\bibinfo
  {title} {Thermal conductivity during phase transitions},\ }\href
  {https://doi.org/10.1002/adma.201806518} {\bibfield  {journal} {\bibinfo
  {journal} {Adv. Mater.}\ ,\ \bibinfo {pages} {1806518}} (\bibinfo {year}
  {2018})}\BibitemShut {NoStop}%
\bibitem [{\citenamefont {He}\ \emph {et~al.}(2016)\citenamefont {He},
  \citenamefont {Amsler}, \citenamefont {Xia}, \citenamefont {Naghavi},
  \citenamefont {Hegde}, \citenamefont {Hao}, \citenamefont {Goedecker},
  \citenamefont {Ozoli\ifmmode \mbox{\c{n}}\else
  \c{n}\fi{}\ifmmode~\check{s}\else \v{s}\fi{}},\ and\ \citenamefont
  {Wolverton}}]{Ba2BiAuLowLTC}%
  \BibitemOpen
  \bibfield  {author} {\bibinfo {author} {\bibfnamefont {J.}~\bibnamefont
  {He}}, \bibinfo {author} {\bibfnamefont {M.}~\bibnamefont {Amsler}}, \bibinfo
  {author} {\bibfnamefont {Y.}~\bibnamefont {Xia}}, \bibinfo {author}
  {\bibfnamefont {S.~S.}\ \bibnamefont {Naghavi}}, \bibinfo {author}
  {\bibfnamefont {V.~I.}\ \bibnamefont {Hegde}}, \bibinfo {author}
  {\bibfnamefont {S.}~\bibnamefont {Hao}}, \bibinfo {author} {\bibfnamefont
  {S.}~\bibnamefont {Goedecker}}, \bibinfo {author} {\bibfnamefont
  {V.}~\bibnamefont {Ozoli\ifmmode \mbox{\c{n}}\else
  \c{n}\fi{}\ifmmode~\check{s}\else \v{s}\fi{}}},\ and\ \bibinfo {author}
  {\bibfnamefont {C.}~\bibnamefont {Wolverton}},\ }\bibfield  {title} {\bibinfo
  {title} {Ultralow thermal conductivity in full-{H}eusler semiconductors},\
  }\href {https://doi.org/10.1103/PhysRevLett.117.046602} {\bibfield  {journal}
  {\bibinfo  {journal} {Phys. Rev. Lett.}\ }\textbf {\bibinfo {volume} {117}},\
  \bibinfo {pages} {046602} (\bibinfo {year} {2016})}\BibitemShut {NoStop}%
\bibitem [{\citenamefont {Park}\ \emph {et~al.}(2019)\citenamefont {Park},
  \citenamefont {Xia},\ and\ \citenamefont {Ozoli\ifmmode \mbox{\c{n}}\else
  \c{n}\fi{}\ifmmode~\check{s}\else \v{s}\fi{}}}]{Ba2BiAuLowLTCComp}%
  \BibitemOpen
  \bibfield  {author} {\bibinfo {author} {\bibfnamefont {J.}~\bibnamefont
  {Park}}, \bibinfo {author} {\bibfnamefont {Y.}~\bibnamefont {Xia}},\ and\
  \bibinfo {author} {\bibfnamefont {V.}~\bibnamefont {Ozoli\ifmmode
  \mbox{\c{n}}\else \c{n}\fi{}\ifmmode~\check{s}\else \v{s}\fi{}}},\ }\bibfield
   {title} {\bibinfo {title} {High thermoelectric power factor and efficiency
  from a highly dispersive band in $\mathrm{Ba}_{2}\mathrm{Bi}\mathrm{Au}$},\
  }\href {https://doi.org/10.1103/PhysRevApplied.11.014058} {\bibfield
  {journal} {\bibinfo  {journal} {Phys. Rev. Appl.}\ }\textbf {\bibinfo
  {volume} {11}},\ \bibinfo {pages} {014058} (\bibinfo {year}
  {2019})}\BibitemShut {NoStop}%
\bibitem [{\citenamefont {Wang}\ \emph {et~al.}(2021)\citenamefont {Wang},
  \citenamefont {Dai}, \citenamefont {Wang}, \citenamefont {Zhong},
  \citenamefont {Zhao},\ and\ \citenamefont {Meng}}]{Ba2SbAu}%
  \BibitemOpen
  \bibfield  {author} {\bibinfo {author} {\bibfnamefont {W.}~\bibnamefont
  {Wang}}, \bibinfo {author} {\bibfnamefont {Z.}~\bibnamefont {Dai}}, \bibinfo
  {author} {\bibfnamefont {X.}~\bibnamefont {Wang}}, \bibinfo {author}
  {\bibfnamefont {Q.}~\bibnamefont {Zhong}}, \bibinfo {author} {\bibfnamefont
  {Y.}~\bibnamefont {Zhao}},\ and\ \bibinfo {author} {\bibfnamefont
  {S.}~\bibnamefont {Meng}},\ }\bibfield  {title} {\bibinfo {title} {Low
  lattice thermal conductivity and high figure of merit in n-type doped
  full-{H}eusler compounds $\rm{X}_{2}\rm{YAu}$ {(X=Sr, Ba; Y=As, Sb)}},\
  }\href {https://doi.org/https://doi.org/10.1002/er.7154} {\bibfield
  {journal} {\bibinfo  {journal} {Int. J. Energy Res.}\ }\textbf {\bibinfo
  {volume} {45}},\ \bibinfo {pages} {20949} (\bibinfo {year}
  {2021})}\BibitemShut {NoStop}%
\bibitem [{\citenamefont {Jana}\ and\ \citenamefont
  {Biswas}(2018)}]{HighMassLowVelocity}%
  \BibitemOpen
  \bibfield  {author} {\bibinfo {author} {\bibfnamefont {M.~K.}\ \bibnamefont
  {Jana}}\ and\ \bibinfo {author} {\bibfnamefont {K.}~\bibnamefont {Biswas}},\
  }\bibfield  {title} {\bibinfo {title} {Crystalline solids with intrinsically
  low lattice thermal conductivity for thermoelectric energy conversion},\
  }\href {https://doi.org/10.1021/acsenergylett.8b00435} {\bibfield  {journal}
  {\bibinfo  {journal} {ACS Energy Lett.}\ }\textbf {\bibinfo {volume} {3}},\
  \bibinfo {pages} {1315} (\bibinfo {year} {2018})}\BibitemShut {NoStop}%
\bibitem [{\citenamefont {Wang}\ \emph
  {et~al.}(2022{\natexlab{b}})\citenamefont {Wang}, \citenamefont {Zhang},
  \citenamefont {Wang}, \citenamefont {Zhang},\ and\ \citenamefont
  {Wang}}]{Ba2AgSbLowLTC}%
  \BibitemOpen
  \bibfield  {author} {\bibinfo {author} {\bibfnamefont {S.-F.}\ \bibnamefont
  {Wang}}, \bibinfo {author} {\bibfnamefont {Z.-G.}\ \bibnamefont {Zhang}},
  \bibinfo {author} {\bibfnamefont {B.-T.}\ \bibnamefont {Wang}}, \bibinfo
  {author} {\bibfnamefont {J.-R.}\ \bibnamefont {Zhang}},\ and\ \bibinfo
  {author} {\bibfnamefont {F.-W.}\ \bibnamefont {Wang}},\ }\bibfield  {title}
  {\bibinfo {title} {Intrinsic ultralow lattice thermal conductivity in the
  full-{H}eusler compound $\mathrm{Ba}_{2}\mathrm{Ag}\mathrm{Sb}$},\ }\href
  {https://doi.org/10.1103/PhysRevApplied.17.034023} {\bibfield  {journal}
  {\bibinfo  {journal} {Phys. Rev. Applied}\ }\textbf {\bibinfo {volume}
  {17}},\ \bibinfo {pages} {034023} (\bibinfo {year}
  {2022}{\natexlab{b}})}\BibitemShut {NoStop}%
\bibitem [{\citenamefont {Berland}\ \emph {et~al.}(2021)\citenamefont
  {Berland}, \citenamefont {Løvvik},\ and\ \citenamefont
  {Tranås}}]{DiscardedGemsBerland}%
  \BibitemOpen
  \bibfield  {author} {\bibinfo {author} {\bibfnamefont {K.}~\bibnamefont
  {Berland}}, \bibinfo {author} {\bibfnamefont {O.~M.}\ \bibnamefont
  {Løvvik}},\ and\ \bibinfo {author} {\bibfnamefont {R.}~\bibnamefont
  {Tranås}},\ }\bibfield  {title} {\bibinfo {title} {Discarded gems:
  Thermoelectric performance of materials with band gap emerging at the
  hybrid-functional level},\ }\href {https://doi.org/10.1063/5.0058685}
  {\bibfield  {journal} {\bibinfo  {journal} {Appl. Phys. Lett.}\ }\textbf
  {\bibinfo {volume} {119}},\ \bibinfo {pages} {081902} (\bibinfo {year}
  {2021})}\BibitemShut {NoStop}%
\bibitem [{\citenamefont {Knoop}\ \emph {et~al.}(2020)\citenamefont {Knoop},
  \citenamefont {Purcell}, \citenamefont {Scheffler},\ and\ \citenamefont
  {Carbogno}}]{AnharmMeasureForMaterials}%
  \BibitemOpen
  \bibfield  {author} {\bibinfo {author} {\bibfnamefont {F.}~\bibnamefont
  {Knoop}}, \bibinfo {author} {\bibfnamefont {T.~A.~R.}\ \bibnamefont
  {Purcell}}, \bibinfo {author} {\bibfnamefont {M.}~\bibnamefont {Scheffler}},\
  and\ \bibinfo {author} {\bibfnamefont {C.}~\bibnamefont {Carbogno}},\
  }\bibfield  {title} {\bibinfo {title} {Anharmonicity measure for materials},\
  }\href {https://doi.org/10.1103/PhysRevMaterials.4.083809} {\bibfield
  {journal} {\bibinfo  {journal} {Phys. Rev. Mater.}\ }\textbf {\bibinfo
  {volume} {4}},\ \bibinfo {pages} {083809} (\bibinfo {year}
  {2020})}\BibitemShut {NoStop}%
\bibitem [{\citenamefont {Legrain}\ \emph {et~al.}(2018)\citenamefont
  {Legrain}, \citenamefont {Carrete}, \citenamefont {van Roekeghem},
  \citenamefont {Madsen},\ and\ \citenamefont
  {Mingo}}]{CarreteHHStabilityScreening}%
  \BibitemOpen
  \bibfield  {author} {\bibinfo {author} {\bibfnamefont {F.}~\bibnamefont
  {Legrain}}, \bibinfo {author} {\bibfnamefont {J.}~\bibnamefont {Carrete}},
  \bibinfo {author} {\bibfnamefont {A.}~\bibnamefont {van Roekeghem}}, \bibinfo
  {author} {\bibfnamefont {G.~K.}\ \bibnamefont {Madsen}},\ and\ \bibinfo
  {author} {\bibfnamefont {N.}~\bibnamefont {Mingo}},\ }\bibfield  {title}
  {\bibinfo {title} {Materials screening for the discovery of new
  half-{H}euslers: Machine learning versus ab initio methods},\ }\href
  {https://doi.org/10.1021/acs.jpcb.7b05296} {\bibfield  {journal} {\bibinfo
  {journal} {J. Phys. Chem. B}\ }\textbf {\bibinfo {volume} {122}},\ \bibinfo
  {pages} {625} (\bibinfo {year} {2018})},\ \bibinfo {note} {pMID:
  28742351}\BibitemShut {NoStop}%
\bibitem [{\citenamefont {Jaafreh}\ \emph {et~al.}(2022)\citenamefont
  {Jaafreh}, \citenamefont {{Yoo Seong}}, \citenamefont {Kim},\ and\
  \citenamefont {Hamad}}]{JaafrehZTDeepLearning}%
  \BibitemOpen
  \bibfield  {author} {\bibinfo {author} {\bibfnamefont {R.}~\bibnamefont
  {Jaafreh}}, \bibinfo {author} {\bibfnamefont {K.}~\bibnamefont {{Yoo
  Seong}}}, \bibinfo {author} {\bibfnamefont {J.-G.}\ \bibnamefont {Kim}},\
  and\ \bibinfo {author} {\bibfnamefont {K.}~\bibnamefont {Hamad}},\ }\bibfield
   {title} {\bibinfo {title} {A deep learning perspective into the
  figure-of-merit of thermoelectric materials},\ }\href
  {https://doi.org/https://doi.org/10.1016/j.matlet.2022.132299} {\bibfield
  {journal} {\bibinfo  {journal} {Mater. Lett.}\ }\textbf {\bibinfo {volume}
  {319}},\ \bibinfo {pages} {132299} (\bibinfo {year} {2022})}\BibitemShut
  {NoStop}%
\bibitem [{\citenamefont {Shulumba}\ \emph {et~al.}(2017)\citenamefont
  {Shulumba}, \citenamefont {Hellman},\ and\ \citenamefont
  {Minnich}}]{shulumbaIntrinsicLocalized}%
  \BibitemOpen
  \bibfield  {author} {\bibinfo {author} {\bibfnamefont {N.}~\bibnamefont
  {Shulumba}}, \bibinfo {author} {\bibfnamefont {O.}~\bibnamefont {Hellman}},\
  and\ \bibinfo {author} {\bibfnamefont {A.~J.}\ \bibnamefont {Minnich}},\
  }\bibfield  {title} {\bibinfo {title} {Intrinsic localized mode and low
  thermal conductivity of {PbSe}},\ }\href
  {https://doi.org/10.1103/PhysRevB.95.014302} {\bibfield  {journal} {\bibinfo
  {journal} {Phys. Rev. B}\ }\textbf {\bibinfo {volume} {95}},\ \bibinfo
  {pages} {014302} (\bibinfo {year} {2017})}\BibitemShut {NoStop}%
\bibitem [{\citenamefont {{Anderson}}(1963)}]{AndersonSimplifiedMethodDebye}%
  \BibitemOpen
  \bibfield  {author} {\bibinfo {author} {\bibfnamefont {O.}~\bibnamefont
  {{Anderson}}},\ }\bibfield  {title} {\bibinfo {title} {{A simplified method
  for calculating the debye temperature from elastic constants}},\ }\href
  {https://doi.org/10.1016/0022-3697(63)90067-2} {\bibfield  {journal}
  {\bibinfo  {journal} {J. Phys. Chem. Solids}\ }\textbf {\bibinfo {volume}
  {24}},\ \bibinfo {pages} {909} (\bibinfo {year} {1963})}\BibitemShut
  {NoStop}%
\bibitem [{\citenamefont {Kresse}\ and\ \citenamefont
  {Hafner}(1993)}]{kresseInitioMolecularDynamics1993}%
  \BibitemOpen
  \bibfield  {author} {\bibinfo {author} {\bibfnamefont {G.}~\bibnamefont
  {Kresse}}\ and\ \bibinfo {author} {\bibfnamefont {J.}~\bibnamefont
  {Hafner}},\ }\bibfield  {title} {\bibinfo {title} {\textit{{Ab} initio}
  molecular dynamics for liquid metals},\ }\href
  {https://doi.org/10.1103/PhysRevB.47.558} {\bibfield  {journal} {\bibinfo
  {journal} {Phys. Rev. B}\ }\textbf {\bibinfo {volume} {47}},\ \bibinfo
  {pages} {558} (\bibinfo {year} {1993})}\BibitemShut {NoStop}%
\bibitem [{\citenamefont {Kresse}\ and\ \citenamefont
  {Furthmüller}(1996{\natexlab{a}})}]{kresseEfficientIterativeSchemes1996}%
  \BibitemOpen
  \bibfield  {author} {\bibinfo {author} {\bibfnamefont {G.}~\bibnamefont
  {Kresse}}\ and\ \bibinfo {author} {\bibfnamefont {J.}~\bibnamefont
  {Furthmüller}},\ }\bibfield  {title} {\bibinfo {title} {Efficient iterative
  schemes for \textit{ab initio} total-energy calculations using a plane-wave
  basis set},\ }\href {https://doi.org/10.1103/PhysRevB.54.11169} {\bibfield
  {journal} {\bibinfo  {journal} {Phys. Rev. B}\ }\textbf {\bibinfo {volume}
  {54}},\ \bibinfo {pages} {11169} (\bibinfo {year}
  {1996}{\natexlab{a}})}\BibitemShut {NoStop}%
\bibitem [{\citenamefont {Kresse}\ and\ \citenamefont
  {Furthmüller}(1996{\natexlab{b}})}]{kresseEfficiencyAbinitioTotal1996}%
  \BibitemOpen
  \bibfield  {author} {\bibinfo {author} {\bibfnamefont {G.}~\bibnamefont
  {Kresse}}\ and\ \bibinfo {author} {\bibfnamefont {J.}~\bibnamefont
  {Furthmüller}},\ }\bibfield  {title} {\bibinfo {title} {Efficiency of
  ab-initio total energy calculations for metals and semiconductors using a
  plane-wave basis set},\ }\href {https://doi.org/10.1016/0927-0256(96)00008-0}
  {\bibfield  {journal} {\bibinfo  {journal} {Comput. Mater. Sci.}\ }\textbf
  {\bibinfo {volume} {6}},\ \bibinfo {pages} {15} (\bibinfo {year}
  {1996}{\natexlab{b}})}\BibitemShut {NoStop}%
\bibitem [{\citenamefont {Perdew}\ \emph {et~al.}(2008)\citenamefont {Perdew},
  \citenamefont {Ruzsinszky}, \citenamefont {Csonka}, \citenamefont {Vydrov},
  \citenamefont {Scuseria}, \citenamefont {Constantin}, \citenamefont {Zhou},\
  and\ \citenamefont {Burke}}]{perdewRestoringDensityGradientExpansion2008}%
  \BibitemOpen
  \bibfield  {author} {\bibinfo {author} {\bibfnamefont {J.~P.}\ \bibnamefont
  {Perdew}}, \bibinfo {author} {\bibfnamefont {A.}~\bibnamefont {Ruzsinszky}},
  \bibinfo {author} {\bibfnamefont {G.~I.}\ \bibnamefont {Csonka}}, \bibinfo
  {author} {\bibfnamefont {O.~A.}\ \bibnamefont {Vydrov}}, \bibinfo {author}
  {\bibfnamefont {G.~E.}\ \bibnamefont {Scuseria}}, \bibinfo {author}
  {\bibfnamefont {L.~A.}\ \bibnamefont {Constantin}}, \bibinfo {author}
  {\bibfnamefont {X.}~\bibnamefont {Zhou}},\ and\ \bibinfo {author}
  {\bibfnamefont {K.}~\bibnamefont {Burke}},\ }\bibfield  {title} {\bibinfo
  {title} {Restoring the {density}-{gradient} {expansion} for {exchange} in
  {solids} and {surfaces}},\ }\href
  {https://doi.org/10.1103/PhysRevLett.100.136406} {\bibfield  {journal}
  {\bibinfo  {journal} {Phys. Rev. Lett.}\ }\textbf {\bibinfo {volume} {100}},\
  \bibinfo {pages} {136406} (\bibinfo {year} {2008})}\BibitemShut {NoStop}%
\bibitem [{\citenamefont {Csonka}\ \emph {et~al.}(2009)\citenamefont {Csonka},
  \citenamefont {Perdew}, \citenamefont {Ruzsinszky}, \citenamefont
  {Philipsen}, \citenamefont {Lebègue}, \citenamefont {Paier}, \citenamefont
  {Vydrov},\ and\ \citenamefont
  {Ángyán}}]{csonkaAssessingPerformanceRecent2009}%
  \BibitemOpen
  \bibfield  {author} {\bibinfo {author} {\bibfnamefont {G.~I.}\ \bibnamefont
  {Csonka}}, \bibinfo {author} {\bibfnamefont {J.~P.}\ \bibnamefont {Perdew}},
  \bibinfo {author} {\bibfnamefont {A.}~\bibnamefont {Ruzsinszky}}, \bibinfo
  {author} {\bibfnamefont {P.~H.~T.}\ \bibnamefont {Philipsen}}, \bibinfo
  {author} {\bibfnamefont {S.}~\bibnamefont {Lebègue}}, \bibinfo {author}
  {\bibfnamefont {J.}~\bibnamefont {Paier}}, \bibinfo {author} {\bibfnamefont
  {O.~A.}\ \bibnamefont {Vydrov}},\ and\ \bibinfo {author} {\bibfnamefont
  {J.~G.}\ \bibnamefont {Ángyán}},\ }\bibfield  {title} {\bibinfo {title}
  {Assessing the performance of recent density functionals for bulk solids},\
  }\href {https://doi.org/10.1103/PhysRevB.79.155107} {\bibfield  {journal}
  {\bibinfo  {journal} {Phys. Rev. B}\ }\textbf {\bibinfo {volume} {79}},\
  \bibinfo {pages} {155107} (\bibinfo {year} {2009})}\BibitemShut {NoStop}%
\bibitem [{\citenamefont {Berland}\ and\ \citenamefont
  {Hyldgaard}(2014)}]{cx1}%
  \BibitemOpen
  \bibfield  {author} {\bibinfo {author} {\bibfnamefont {K.}~\bibnamefont
  {Berland}}\ and\ \bibinfo {author} {\bibfnamefont {P.}~\bibnamefont
  {Hyldgaard}},\ }\bibfield  {title} {\bibinfo {title} {Exchange functional
  that tests the robustness of the plasmon description of the van der {W}aals
  density functional},\ }\href {https://doi.org/10.1103/PhysRevB.89.035412}
  {\bibfield  {journal} {\bibinfo  {journal} {Phys. Rev. B}\ }\textbf {\bibinfo
  {volume} {89}},\ \bibinfo {pages} {035412} (\bibinfo {year}
  {2014})}\BibitemShut {NoStop}%
\bibitem [{\citenamefont {Berland}\ \emph {et~al.}(2014)\citenamefont
  {Berland}, \citenamefont {Arter}, \citenamefont {Cooper}, \citenamefont
  {Lee}, \citenamefont {Lundqvist}, \citenamefont {Schröder}, \citenamefont
  {Thonhauser},\ and\ \citenamefont {Hyldgaard}}]{cx2}%
  \BibitemOpen
  \bibfield  {author} {\bibinfo {author} {\bibfnamefont {K.}~\bibnamefont
  {Berland}}, \bibinfo {author} {\bibfnamefont {C.~A.}\ \bibnamefont {Arter}},
  \bibinfo {author} {\bibfnamefont {V.~R.}\ \bibnamefont {Cooper}}, \bibinfo
  {author} {\bibfnamefont {K.}~\bibnamefont {Lee}}, \bibinfo {author}
  {\bibfnamefont {B.~I.}\ \bibnamefont {Lundqvist}}, \bibinfo {author}
  {\bibfnamefont {E.}~\bibnamefont {Schröder}}, \bibinfo {author}
  {\bibfnamefont {T.}~\bibnamefont {Thonhauser}},\ and\ \bibinfo {author}
  {\bibfnamefont {P.}~\bibnamefont {Hyldgaard}},\ }\bibfield  {title} {\bibinfo
  {title} {van der {W}aals density functionals built upon the electron-gas
  tradition: Facing the challenge of competing interactions},\ }\href
  {https://doi.org/10.1063/1.4871731} {\bibfield  {journal} {\bibinfo
  {journal} {J. Chem. Phys.}\ }\textbf {\bibinfo {volume} {140}},\ \bibinfo
  {pages} {18A539} (\bibinfo {year} {2014})}\BibitemShut {NoStop}%
\bibitem [{\citenamefont {Frostenson}\ \emph {et~al.}(2022)\citenamefont
  {Frostenson}, \citenamefont {Granhed}, \citenamefont {Shukla}, \citenamefont
  {Olsson}, \citenamefont {Schr\"{o}der},\ and\ \citenamefont
  {Hyldgaard}}]{VDWAccuracyPerovskite}%
  \BibitemOpen
  \bibfield  {author} {\bibinfo {author} {\bibfnamefont {C.~M.}\ \bibnamefont
  {Frostenson}}, \bibinfo {author} {\bibfnamefont {E.~J.}\ \bibnamefont
  {Granhed}}, \bibinfo {author} {\bibfnamefont {V.}~\bibnamefont {Shukla}},
  \bibinfo {author} {\bibfnamefont {P.~A.~T.}\ \bibnamefont {Olsson}}, \bibinfo
  {author} {\bibfnamefont {E.}~\bibnamefont {Schr\"{o}der}},\ and\ \bibinfo
  {author} {\bibfnamefont {P.}~\bibnamefont {Hyldgaard}},\ }\bibfield  {title}
  {\bibinfo {title} {Hard and soft materials: putting consistent van der waals
  density functionals to work},\ }\href
  {https://doi.org/10.1088/2516-1075/ac4468} {\bibfield  {journal} {\bibinfo
  {journal} {Electron. Struct.}\ }\textbf {\bibinfo {volume} {4}},\ \bibinfo
  {pages} {014001} (\bibinfo {year} {2022})}\BibitemShut {NoStop}%
\bibitem [{\citenamefont {Lindroth}\ and\ \citenamefont
  {Erhart}(2016)}]{CXLatticeParameter}%
  \BibitemOpen
  \bibfield  {author} {\bibinfo {author} {\bibfnamefont {D.~O.}\ \bibnamefont
  {Lindroth}}\ and\ \bibinfo {author} {\bibfnamefont {P.}~\bibnamefont
  {Erhart}},\ }\bibfield  {title} {\bibinfo {title} {Thermal transport in van
  der waals solids from first-principles calculations},\ }\href
  {https://doi.org/10.1103/PhysRevB.94.115205} {\bibfield  {journal} {\bibinfo
  {journal} {Phys. Rev. B}\ }\textbf {\bibinfo {volume} {94}},\ \bibinfo
  {pages} {115205} (\bibinfo {year} {2016})}\BibitemShut {NoStop}%
\bibitem [{\citenamefont {Gharaee}\ \emph {et~al.}(2017)\citenamefont
  {Gharaee}, \citenamefont {Erhart},\ and\ \citenamefont
  {Hyldgaard}}]{VDWAccuracy}%
  \BibitemOpen
  \bibfield  {author} {\bibinfo {author} {\bibfnamefont {L.}~\bibnamefont
  {Gharaee}}, \bibinfo {author} {\bibfnamefont {P.}~\bibnamefont {Erhart}},\
  and\ \bibinfo {author} {\bibfnamefont {P.}~\bibnamefont {Hyldgaard}},\
  }\bibfield  {title} {\bibinfo {title} {Finite-temperature properties of
  nonmagnetic transition metals: Comparison of the performance of
  constraint-based semilocal and nonlocal functionals},\ }\href
  {https://doi.org/10.1103/PhysRevB.95.085147} {\bibfield  {journal} {\bibinfo
  {journal} {Phys. Rev. B}\ }\textbf {\bibinfo {volume} {95}},\ \bibinfo
  {pages} {085147} (\bibinfo {year} {2017})}\BibitemShut {NoStop}%
\bibitem [{\citenamefont {He}\ \emph {et~al.}(2014)\citenamefont {He},
  \citenamefont {Liu}, \citenamefont {Hautier}, \citenamefont {Oliveira},
  \citenamefont {Marques}, \citenamefont {Vila}, \citenamefont {Rehr},
  \citenamefont {Rignanese},\ and\ \citenamefont {Zhou}}]{PBESolAccuracy}%
  \BibitemOpen
  \bibfield  {author} {\bibinfo {author} {\bibfnamefont {L.}~\bibnamefont
  {He}}, \bibinfo {author} {\bibfnamefont {F.}~\bibnamefont {Liu}}, \bibinfo
  {author} {\bibfnamefont {G.}~\bibnamefont {Hautier}}, \bibinfo {author}
  {\bibfnamefont {M.~J.~T.}\ \bibnamefont {Oliveira}}, \bibinfo {author}
  {\bibfnamefont {M.~A.~L.}\ \bibnamefont {Marques}}, \bibinfo {author}
  {\bibfnamefont {F.~D.}\ \bibnamefont {Vila}}, \bibinfo {author}
  {\bibfnamefont {J.~J.}\ \bibnamefont {Rehr}}, \bibinfo {author}
  {\bibfnamefont {G.-M.}\ \bibnamefont {Rignanese}},\ and\ \bibinfo {author}
  {\bibfnamefont {A.}~\bibnamefont {Zhou}},\ }\bibfield  {title} {\bibinfo
  {title} {Accuracy of generalized gradient approximation functionals for
  density-functional perturbation theory calculations},\ }\href
  {https://doi.org/10.1103/PhysRevB.89.064305} {\bibfield  {journal} {\bibinfo
  {journal} {Phys. Rev. B}\ }\textbf {\bibinfo {volume} {89}},\ \bibinfo
  {pages} {064305} (\bibinfo {year} {2014})}\BibitemShut {NoStop}%
\bibitem [{\citenamefont {Noack}\ \emph {et~al.}(2020)\citenamefont {Noack},
  \citenamefont {Doerk}, \citenamefont {Li}, \citenamefont {Streit},
  \citenamefont {Vaia}, \citenamefont {Yager},\ and\ \citenamefont
  {Fukuto}}]{GPRMaterialsDiscovery}%
  \BibitemOpen
  \bibfield  {author} {\bibinfo {author} {\bibfnamefont {M.~M.}\ \bibnamefont
  {Noack}}, \bibinfo {author} {\bibfnamefont {G.~S.}\ \bibnamefont {Doerk}},
  \bibinfo {author} {\bibfnamefont {R.}~\bibnamefont {Li}}, \bibinfo {author}
  {\bibfnamefont {J.~K.}\ \bibnamefont {Streit}}, \bibinfo {author}
  {\bibfnamefont {R.~A.}\ \bibnamefont {Vaia}}, \bibinfo {author}
  {\bibfnamefont {K.~G.}\ \bibnamefont {Yager}},\ and\ \bibinfo {author}
  {\bibfnamefont {M.}~\bibnamefont {Fukuto}},\ }\bibfield  {title} {\bibinfo
  {title} {Autonomous materials discovery driven by gaussian process regression
  with inhomogeneous measurement noise and anisotropic kernels},\ }\bibfield
  {journal} {\bibinfo  {journal} {Sci. Rep.}\ }\textbf {\bibinfo {volume}
  {10}},\ \href {https://doi.org/10.1038/s41598-020-74394-1}
  {10.1038/s41598-020-74394-1} (\bibinfo {year} {2020})\BibitemShut {NoStop}%
\bibitem [{\citenamefont {Goodarzi}\ and\ \citenamefont
  {Bahramian}(2021)}]{GPRAerogels}%
  \BibitemOpen
  \bibfield  {author} {\bibinfo {author} {\bibfnamefont {B.}~\bibnamefont
  {Goodarzi}}\ and\ \bibinfo {author} {\bibfnamefont {A.~R.}\ \bibnamefont
  {Bahramian}},\ }\bibfield  {title} {\bibinfo {title} {Applying machine
  learning for predicting thermal conductivity coefficient of polymeric
  aerogels},\ }\href@noop {} {\bibfield  {journal} {\bibinfo  {journal} {J.
  Therm. Anal. Calorim.}\ ,\ \bibinfo {pages} {1 }} (\bibinfo {year}
  {2021})}\BibitemShut {NoStop}%
\bibitem [{\citenamefont {Deringer}\ \emph {et~al.}(2021)\citenamefont
  {Deringer}, \citenamefont {Bartók}, \citenamefont {Bernstein}, \citenamefont
  {Wilkins}, \citenamefont {Ceriotti},\ and\ \citenamefont
  {Csányi}}]{GPRForMaterialsAndMolecules}%
  \BibitemOpen
  \bibfield  {author} {\bibinfo {author} {\bibfnamefont {V.~L.}\ \bibnamefont
  {Deringer}}, \bibinfo {author} {\bibfnamefont {A.~P.}\ \bibnamefont
  {Bartók}}, \bibinfo {author} {\bibfnamefont {N.}~\bibnamefont {Bernstein}},
  \bibinfo {author} {\bibfnamefont {D.~M.}\ \bibnamefont {Wilkins}}, \bibinfo
  {author} {\bibfnamefont {M.}~\bibnamefont {Ceriotti}},\ and\ \bibinfo
  {author} {\bibfnamefont {G.}~\bibnamefont {Csányi}},\ }\bibfield  {title}
  {\bibinfo {title} {Gaussian process regression for materials and molecules},\
  }\href {https://doi.org/10.1021/acs.chemrev.1c00022} {\bibfield  {journal}
  {\bibinfo  {journal} {Chem. Rev.}\ }\textbf {\bibinfo {volume} {121}},\
  \bibinfo {pages} {10073} (\bibinfo {year} {2021})}\BibitemShut {NoStop}%
\bibitem [{\citenamefont {Seko}\ \emph {et~al.}(2017)\citenamefont {Seko},
  \citenamefont {Hayashi}, \citenamefont {Nakayama}, \citenamefont
  {Takahashi},\ and\ \citenamefont {Tanaka}}]{GPRCompundRepresentation}%
  \BibitemOpen
  \bibfield  {author} {\bibinfo {author} {\bibfnamefont {A.}~\bibnamefont
  {Seko}}, \bibinfo {author} {\bibfnamefont {H.}~\bibnamefont {Hayashi}},
  \bibinfo {author} {\bibfnamefont {K.}~\bibnamefont {Nakayama}}, \bibinfo
  {author} {\bibfnamefont {A.}~\bibnamefont {Takahashi}},\ and\ \bibinfo
  {author} {\bibfnamefont {I.}~\bibnamefont {Tanaka}},\ }\bibfield  {title}
  {\bibinfo {title} {Representation of compounds for machine-learning
  prediction of physical properties},\ }\href
  {https://doi.org/10.1103/PhysRevB.95.144110} {\bibfield  {journal} {\bibinfo
  {journal} {Phys. Rev. B}\ }\textbf {\bibinfo {volume} {95}},\ \bibinfo
  {pages} {144110} (\bibinfo {year} {2017})}\BibitemShut {NoStop}%
\bibitem [{\citenamefont {Bisbo}\ and\ \citenamefont
  {Hammer}(2020)}]{GPRSurrogateModel}%
  \BibitemOpen
  \bibfield  {author} {\bibinfo {author} {\bibfnamefont {M.~K.}\ \bibnamefont
  {Bisbo}}\ and\ \bibinfo {author} {\bibfnamefont {B.}~\bibnamefont {Hammer}},\
  }\bibfield  {title} {\bibinfo {title} {Efficient global structure
  optimization with a machine-learned surrogate model},\ }\href
  {https://doi.org/10.1103/PhysRevLett.124.086102} {\bibfield  {journal}
  {\bibinfo  {journal} {Phys. Rev. Lett.}\ }\textbf {\bibinfo {volume} {124}},\
  \bibinfo {pages} {086102} (\bibinfo {year} {2020})}\BibitemShut {NoStop}%
\bibitem [{\citenamefont {Wei}\ \emph {et~al.}(2022)\citenamefont {Wei},
  \citenamefont {Bao},\ and\ \citenamefont
  {Ruan}}]{PerspectiveMLThermalTransport}%
  \BibitemOpen
  \bibfield  {author} {\bibinfo {author} {\bibfnamefont {H.}~\bibnamefont
  {Wei}}, \bibinfo {author} {\bibfnamefont {H.}~\bibnamefont {Bao}},\ and\
  \bibinfo {author} {\bibfnamefont {X.}~\bibnamefont {Ruan}},\ }\bibfield
  {title} {\bibinfo {title} {Perspective: Predicting and optimizing thermal
  transport properties with machine learning methods},\ }\href
  {https://doi.org/https://doi.org/10.1016/j.egyai.2022.100153} {\bibfield
  {journal} {\bibinfo  {journal} {Energy AI}\ }\textbf {\bibinfo {volume}
  {8}},\ \bibinfo {pages} {100153} (\bibinfo {year} {2022})}\BibitemShut
  {NoStop}%
\bibitem [{\citenamefont {Pedregosa}\ \emph {et~al.}(2011)\citenamefont
  {Pedregosa}, \citenamefont {Varoquaux}, \citenamefont {Gramfort},
  \citenamefont {Michel}, \citenamefont {Thirion}, \citenamefont {Grisel},
  \citenamefont {Blondel}, \citenamefont {Prettenhofer}, \citenamefont {Weiss},
  \citenamefont {Dubourg}, \citenamefont {Vanderplas}, \citenamefont {Passos},\
  and\ \citenamefont {Cournapeau}}]{pedregosaScikitlearnMachineLearning2011}%
  \BibitemOpen
  \bibfield  {author} {\bibinfo {author} {\bibfnamefont {F.}~\bibnamefont
  {Pedregosa}}, \bibinfo {author} {\bibfnamefont {G.}~\bibnamefont
  {Varoquaux}}, \bibinfo {author} {\bibfnamefont {A.}~\bibnamefont {Gramfort}},
  \bibinfo {author} {\bibfnamefont {V.}~\bibnamefont {Michel}}, \bibinfo
  {author} {\bibfnamefont {B.}~\bibnamefont {Thirion}}, \bibinfo {author}
  {\bibfnamefont {O.}~\bibnamefont {Grisel}}, \bibinfo {author} {\bibfnamefont
  {M.}~\bibnamefont {Blondel}}, \bibinfo {author} {\bibfnamefont
  {P.}~\bibnamefont {Prettenhofer}}, \bibinfo {author} {\bibfnamefont
  {R.}~\bibnamefont {Weiss}}, \bibinfo {author} {\bibfnamefont
  {V.}~\bibnamefont {Dubourg}}, \bibinfo {author} {\bibfnamefont
  {J.}~\bibnamefont {Vanderplas}}, \bibinfo {author} {\bibfnamefont
  {A.}~\bibnamefont {Passos}},\ and\ \bibinfo {author} {\bibfnamefont
  {D.}~\bibnamefont {Cournapeau}},\ }\bibfield  {title} {\bibinfo {title}
  {Scikit-learn: {machine} {learning} in {Python}},\ }\href@noop {} {\bibfield
  {journal} {\bibinfo  {journal} {J. Mach. Learn. Res.}\ }\textbf {\bibinfo
  {volume} {12}},\ \bibinfo {pages} {2825} (\bibinfo {year}
  {2011})}\BibitemShut {NoStop}%
\bibitem [{\citenamefont {Pudil}\ \emph {et~al.}(1994)\citenamefont {Pudil},
  \citenamefont {Novovičová},\ and\ \citenamefont {Kittler}}]{SFFS}%
  \BibitemOpen
  \bibfield  {author} {\bibinfo {author} {\bibfnamefont {P.}~\bibnamefont
  {Pudil}}, \bibinfo {author} {\bibfnamefont {J.}~\bibnamefont
  {Novovičová}},\ and\ \bibinfo {author} {\bibfnamefont {J.}~\bibnamefont
  {Kittler}},\ }\bibfield  {title} {\bibinfo {title} {Floating search methods
  in feature selection},\ }\href
  {https://doi.org/https://doi.org/10.1016/0167-8655(94)90127-9} {\bibfield
  {journal} {\bibinfo  {journal} {Pattern Recognit. Lett.}\ }\textbf {\bibinfo
  {volume} {15}},\ \bibinfo {pages} {1119} (\bibinfo {year}
  {1994})}\BibitemShut {NoStop}%
\bibitem [{\citenamefont {Raschka}(2018)}]{raschkaMLxtendProvidingMachine2018}%
  \BibitemOpen
  \bibfield  {author} {\bibinfo {author} {\bibfnamefont {S.}~\bibnamefont
  {Raschka}},\ }\bibfield  {title} {\bibinfo {title} {{MLxtend}: {Providing}
  machine learning and data science utilities and extensions to {Python}’s
  scientific computing stack},\ }\href {https://doi.org/10.21105/joss.00638}
  {\bibfield  {journal} {\bibinfo  {journal} {J. Open Source Softw.}\ }\textbf
  {\bibinfo {volume} {3}},\ \bibinfo {pages} {638} (\bibinfo {year}
  {2018})}\BibitemShut {NoStop}%
\bibitem [{\citenamefont {Ganose}\ and\ \citenamefont {Jain}(2019)}]{robocrys}%
  \BibitemOpen
  \bibfield  {author} {\bibinfo {author} {\bibfnamefont {A.~M.}\ \bibnamefont
  {Ganose}}\ and\ \bibinfo {author} {\bibfnamefont {A.}~\bibnamefont {Jain}},\
  }\bibfield  {title} {\bibinfo {title} {Robocrystallographer: automated
  crystal structure text descriptions and analysis},\ }\href
  {https://doi.org/10.1557/mrc.2019.94} {\bibfield  {journal} {\bibinfo
  {journal} {{MRS} Commun.}\ }\textbf {\bibinfo {volume} {9}},\ \bibinfo
  {pages} {874} (\bibinfo {year} {2019})}\BibitemShut {NoStop}%
\bibitem [{\citenamefont {Schwerdtfeger}\ and\ \citenamefont
  {Nagle}(2019)}]{dipolepolarizability}%
  \BibitemOpen
  \bibfield  {author} {\bibinfo {author} {\bibfnamefont {P.}~\bibnamefont
  {Schwerdtfeger}}\ and\ \bibinfo {author} {\bibfnamefont {J.~K.}\ \bibnamefont
  {Nagle}},\ }\bibfield  {title} {\bibinfo {title} {2018 table of static dipole
  polarizabilities of the neutral elements in the periodic table},\ }\href
  {https://doi.org/10.1080/00268976.2018.1535143} {\bibfield  {journal}
  {\bibinfo  {journal} {Mol. Phys.}\ }\textbf {\bibinfo {volume} {117}},\
  \bibinfo {pages} {1200} (\bibinfo {year} {2019})}\BibitemShut {NoStop}%
\bibitem [{\citenamefont {Voronoi}(1908)}]{VoronoiOriginal}%
  \BibitemOpen
  \bibfield  {author} {\bibinfo {author} {\bibfnamefont {G.}~\bibnamefont
  {Voronoi}},\ }\bibfield  {title} {\bibinfo {title} {Nouvelles applications
  des paramètres continus à la théorie des formes quadratiques. deuxième
  mémoire. recherches sur les parallélloèdres primitifs.},\ }\href
  {https://doi.org/doi:10.1515/crll.1908.134.198} {\bibfield  {journal}
  {\bibinfo  {journal} {J. Reine Angew. Math.}\ }\textbf {\bibinfo {volume}
  {1908}},\ \bibinfo {pages} {198} (\bibinfo {year} {1908})}\BibitemShut
  {NoStop}%
\bibitem [{\citenamefont {Ward}\ \emph {et~al.}(2017)\citenamefont {Ward},
  \citenamefont {Liu}, \citenamefont {Krishna}, \citenamefont {Hegde},
  \citenamefont {Agrawal}, \citenamefont {Choudhary},\ and\ \citenamefont
  {Wolverton}}]{VoronoiAsMlFeatures}%
  \BibitemOpen
  \bibfield  {author} {\bibinfo {author} {\bibfnamefont {L.}~\bibnamefont
  {Ward}}, \bibinfo {author} {\bibfnamefont {R.}~\bibnamefont {Liu}}, \bibinfo
  {author} {\bibfnamefont {A.}~\bibnamefont {Krishna}}, \bibinfo {author}
  {\bibfnamefont {V.~I.}\ \bibnamefont {Hegde}}, \bibinfo {author}
  {\bibfnamefont {A.}~\bibnamefont {Agrawal}}, \bibinfo {author} {\bibfnamefont
  {A.}~\bibnamefont {Choudhary}},\ and\ \bibinfo {author} {\bibfnamefont
  {C.}~\bibnamefont {Wolverton}},\ }\bibfield  {title} {\bibinfo {title}
  {Including crystal structure attributes in machine learning models of
  formation energies via voronoi tessellations},\ }\href
  {https://doi.org/10.1103/PhysRevB.96.024104} {\bibfield  {journal} {\bibinfo
  {journal} {Phys. Rev. B}\ }\textbf {\bibinfo {volume} {96}},\ \bibinfo
  {pages} {024104} (\bibinfo {year} {2017})}\BibitemShut {NoStop}%
\bibitem [{\citenamefont {Menon}\ \emph {et~al.}(2019)\citenamefont {Menon},
  \citenamefont {Leines},\ and\ \citenamefont {Rogal}}]{pyscal}%
  \BibitemOpen
  \bibfield  {author} {\bibinfo {author} {\bibfnamefont {S.}~\bibnamefont
  {Menon}}, \bibinfo {author} {\bibfnamefont {G.~D.}\ \bibnamefont {Leines}},\
  and\ \bibinfo {author} {\bibfnamefont {J.}~\bibnamefont {Rogal}},\ }\bibfield
   {title} {\bibinfo {title} {pyscal: A python module for structural analysis
  of atomic environments},\ }\href {https://doi.org/10.21105/joss.01824}
  {\bibfield  {journal} {\bibinfo  {journal} {Journal of Open Source Software}\
  }\textbf {\bibinfo {volume} {4}},\ \bibinfo {pages} {1824} (\bibinfo {year}
  {2019})}\BibitemShut {NoStop}%
\end{thebibliography}%

\begin{acknowledgments}
The computations in this work were funded by The Norwegian e-infrastructure for research and education, Sigma2, through grants No. nn9711k and nn2615k. This work is part of the Allotherm project (Project No. 314778) supported by the Research Council of Norway. 
\end{acknowledgments}

\end{document}